\documentclass[aps,prb,twocolumn,nofootinbib,superscriptaddress,10pt]{revtex4-2}
\pdfoutput=1 
\usepackage[utf8]{inputenc}
\usepackage[english]{babel}
\usepackage[T1]{fontenc}
\usepackage[pdftex]{graphicx}
\usepackage{amsmath}
\usepackage{cancel}
\usepackage{amsthm}
\usepackage{amssymb}
\usepackage{bbm}
\usepackage{braket}

\newcommand{\TTbar}{\texorpdfstring{$T\!\bar{T}$}{T\bar{T}}}

\usepackage{xcolor}
\usepackage{dcolumn}
\usepackage{bbm}
\usepackage{bbold}
\usepackage[normalem]{ulem}
\usepackage{footmisc}
\usepackage{xcolor}
\usepackage{booktabs}
\usepackage{comment}
\usepackage{multirow}
\usepackage[colorlinks=true,linkcolor=blue,citecolor=blue,urlcolor=blue,unicode]{hyperref}

\newcommand{\Tr}{\mathrm{Tr}}

\begin{document}
\title{Quasiconservation Laws and Suppressed Transport in Weakly Interacting Localized Models}
\author{Jessica K. Jiang}
\author{Federica M. Surace}
\author{Olexei I. Motrunich}
\affiliation{Department of Physics and Institute for Quantum Information and Matter,
California Institute of Technology, Pasadena, California 91125, USA}
\date{\today}

\begin{abstract}
The stability of localization in the presence of interactions remains an open problem, with finite-size effects posing significant challenges to numerical studies.
In this work, we investigate the perturbative stability of noninteracting localization under weak interactions, which allows us to analyze much larger system sizes. 
Focusing on disordered Anderson and quasiperiodic Aubry-André models in one dimension, 
and using the adiabatic gauge potential (AGP) at first order in perturbation theory, we compute first-order corrections to noninteracting local integrals of motion (LIOMs).
We find that for at least an $O(1)$ fraction of the LIOMs, the corrections are well-controlled and converge at large system sizes, while others suffer from resonances.
Additionally, we introduce and study the charge-transport capacity of this weakly interacting model.
To first order, we find that the charge transport capacity remains bounded in the presence of interactions. 
Taken together, these results demonstrate that localization is perturbatively stable to weak interactions at first order, implying that, at the very least, localization persists for parametrically long times in the inverse interaction strength.
We expect this perturbative stability to extend to all orders at sufficiently strong disorder, where the localization length is short, representing the true localized phase.
Conversely, our findings suggest that the previously proposed interaction-induced avalanche instability, namely in the weakly localized regime of the Anderson and Aubry-André models, is a more subtle phenomenon arising only at higher orders in perturbation theory or through nonperturbative effects.
\end{abstract}

\maketitle

\section{Introduction}
One of the most profound predictions of classical statistical mechanics is that the microstates of an isolated system evolving under its own dynamics will, over time, approach an equilibrium ensemble characterized by maximal entropy. 
This basic process, which underpins much of our intuition about natural processes, becomes notably nontrivial to characterize when quantum mechanics enters the picture.
The Eigenstate Thermalization Hypothesis (ETH) offers one resolution, proposing that individual eigenstates of a many-body quantum Hamiltonian yield expectation values of local observables consistent with those predicted by statistical mechanics for a thermal ensemble~\cite{deutsch_quantum_1991, srednicki_chaos_1994}.
However, violations of the ETH are possible.
In particular, interacting quantum systems, subject to strong quenched disorder, give rise to the many-body localized (MBL) phase, which is the only known stable phase of matter that avoids thermalization.
In this phase, the system avoids thermalization by acting as a perfect insulator, even at nonzero temperatures. For a more detailed overview of MBL, see reviews \cite{nandkishore_many-body_2015, abanin_colloquium_2019,sierant_many-body_2025}.

The MBL phase is an exciting phase of matter whose emergent integrability protects quantum order~\cite{huse_localization-protected_2013}.
Much of the phenomenology of MBL follows from the conjecture that an MBL Hamiltonian can be diagonalized in terms of a set of commuting operators known as local integrals of motion (LIOMs) \cite{huse_phenomenology_2014}.
This emergent integrability also gives rise to several key properties unique to MBL, such as zero DC transport~\cite{znidaric_many-body_2008}, area-law entanglement of its eigenstates~\cite{bauer_area_2013}, Poisson level statistics (i.e., an absence of level repulsion)~\cite{pal_many-body_2010}, and logarithmic growth of entanglement following a quantum quench~\cite{bardarson_unbounded_2012}.

Showing the existence of MBL in the thermodynamic limit is a daunting task due to the interacting nature of the system.
On the one hand, there is a wealth of numerical evidence supporting the existence of an MBL phase.
Signatures of MBL have been found in studies of, for instance, disordered spin chains and various quasiperiodic models of small system sizes $L$, on the order of $L\approx 20$~\cite{oganesyan_localization_2007,znidaric_many-body_2008,pal_many-body_2010,bauer_area_2013, iyer_many-body_2013,kjall_many-body_2014,luitz_many-body_2015}. 
Experimental evidence for MBL also exists in platforms with cold atom and trapped ion systems~\cite{schreiber_observation_2015, smith_many-body_2016}.
Additionally, a rigorous proof of MBL in 1D disordered spin chains was formulated under the assumption that there is no attraction between energy levels~\cite{imbrie_many-body_2016, imbrie_diagonalization_2016}.
Very recently, the vanishing of conductivity was proven for strongly disordered, interacting quantum chains \cite{de_roeck_absence_2024} without making any assumptions.
On the other side of the debate, a series of works recently emerged that challenged the existence of the MBL phase transition in the thermodynamic limit. 
These works argued that the MBL transition diverges (i.e., requires infinite microscopic disorder) in the thermodynamic limit based on scaling analysis of numerical studies of small system sizes~\cite{suntajs_quantum_2020, sels_bath-induced_2022}.
These claims have, in turn, been contested, where these arguments against MBL have been attributed to finite-size effects~\cite{abanin_distinguishing_2021}.
To this day, the debate over the existence of the MBL phase remains unresolved, clouded by the technical complexity of rigorous proofs and the pronounced finite-size effects that plague numerical studies.
These limitations obscure the true nature of localization and raise serious concerns about the validity of extrapolating results from the small system sizes accessible via exact diagonalization (ED).
At times, the challenges involved in studying MBL appear nearly insurmountable.

Standing in this uncertain landscape, one may ask whether there are alternative approaches to studying MBL that go beyond the exact diagonalization of small system sizes. 
One promising direction is to develop perturbative insights, starting from \emph{noninteracting} localized systems where localization is rigorously established.
A paradigmatic example of such a system is single-particle Anderson localization, proposed in Anderson’s seminal 1958 paper~\cite{anderson_absence_1958}, for which many subsequent works then developed firm physical and mathematical foundations (see, e.g.,~\cite{kramer_localization_1993,lagendijk_fifty_2009} for a review).
Another prominent example involving a localization transition appears in quasiperiodic systems, such as the Aubry-André model, where the underlying potential is incommensurate with the lattice~\cite{aubry_analyticity_1980}.
Using these single-particle localized models as a starting point, the foundations of MBL could then potentially be built with perturbative arguments in the interaction strength.
Specifically, early arguments provided evidence that highly excited states in interacting systems can show localization, even up to infinite orders in perturbation theory~\cite{fleishman_interactions_1980,altshuler_quasiparticle_1997,gornyi_interacting_2005, basko_metalinsulator_2006}. 
These perturbative arguments established initial analytical foundations for the existence of MBL.
Later, following the important conjecture that MBL Hamiltonians can be written in terms of LIOMs~\cite{huse_phenomenology_2014}, perturbative methods once again proved useful:
Specifically, perturbative approaches in the hopping~\cite{serbyn_local_2013,imbrie_many-body_2016,de_roeck_absence_2024} can be used to ``dress'' the trivially structured LIOMs in the noninteracting cases and construct the quasi-local unitary transformations that diagonalize the Hamiltonian.

Treating interactions as a perturbation has its advantages and disadvantages in studying the structure of the LIOMs. 
Indeed, the important work of Ref.~\cite{ros_integrals_2015} established a direct connection between the original perturbative treatment of weak interactions of Ref.~\cite{basko_metalinsulator_2006} and the LIOM phenomenology.
When starting from the single-particle localized models and adding interactions perturbatively, the starting point already has a non-trivial structure to the LIOMs due to the non-zero localization length.
As a result, the stability of localization against interactions is a rather subtle question, even at low orders in perturbation theory. 
Previous studies suggest that the fate of the noninteracting localization under interactions depends sensitively on the parameters of the model.
For instance, recent numerical studies suggest that in the Anderson model, localization is unstable against interactions at weak disorder but remains stable at sufficiently strong disorder~\cite{crowley_avalanche_2020,leblond_universality_2021,colbois_interaction-driven_2024}.
Earlier results on the quasiperiodic Aubry-André model paint a similar picture, suggesting that diffusion is restored when weak interactions are added to the localized regime of the model, at least within the lower end of the potential depth range where localization occurs~\cite{znidaric_interaction_2018}.
In this regard, a systematic large-scale analysis of the perturbative impact of interactions can offer valuable insight.
In particular, numerical methods based on perturbative treatments of interactions have an advantage since they only require the diagonalization of the noninteracting localized problem.
Thus, these methods can probe very large system sizes, far surpassing what is achievable with exact diagonalization of the interacting problem.

In a parallel line of research, substantial progress has been made in understanding the dynamics of quantum integrable systems, particularly under the influence of integrability-breaking perturbations. 
These developments offer promising frameworks to study MBL perturbatively.
Integrable systems are characterized by an extensive set of conserved quantities, which—similar to the role played by LIOMs in MBL systems—profoundly influence their dynamics, resulting in Poisson level statistics and preventing thermalization \cite{ilievski_complete_2015, rigol_relaxation_2007, cassidy_generalized_2011, rigol_breakdown_2009}.
It is believed that introducing a generic integrability-breaking perturbation to such systems causes thermalization to eventually resume, on a timescale of $\sim\!\! \lambda^{-2}$ in the perturbation strength $\lambda$, predicted by Fermi's golden rule (FGR) type arguments~\cite{mallayya_prethermalization_2019, mallayya_prethermalization_2021}.
However, recent studies have found a class of so-called weak integrability-breaking perturbations, which induce thermalization on timescales significantly longer than predicted by the naive FGR, with different parametric scaling $\sim\!\! \lambda^{-p}$, $p > 2$~\cite{pozsgay_tt-deformation_2020,vanovac_finite-size_2024,orlov_adiabatic_2023,doyon_space_2022,szasz-schagrin_weak_2021,pozsgay_current_2020, pozsgay_adiabatic_2024,szasz-schagrin_weak_2021, durnin_nonequilibrium_2021, yan_duality_2024, kuo_energy_2024,surace_weak_2023,vanovac_finite-size_2024}.
This slow approach to equilibrium is attributed to the emergence of quasiconserved charges—quantities that remain approximately conserved up to some order in the perturbation strength $\lambda$~\cite{jung_transport_2006, kurlov_quasiconserved_2022, surace_weak_2023}.
These quasiconserved quantities have also been attributed to other abnormal dynamical behavior, such as high heat and spin conductivity~\cite{jung_transport_2006, Jung2007spin}.

The theoretical frameworks developed to describe these slow thermalization processes in nearly integrable systems offer promising tools for understanding MBL.
Although disordered systems are not strictly integrable in the same sense as integrable clean chains, noninteracting localized systems, like integrable ones, possess an extensive number of conserved quantities---the occupations of localized eigenstates, which we refer to in this work as noninteracting LIOMs. 
There are many open questions around what happens to these single-particle localized models when interactions are introduced.
In particular, how do these noninteracting LIOMs change under perturbation?
Do quasiconserved quantities, analogous to those in nearly integrable systems, also emerge in MBL?
If so, how can one connect this to the LIOM picture of MBL?
Addressing these questions could shed light on the emergence of long time scales in the dynamics of putative MBL regimes.

In this work, we seek answers to these questions by studying the robustness of localization in the Anderson and quasiperiodic Aubry-André models, using a perturbative approach in the interaction strength. 
In addition to helping address the questions above, the advantage of our approach is that it can be done for system sizes up to $L=100$, which suffer from significantly less finite-size effects that plague ED studies of the fully interacting model.
By examining both the LIOMs and a newly introduced diagnostic of transport---dubbed the \emph{charge transport capacity}---our numerics show that localization, to first order, is perturbatively robust in both the Anderson and quasiperiodic Aubry-André models.

First, we consider the adiabatic gauge potential (AGP), a generator of adiabatic transformations of eigenstates.
This quantity was originally introduced as a probe for quantum chaos~\cite{kolodrubetz_geometry_2017,sierant_fidelity_2019, pandey_adiabatic_2020,leblond_universality_2021}
We find that the addition of weak interactions indeed preserves at least an $O(1)$ fraction of noninteracting LIOMs up to the first order, resulting in an extensive number of quasiconserved quantities, i.e., that commute with the Hamiltonian to $O(\lambda^2)$; the corresponding LIOMs hence survive until time scales of $O(\lambda^{-4})$. 
On the other hand, a fraction of the original LIOMs suffer from resonances and cannot be treated perturbatively. We find that these resonances are local and in principle can be treated locally, upon which one could potentially expand this perturbation theory to higher orders.

In addition to studying the stability of LIOMs, we also investigate how interactions perturbatively affect charge transport in the system.
Analyzing charge transport offers a particularly robust diagnostic of localization, as the absence of transport is a more controlled and physically well-defined criterion than the stability of all LIOMs. 
To this end, we introduce a new quantity---the charge transport capacity (for short, transport capacity)---which provides an upper bound on the amount of charge that can be transferred across a link at arbitrary times.
To demonstrate localization, it suffices to show that the transport capacity across the center of a chain of length $L$ remains bounded as $L$ increases.
For both the Anderson and the quasiperiodic Aubry-André models, we find that the first-order correction to the charge transport capacity does not grow with $L$ when interactions are added to the localized regimes, indicating that localization is perturbatively robust to the first order.

For the Anderson model, our results are consistent with the perturbative robustness of localization against interactions across the entire range of disorder strengths we consider. 
This result seems to conflict with the prediction of an avalanche instability at small enough disorder~\cite{de_roeck_stability_2017, colbois_interaction-driven_2024}, which could mean that capturing an avalanche-like effect is beyond the reach of our perturbative method.
We reach similar conclusions with the quasiperiodic Aubry-André model, where we again do not see any instability of localization against interactions, in variance with proposals in some previous works~\cite{znidaric_interaction_2018}.
Our results in this case are similar to those obtained when using the single-particle Anderson model as a starting point. 
They suggest that the different spatial structure of resonances in the quasiperiodic case does not significantly affect the perturbative calculations. 
This indicates that any potential instability would need to be a nonperturbative effect \cite{znidaric_interaction_2018}.
Further work is needed to reconcile this with some proposals suggesting more perturbative-like instability.

The paper is organized as follows. 
We introduce the models considered in this work in Sec.~\ref{sec:preliminaries}.
We then outline the analytical framework for the perturbation theory of LIOMs and present numerical results for the AGP and corrections to the noninteracting LIOMs in Sec.~\ref{sec:quasiLIOMs}.
We then introduce the charge transport capacity and present perturbative numerical calculations of its Frobenius norm in Sec.~\ref{sec:transport}, in both noninteracting and interacting localized systems.
Finally, we summarize and discuss future directions in Sec.~\ref{sec:conclusion}.

\section{Preliminaries
}\label{sec:preliminaries}

\subsection{The Models}\label{sec:models}
We consider a quantum system described by a free-fermion Hamiltonian $H_0$ perturbed by an interaction term $\hat V$ with strength $\lambda\in\mathbb{R}$:
\begin{equation}
 \hat H = \hat H_0 + \lambda \hat V~.
\end{equation}
$\hat H_0$ describes a spinless fermion hopping on a one-dimensional (1D) and finite lattice with sites $i = 1, 2, \dots, L$:
\begin{equation}\label{eq:hamiltonian}
    \hat H_0=-t\sum_{j=1}^{L_{\mathrm{max}}} \left(\hat{c}_j^\dagger \hat{c}_{j+1}+\hat{c}_{j+1}^
    \dagger \hat{c}_{j}\right)+\sum_{j=1}^L h_j \hat{c}_j^\dagger \hat{c}_j=\sum_{jk} A_{jk}\hat{c}_j^\dagger \hat{c}_k.
\end{equation}
Here, $\hat{c}_j^\dagger$ and $\hat{c}_j$ represent the fermion creation and annihilation operators at site $j$, respectively.
$t\in\mathrm{R}^{+}$ is the hopping parameter, and $h_j\in \mathbb{R}$ represent the onsite potentials.
$L_{\mathrm{max}}=L$ for periodic boundary conditions (with $c_{L+1}\equiv c_1$), and $L_{\mathrm{max}}=L-1$ for open boundary conditions.
In this paper, we fix $t=1$ uniformly across all sites. 
Without any perturbations, this describes free fermions with nearest-neighbor hoppings and onsite potentials $h_j$.
In the last equality, $A_{jk} = -t$ if $j=k \pm 1$, and $A_{jk} = h_j$ if $j=k$.
The Hamiltonian can be diagonalized by solving the eigenproblem $A\vec{\phi} = \epsilon \vec{\phi}$.

The perturbation $\hat V$ consists of interactions between neighboring electrons:
\begin{equation}\label{eq:interaction}
\hat V = \sum_{j=1}^{L_{\mathrm{max}}} \hat{n}_j \hat{n}_{j+1} .
\end{equation}
Here, $\hat{n}_j\equiv \hat{c}_j^\dagger \hat{c}_j$ is the number operator at site $j$. 

In this work, we consider two distributions of $h_j$ that are known to lead to localization in 1D: a disordered (Anderson) potential landscape and a quasiperiodic (Aubry-André) potential landscape.
These models are well-studied examples of localization, with the quasiperiodic model featuring a localization transition.
In the next section, we will briefly summarize the important physical properties of these models, with and without interactions.

\subsubsection{The Interacting Anderson Model}
We first consider the model where $h_j$ is randomly and independently sampled from a uniform distribution $\mathrm{Unif}[-W,W]$, where $W\in \mathbb{R}$ is the disorder strength. With this choice, $H_0$ is the well-known 1D Anderson model.

The most well-known property of the noninteracting Anderson model in 1D is that all of its eigenstates are exponentially localized at any disorder strength \cite{anderson_absence_1958}. When interactions are added, the Anderson model is expected to exhibit many-body localization for strong enough disorder.
For a fixed interaction strength, a many-body delocalized-to-localized phase transition occurs when one increases the disorder strength.
Unlike in the noninteracting case, where localization occurs for any disorder strength, MBL generally occurs at large disorder.
This is because interactions can push the system towards ergodicity, as they cause the exchange of information and energy among the particles in the system, leading to the prediction of the ETH. 
Thus, a stronger disorder may be needed to counteract this process\footnote{Some works also suggest that the localized phase is reentrant in the interaction strength, where large enough finite interaction strength can act to restore localization even at infinite temperatures~\cite{bar_lev_absence_2015}.
Our work is far from the regime where this physics occurs.}.

Finally, we note that the stability of the single-particle localized phase to added interactions is an open question.
Some previous work suggests that the Anderson model, at low disorders, is unstable to added interactions due to avalanche-like mechanisms~\cite{de_roeck_stability_2017}.
There has also been past numerical work that explores the phase diagram of the model~\cite{colbois_interaction-driven_2024}, but the question remains difficult to fully solve due to the interacting nature of the problem.

\subsubsection{The Interacting Aubry-André Model}
In this work, we also study the interacting Aubry-André model, where we set $h_j$ to be a deterministic quasiperiodic potential:
\begin{equation}\label{eq:quasiperiodic_potential}
    h_j=h\cos{\left[2\pi k\left( j+\delta\right)\right]}.
\end{equation}
Here, the wavenumber $k$ is some irrational number, $\delta \in [0,1/k)$ is a phase shift, and $h\in \mathbb{R}$ is the potential depth. In the rest of this paper, unless specified, we fix the irrational wavenumber $k=\sqrt{2}$.

If the perturbation is turned off, this model is simply the noninteracting Aubry-André model. 
To discuss the phases in this model, we fix $t=1$ and observe a phase transition when we tune the potential depth $h$. 
When $h<2$, the states are extended in real space.
When $h>2$, the states are localized in real space.
At exactly $h_c = 2$, the eigenstates are multifractal~\cite{aubry_analyticity_1980,siebesma_multifractal_1987}.

In the presence of interactions, we expect a many-body localization transition.
Starting in the localized single-particle phase at large $h$, increasing the interaction strength will eventually destroy the localization.
The proposed phase diagram for this interacting Aubry-André model shows that at larger interactions, a stronger potential depth is needed for localization to occur~\cite{iyer_many-body_2013, znidaric_interaction_2018}.
Close to $h_c$ when the localization length is large, previous studies again predict instability for arbitrarily small interactions.
Finally, starting in the delocalized single-particle phase, we expect interactions to further promote the delocalization (at least for generic weak to moderate interactions away from special limits).

\subsection{Local Integrals of Motion}
The models introduced in section \ref{sec:models} are single particle models that can be diagonalized by solving the eigenequation $A\vec{\phi}_{\alpha} = \epsilon_\alpha \vec{\phi}_\alpha$, where $\vec{\phi}_{\alpha}$ is a column vector representing the orbital $\alpha$ with site $j$ amplitude $\phi_\alpha(j)$ and energy $\epsilon_\alpha$.
We define the localized orbital annihilation operator $d_\alpha$, which destroys a particle in the localized orbital $\alpha$:
\begin{equation}
    \hat{d}_\alpha= \sum_j \phi_\alpha^\star(j) \hat c_j.
\end{equation}
From now on, we define $\phi_\alpha^\star(j)\equiv \phi_{j\alpha}^\star$.
We can rewrite the noninteracting Hamiltonian as:
\begin{equation}
    \hat H_0 = \sum_{\alpha}\epsilon_{\alpha}\hat{d}^{\dagger}_{\alpha}\hat{d}_{\alpha} = \sum_\alpha \epsilon_\alpha\tilde{n}_\alpha^{(0)}.
\end{equation}
We note that $\tilde{n}_\alpha^{(0)} \equiv \hat{d}^{\dagger}_\alpha\hat{d}_\alpha$, which represent the occupations of the localized orbitals, are conserved quantities of $\hat H_0$, so that $[H_0, \tilde n_{\alpha}^{(0)}]=0$. This gives $\hat H_0$ a noninteracting integrability, where $\hat H_0$ can be diagonalized with these occupations.

It is conjectured that for MBL systems, there exists a quasilocal transformation that completely diagonalizes the Hamiltonian $\hat H(\lambda)$. Under such a transformation, the Hamiltonian can be expressed as:
\begin{equation}\label{eq:liom_equation}
    \hat H(\lambda) = \sum_{\alpha} \tilde \epsilon_\alpha \tilde n_\alpha + \sum_{\alpha>\beta} J_{\alpha,\beta} \tilde n_\alpha \tilde n_\beta+\sum_{\alpha> \beta>\gamma} J_{\alpha, \beta, \gamma} \tilde n_\alpha \tilde n_\beta \tilde n_\gamma + ... 
\end{equation}
Here, the coefficients $\tilde \epsilon_{\alpha}$ and $J_{\alpha\beta\gamma...}$ depend on $\lambda$ and $\tilde \epsilon_\alpha(\lambda=0)=\epsilon_\alpha$.
The operators $\tilde{n}_\alpha$ also depend on $\lambda$, and again $\tilde{n}_\alpha(\lambda=0)=\tilde n_\alpha^{(0)}$.

These $\tilde n_\alpha$ are known as \emph{local integrals of motion} (LIOMs)\footnote{We note that LIOMs in the more conventional Ising-like (two-level) language, denoted as $\hat{\tau}_\alpha^z$, are related to $\tilde n_\alpha$ by $\hat{\tau}_\alpha^z = 2{\tilde{n}}_\alpha - 1$.}.
These LIOMs have several important properties.
To name a few, they are quasilocal, are two-level operators, commute with the Hamiltonian, and form a complete set of generators of all conserved quantities \cite{serbyn_local_2013}.
The ability to rewrite an MBL Hamiltonian in this form is an important property of MBL systems, since the rest of the phenomenology of MBL directly follows from Eq.~(\ref{eq:liom_equation}). Naturally, then, LIOMs are some of the most interesting objects to study in the context of MBL systems. For a more in-depth discussion of LIOMs and MBL, see reviews~\cite{nandkishore_many-body_2015,sierant_many-body_2025,abanin_colloquium_2019}.

When arbitrarily weak interactions are added, the conservation laws that exist in the noninteracting Hamiltonians are no longer preserved.
Still, it is intriguing to observe how the LIOMs in the noninteracting localized problem, $\tilde{n}_{\alpha}^{(0)}$, are perturbatively affected by interactions. 
Furthermore, it is of interest to connect the perturbatively corrected single-particle LIOMs to the many-body LIOMs that are conjectured to exist in MBL Hamiltonians.
The goal of this analysis is to obtain further insights into the nature of MBL. 

\section{Approximate LIOMs}
\label{sec:quasiLIOMs}

In this section, we present a framework for perturbatively analyzing how single-particle LIOMs change under the influence of added interactions. 
Here, we borrow ideas from previous works proposing iterative constructions of the LIOMs in perturbation theory~\cite{modak_integrals_2016, ros_integrals_2015}, as well as the analytical treatment of generic~
\cite{durnin_nonequilibrium_2021,Brandino2015glimmers,Bertini2015Prethermalization,Essler2014Quench,Schmiedmayer2008Breakdown,Mierzejewski2015Approximate,leblond_universality_2021,pandey_adiabatic_2020,mallayya_prethermalization_2019,mallayya_prethermalization_2021} and special so-called weak-integrability breaking~\cite{pozsgay_current_2020,szasz-schagrin_weak_2021,durnin_nonequilibrium_2021,kurlov_quasiconserved_2022, surace_weak_2023, orlov_adiabatic_2023, vanovac_finite-size_2024,bahovadinov_tomonaga-luttinger_2024,pozsgay_adiabatic_2024} perturbations.

The noninteracting Hamiltonian on a finite lattice of size $L$ features $L$ conservation laws:
\begin{equation}
    [\hat H_0, \hat{d}^{\dagger}_\alpha \hat{d}_{\alpha}] = [\hat H_0, \tilde{n}^{(0)}_{\alpha}] = 0.
\end{equation}
Here, $\alpha=1,\dots,L$, and for any $\alpha, \beta$, $[\tilde{n}^{(0)}_{\alpha},\tilde{n}^{(0)}_{\beta}]=0$.
These conservation laws are broken when the perturbation given in  Eq.~(\ref{eq:interaction}) is added. 
We consider the problem of finding corrections $\tilde{n}^{(1)}_\alpha$ to $\tilde{n}^{(0)}_\alpha$, such that the quantity $\tilde{n}^{(0)}_\alpha+\lambda \tilde{n}^{(1)}$ still commutes with $H_0+\lambda V$ up to second order in $\lambda$:
\begin{equation}
    \left[\hat H_0+\lambda \hat V, \tilde{n}^{(0)}_\alpha+\lambda \tilde{n}^{(1)}_\alpha\right] = O(\lambda^2).
\end{equation}
The quantity $\tilde{n}^{(0)}_\alpha+\lambda \tilde{n}^{(1)}$ is an ``approximate LIOM''.
The quasi-conservation law above is equivalent to the relation:
\begin{equation}\label{eq:quasiconservation_law}
    \left[\hat V, \tilde{n}^{(0)}_\alpha
    \right] = -\left[\hat H_0,\tilde{n}^{(1)}_\alpha\right].
\end{equation}
Here, we seek a bounded solution for $\tilde{n}^{(1)}_\alpha$ for each $\tilde{n}^{(0)}_\alpha$ such that Eq.~(\ref{eq:quasiconservation_law}) holds.
If this is possible for at least a subset of $\tilde{n}^{(1)}_\alpha$, $\hat V$ is a special type of ``weak'' perturbation in the sense that it perturbatively preserves the corresponding conservation laws.
In the context of quantum integrable models, which feature a large number of conserved charges, a perturbation $V$ that is ``weak'' in the sense introduced above is associated with longer thermalization times $\tau$, that depend on $\lambda$ as a power-law $\tau \sim \lambda^{-4}$ instead of the typical $\tau \sim \lambda^{-2}$~\cite{mallayya_prethermalization_2019,szasz-schagrin_weak_2021, surace_weak_2023}. 
Just as above, weak integrability breaking in quantum systems means a large number of quasi-charges. 

One important note is that one could, in principle, analyze higher $l-$order LIOM corrections $\tilde n^{(l)}_\alpha$, such that the quasiconservation law holds up to $O(\lambda^l)$ for some $l\in \mathbb{Z^+}$.
These corrections will be $(2l+2)$--fermion terms, and their numerical calculations will scale at least as $L^{2l+2}$, becoming very expensive. Furthermore, the energy resonances that arise from perturbation theory will grow increasingly complex.
As a result, we do not attempt higher-order calculations in this work.

One formal solution to the problem of finding approximate LIOMs is given by the Adiabatic Gauge Potential (AGP), $\hat X$.
We split the perturbation $\hat V$ into a diagonal and off-diagonal part,
\begin{equation}
    \hat V= \hat V_{\text{diag}}+\hat V_{\text{off-diag}},
\end{equation}
with $\left[\hat V_{\text{diag}},\tilde n_\alpha^{(0)}\right]=0$ and $\braket{\tilde n_\alpha^{(0)} \hat V_\text{off-diag}}\propto \Tr[\tilde n_\alpha^{(0)} \hat V_\text{off-diag}]=0$ for every $\alpha$.
Moreover, the diagonal and off-diagonal parts are orthogonal in the Frobenius inner product, i.e., $\braket{\hat V_{\text{diag}} \hat V_\text{off-diag}}\propto \Tr[\hat V_{\text{diag}} \hat V_\text{off-diag}]=0$.
The AGP is a Hermitian operator $\hat X$ that solves the equation
\begin{equation}\label{eq:Xdef}
    \hat V_\text{off-diag}=i\left[\hat X, \hat H_0\right].
\end{equation}
Note that this operator is defined up to additive terms that commute with $H_0$. 
\footnote{Conventionally, the AGP is defined as the generator of adiabatic transformations of the eigenstates $|n(\lambda)\rangle$ of a Hamiltonian $\hat H(\lambda)$ \cite{jarzynski_generating_2013, kolodrubetz_geometry_2017,karve_adiabatic_2025}. In our notation (note that the convention used in some of the literature has the opposite sign for $\hat X$):
\begin{equation}\label{eq:conventional_agp}
    \hat X|n(\lambda)\rangle=-i\partial_\lambda|n(\lambda)\rangle.
\end{equation}
The relation between our definition of the AGP and Eq.~(\ref{eq:conventional_agp}) is seen by observing that Eq.~(\ref{eq:conventional_agp}) obeys the relation
\begin{equation}\label{eq:conventional_agp_relation}
    \partial_\lambda \hat H(\lambda)=i[\hat X, \hat H(\lambda)]+\sum_n \partial_{\lambda}E(\lambda) |n(\lambda)\rangle\langle n(\lambda) |.
\end{equation}
One can verify this property by assuming the definition Eq.~(\ref{eq:conventional_agp_relation}), then evaluating the left-hand side $\partial_\lambda \hat H=\partial_\lambda\sum_n E_n |n\rangle\langle n|$. 
If one sets $\lambda=0$ on the right-hand side of Eq.~(\ref{eq:conventional_agp_relation}), then one recovers Eq.~(\ref{eq:Xdef}).
Specifically, in our case, $\partial_\lambda \hat H(\lambda)=\hat V$, the first term on the right-hand side is $i[\hat X, \hat H_0]=\hat V_\mathrm{off-diag}$ and the second term on the right-hand side is $\hat V_\mathrm{diag}$.}  

Now, using $\hat X$, we can define the corrections $\tilde{n}^{(1)}_\alpha$:
\begin{equation}\label{eqn:liom_correction_def}
    \tilde{n}^{(1)}_\alpha = i\left[\hat X, \tilde{n}^{(0)}_\alpha\right] ~.
\end{equation}
A simple application of the Jacobi identity $[[\hat A, \hat B], \hat C] + [[\hat B, \hat C], \hat A] + [[\hat C, \hat A], \hat B] = 0$ with $\hat{A}, \hat{B}, \hat{C}$ being $\hat X, \tilde{n}^{(0)}, \hat H_0$, respectively, shows that the form of LIOM corrections in Eq.~(\ref{eqn:liom_correction_def}) satisfies Eq.~(\ref{eq:quasiconservation_law}).

For finite $L$, it is always possible to formally define $\hat X$ given any $\hat V$. 
However, the existence of the AGP is not equivalent to the corresponding $\hat V$ being a weak integrability-breaking perturbation, as $\hat X$ may be highly nonlocal, unbounded, and have other undesirable properties.
Therefore, it may not always give finite and quasilocal corrections $\tilde{n}^{(1)}_\alpha$. 
For this work, we are interested in probing the properties of the AGP corresponding to perturbative interactions to the noninteracting localized model. From a well-behaved AGP, we can then find finite corrections to the single-particle LIOMs.

In the following section, we outline the derivation of exact expressions for the AGP and LIOM corrections corresponding to the Hamiltonians introduced in Sec.~\ref{sec:models}.

\subsection{Free Fermion AGP and $\tilde n_{\alpha}^{(1)}$}
We consider the general free fermion Hamiltonian as in Eq.~(\ref{eq:hamiltonian}). Our goal is to find the generator $\hat{X}$ such that:
\begin{equation*}
    \hat V_{\text{off-diag}}=i\left[\hat X,\hat H_0\right].
\end{equation*}
Here, $\hat V$ consists of nearest-neighbor interactions, defined in Eq.~(\ref{eq:interaction}).
In this derivation, there are no restrictions on the form of the fermion bilinear matrix $A$ or even the dimensionality of the system. 
We refer the reader to Appendix~\ref{app:appendix_agp_derivation} for full details of the derivations of expressions given here.

We can perform a change of basis to rewrite $\hat V$ in terms of the localized orbital basis that we defined earlier, $\hat{d}_\alpha= \sum_k \phi_{k\alpha}^* \hat{c}_k$:
\begin{equation}
    \hat V=\sum_{\alpha, \beta,\gamma,\delta}V_{\alpha\beta\gamma\delta}\hat{d}_\alpha^\dagger \hat{d}_\beta^\dagger \hat{d}_\gamma \hat{d}_\delta.
\end{equation}
Here, hermiticity of $\hat{V}$ requires $V^*_{\alpha\beta\gamma\delta}= V_{\delta\gamma\beta\alpha}$, and we choose $V_{\alpha\beta\gamma\delta}=-V_{\beta\alpha\gamma\delta}=-V_{\alpha\beta\delta\gamma}$, consistent with the fermion anticommutation relations. 
The matrix elements are given by
\begin{multline}
    V_{\alpha \beta \gamma \delta}=\frac{1}{4}\sum_j(\phi_{j+1,\alpha}^* \phi_{j,\beta}^* \phi_{j,\gamma} \phi_{j+1,\delta}-\phi_{j,\alpha}^* \phi_{j+1,\beta}^* \phi_{j,\gamma} \phi_{j+1, \delta}\\
    -\phi_{j+1, \alpha}^* \phi_{j, \beta}^* \phi_{j+1, \gamma} \phi_{j, \delta}+\phi_{j, \alpha}^* \phi_{j+1,\beta}^* \phi_{j+1,\gamma} \phi_{j,\delta}).
\end{multline}
With this rewriting, we can immediately identify the diagonal part of the interaction from the terms where either $\alpha=\gamma \neq \beta=\delta$ or $\alpha=\delta \neq \beta=\gamma$:
\begin{equation}
V_{\text{diag}}=\sum_{\alpha \neq \beta}2V_{\alpha\beta\beta\alpha} \tilde n_\alpha^{(0)}\tilde n_{\beta}^{(0)}.
\end{equation}
This diagonal component of $\hat V$ commutes with the unperturbed Hamiltonian $\hat H_0$ and with its LIOMs $\tilde n_\alpha^{(0)}$. Therefore, the ``genuine'' perturbation consists only of the off-diagonal part $\hat V_\text{off-diag}=\hat V-\hat V_{\text{diag}}$. This decomposition was used in Ref.~\cite{krajewski_restoring_2022} to argue that, in the strong-disorder regime -- where $\hat V$ is predominantly diagonal -- the effective perturbation strength is significantly reduced. This leads to distinct scaling behaviors observed in numerical simulations. We aim to build upon these insights and develop a perturbative scheme that removes, to leading order, the residual effects of the off-diagonal term.

We can find an explicit solution to Eq.~(\ref{eq:Xdef}) by considering the following four-fermion operator ansatz for the AGP:
\begin{equation}
\label{eq:Xabcd}
    \hat X=\sum_{\alpha, \beta,\gamma,\delta }X_{\alpha\beta\gamma\delta}\hat{d}_\alpha^\dagger \hat{d}_\beta^\dagger \hat{d}_\gamma \hat{d}_\delta.
\end{equation}
Similarly to the coefficients $V_{\alpha\beta\gamma\delta}$, hermiticity of $\hat X$ imposes the constraint $X_{\alpha\beta\gamma\delta}^*=X_{\delta \gamma \beta \alpha}$. In addition, we choose the coefficients to satisfy the antisymmetry conditions $X_{\alpha\beta\gamma\delta}=-X_{\beta \alpha \gamma \delta}=-X_{\alpha\beta\delta\gamma}$. Since $\hat X$ is defined only up to additive terms that commute with $\hat H_0$, we can, without loss of generality, impose that it has no diagonal part, setting $X_{\alpha\beta\alpha\beta}=X_{\alpha\beta\beta\alpha}=0$.
From Eq.~(\ref{eq:Xabcd}), the remaining (off-diagonal) AGP coefficients $X_{\alpha\beta\gamma\delta}$ can be directly computed by evaluating Eq.~(\ref{eq:Xdef}), leading to the expression
\begin{equation}
\label{eq:Xelements}
X_{\alpha\beta\gamma\delta}=i\frac{V_{\alpha\beta\gamma\delta}}{\epsilon_\alpha+\epsilon_\beta-\epsilon_\gamma-\epsilon_\delta}.
\end{equation}
Note that the restriction to the off-diagonal part, enforced by setting $X_{\alpha\beta\alpha\beta}=X_{\alpha\beta\beta\alpha}=0$, ensures that the denominators never vanish exactly. Nonetheless, the energy differences in the denominator can become arbitrarily small, potentially resulting in very large values for the corresponding matrix elements of $X_{\alpha\beta\gamma\delta}$.

We can evaluate Eq.~(\ref{eqn:liom_correction_def}) to obtain the LIOM corrections $\tilde n^{(1)}_\alpha$, which is once again a four-fermionic operator:
\begin{equation}
    \tilde{n}_{\alpha}^{(1)} =\sum_{\mu\nu\rho\sigma} f^\alpha_{\mu\nu\rho\sigma}\hat{d}_\mu^\dagger \hat{d}_\nu^\dagger \hat{d}_\rho \hat{d}_\sigma,
\end{equation}
and the matrix elements $f^\alpha_{\mu\nu\rho\sigma}$ are obtained from contractions of the rank-4 tensor $X_{\alpha\beta\gamma\delta}$ (see Appendix \ref{app:appendix_agp_derivation}):
\begin{equation}\label{eq:liom_matrix_elements}f^\alpha_{\mu\nu\rho\sigma}=i(\delta_{\alpha \sigma} +\delta_{\alpha \rho} -\delta_{\alpha \mu} -\delta_{\alpha \nu}) X_{\mu\nu\rho \sigma}.
\end{equation}

\subsection{Methods}\label{sec:methods}
All quantities introduced in this paper, such as $\hat X$ and $\tilde n_{\alpha}^{(1)}$, have analytical expressions that only depend on eigenstates and eigenvalues of the noninteracting localization problem. 
The noninteracting eigenproblem is computationally easy to solve, which makes studying $\hat X$ and $\tilde n_{\alpha}^{(1)}$ appealing compared to full-diagonalization of the many-body interacting Hamiltonian. 
In this work, we calculate all quantities up to $L=100$, noting that these system sizes are completely inaccessible to exact diagonalization of the fully interacting models.
After diagonalizing the single-particle problem, we numerically compute these matrix elements using the TensorOperations package in Julia for tensor contractions \cite{bezanson_julia_2017,TensorOperations.jl}.

In this work, we study most quantities (except for noninteracting charge transport) by computing their Frobenius norms, which are computationally more tractable than the operator norms.
Here, we define the Frobenius norm of an operator $\hat{O}$ as the normalized Hilbert-Schmidt inner-product
\begin{equation}
\|\hat O\|_F^2 = \frac{1}{\mathcal D}\Tr[\hat O^\dagger \hat O],
\end{equation}
where $\mathcal D$ is the total Hilbert space dimension.
We are interested in the {\it normalized} Frobenius norm as this results in localized operators having $O(1)$ norm. 
Formulas for Frobenius norms of four-fermionic operators can be found in Appendix \ref{app:a_frobenius_norm_derivation}.
For exponentially localized operators, both Frobenius and operator norms are expected to be finite in the thermodynamic limit.
Here, we are mostly interested in assessing whether $\tilde n_{\alpha}^{(1)}$ is localized. 
Therefore, by verifying whether its Frobenius norm is finite and bounded, we obtain strong evidence of its localization.

Our implementation achieves the optimal asymptotic runtime of $O(L^4)$. We expect this scaling because the number of nonzero and distinct matrix elements to be computed scales as $O(L^4)$. Since each of these elements must be individually evaluated, this constitutes a lower bound on the computational complexity. Our procedure matches this bound, indicating that it is asymptotically optimal.

For all calculations, we fix the hopping parameter $t=1$ and restrict ourselves to open boundary conditions on a 1D lattice. 
In the quasiperiodic model, we fix the irrational wavenumber to $k = \sqrt{2}$. While the specific choice of $k$ affects the commensurability between the potential and the lattice, we find that it does not qualitatively alter any conclusions we draw in this work. 
In particular, see Appendix~\ref{app:additional_quasiperiodic_wavenumbers} for additional results on the charge transport capacity (discussed in Sec.~\ref{sec:transport}) using a different wavenumber $k = \sqrt{3}$.

For the disordered model, calculations are presented with a distribution over independent and randomly generated disorder realizations. 
We also obtain distributions for the quasiperiodic model. Although the model is deterministic, our calculations found that changing the phase shift $\delta$ affects the results. 
To avoid commensuration issues, we found it useful to sample over these variations with a distribution over the phase $\delta$, spanning 500 evenly spaced phases between $0\leq \delta<1/k$, which covers one period of the potential. 
This setup also allows subsequent analysis of interaction corrections to charge transport in both models.

\subsection{AGP Results}\label{sec:AGP_results}
\subsubsection{Disordered Model AGP}\label{sec:agp_resulsts_disordered}

\begin{figure*}[t]
    \centering
    \includegraphics[width=\linewidth]
    {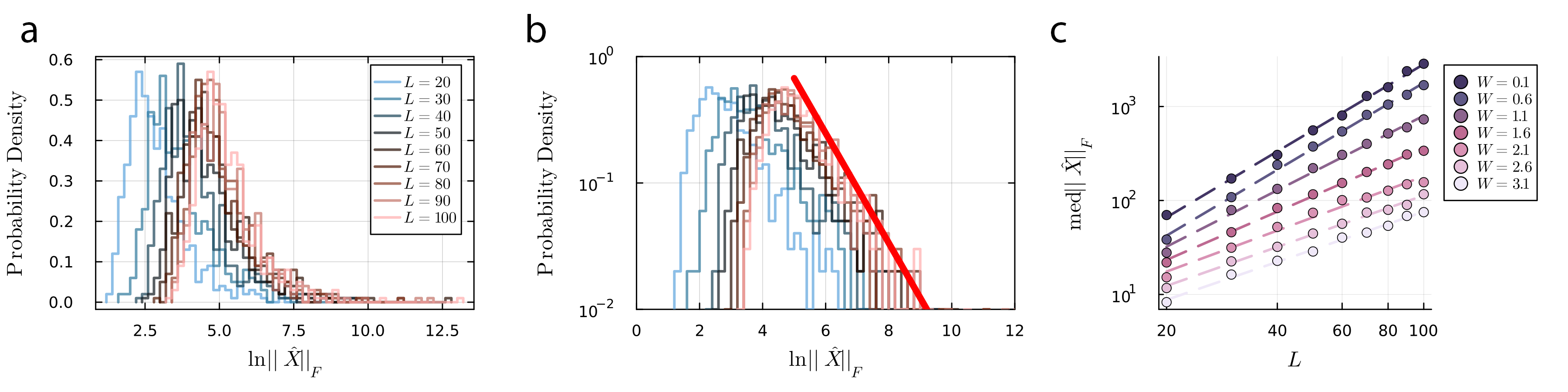}
    \caption{
    Analysis of the AGP Frobenius norm in the interacting Anderson Model. All three figures were generated based on 500 disorder realizations and open boundary conditions. (a) Calculated probability distributions of $\ln\|\hat X\|_F$ with disorder strength $W=2.1$. Each colored curve represents varying system sizes $L$, from $L=20$ to $L=100$ in steps of $10$. 
    (b) Same data as in (a) but with the vertical scale shown on a log scale.
    The red line represents a polynomial $\|\hat X\|_F^{-2}$ fit to the tail of the probability distribution of $\|\hat X\|_F$---see text around Eq.~(\ref{eq:tail_behavior}) for details on the form plotted. 
    (c) Median of the AGP norm as a function of system size $L$. Each colored curve represents a different disorder strength $W$, from $W=0.1$ to $W=3.1$ in step sizes of 0.5.
    Here, dashed lines represent power-law fits to the data, $\|\hat X\|_F = A L^D + C$ with power-law exponents $D$.
    In order of increasing disorder strength, we find $D=2.33, 2.33, 1.99, 1.71, 1.45, 1.38, 1.34$, consistent with expected eventual linear in $L$ scaling for a strictly localized system (see text in Sec.~\ref{sec:agp_resulsts_disordered} for discussion).
    The error bars on the median curves, estimated using a bootstrapping method, are very small and are therefore omitted in this and subsequent plots for clarity.
    }
    \label{fig:agp_disordered_figure}
\end{figure*}

We now present numerical calculations for the Frobenius norm of the AGP, $\|\hat X\|_{F}$, for the weakly-interacting disordered model. 
Here, and in the rest of the paper, $\|\cdot\|_F$ denotes the Frobenius norm.
Formulas for the Frobenius norm for quadratic and quartic fermionic operators are presented in Appendix \ref{app:a_frobenius_norm_derivation}.
We show the sampled probability distribution of $\ln\|\hat X\|_F$ over 500 disorder realizations for $W=2.1$ in Fig.~\ref{fig:agp_disordered_figure}(a), which we found easier to visualize than the distribution of $\|\hat X\|_F$.
The calculations for other disorder strengths can be found in Appendix \ref{appendix_agp_full_results}.

The most significant feature of the calculated probability distribution of the norm of the AGP, $P\left(\|\hat X\|_F\right)$, are the long inverse square power-law tails $\sim \frac{1}{\|\hat X\|_F^2}$. 
The curve corresponding to $P\left(\|\hat X\|_F\right)\sim\|\hat X\|_F^{-2}$ is shown as a comparison in Fig.~\ref{fig:agp_disordered_figure}(b).
We emphasize that this is the tail for the probability distribution of $\|\hat X\|_F$, not for the plotted distribution of $\ln(\|\hat X\|_F)$.
The conversion between the two is straightforward: a power-law $\sim\! x^{-p}$ tail of the probability distribution $P(x)$, with $p\in \mathbb{R}$, translates to the tail 
\begin{equation}\label{eq:tail_behavior}
R(\zeta) \sim e^{-(p-1)\zeta}
\end{equation}
of the probability distribution of $\zeta = \ln(x)$ for large $\zeta$ [with $P(x) dx = R(\zeta) d\zeta$], which we plot in Fig.~\ref{fig:agp_disordered_figure}(b).
These long tails are the result of a fraction of disorder realizations with four-energy resonances (i.e., occurrences with very small denominators $\epsilon_\alpha+\epsilon_\beta-\epsilon_\gamma-\epsilon_\delta$ in Eq.~(\ref{eq:Xelements})), which results in large, unbounded AGP norms.
As a consequence of these tails, the mean of the distributions is divergent. 

The inverse square tail arises naturally from the distribution of the denominators $x \equiv \frac{1}{\epsilon_\alpha+\epsilon_\beta-\epsilon_\delta-\epsilon_\gamma}$.
If one approximates the eigenvalues $\epsilon_\alpha$ as being randomly sampled from a uniform distribution between $-W'$ and $W'$ (the eigenvalues do not exactly follow a uniform distribution, but it is a reasonable approximation), then the distribution of the reciprocal variable $x\equiv 1/(\epsilon_\alpha+\epsilon_\beta-\epsilon_\delta-\epsilon_\gamma)$ follows $P_{1/(\epsilon_\alpha+\epsilon_\beta-\epsilon_\delta-\epsilon_\gamma)}(x)\approx \frac{1}{x^2}$ at very large $x$.
Details of the derivation can be found in Appendix \ref{app:derivation_tails}.
Even though many of the numerators $V_{\alpha\beta\gamma\delta}$ are very small due to the Anderson localization of the eigenstates $\phi_{\alpha}(j)$, we still expect $O(L)$ of them to be $O(1)$. 
Therefore, the distribution of the Frobenius norm of $\hat X$ will also present a tail $\propto 1/\|\hat X\|_F^2$ at large $\|\hat X\|_F$, as seen in the data.
As a result, we expect these tails to exist for any disorder strength $W$. 
Indeed, we performed similar calculations for disorder strengths much higher than the ones presented in this paper ($W=6$) and verified that these inverse power-law tails exist even at high disorders. 
This remark applies to other quantities calculated in this work that regard the disordered Anderson model.

The scaling of the AGP with system size is a useful tool to distinguish between some integrable and thermal models~\cite{pandey_adiabatic_2020, orlov_adiabatic_2023, pozsgay_adiabatic_2024, vanovac_finite-size_2024}.
Here, the distribution's mean diverges because of the inverse-square tails.
Regardless, observing how the median of $\|\hat X\|_F$ grows with $L$ is informative.
We show these results in Fig.~\ref{fig:agp_disordered_figure}(c). 
It is known that the norm of the AGP in all free fermion systems should scale as a polynomial with the system size $L$, $\mathrm{med}\|\hat X\|_F=A L^{D}+C$ \cite{pozsgay_adiabatic_2024}. 
Here, the polynomial growth of the median of $\|\hat X\|_F$, for all $W$ values, alludes to this result.
Interestingly, we note that the power-law exponents $D$, obtained by polynomial regression on the data and written in the caption of Fig.~\ref{fig:agp_disordered_figure}, decrease with increasing $W$, where the Frobenius norm of the AGP possibly trends towards linear scaling with $L$. 
If the orbitals are strictly localized, then the terms in the AGP will only depend on a sum of local terms.
Barring the possibility of resonances in the denominators, each term in the sum is $O(1)$, and the strict localization requires that the number of nontrivial terms would be proportional to $L$, so we generally would expect the median of the AGP norm to scale as $\sim L$.
Therefore, the trend of the power-law coefficient of the AGP in our work towards linear scaling possibly reflects this strictly localized limit.

Finally, we briefly comment on the dependence of the distributions of the AGP norms on $L$. 
Consistent with the power-law scaling of the median of the AGP norms observed in panel (c) of Fig.~\ref{fig:agp_disordered_figure}, the probability distributions in panel (b) remain unconverged even at the largest system sizes studied, and shift rightwards at a linear pace with increasing $L$.
For clarity, we emphasize that the plotted quantity is actually the distribution of $\ln\|\hat X\|_F$, chosen for visibility. This transformation can create the misleading visual impression that the distributions are converging, whereas in reality they are not.

\subsubsection{Quasiperiodic Model AGP}\label{sec:agp_results_quasiperiodic}

\begin{figure*}[t]
    \centering
    \includegraphics[width=\linewidth]
    {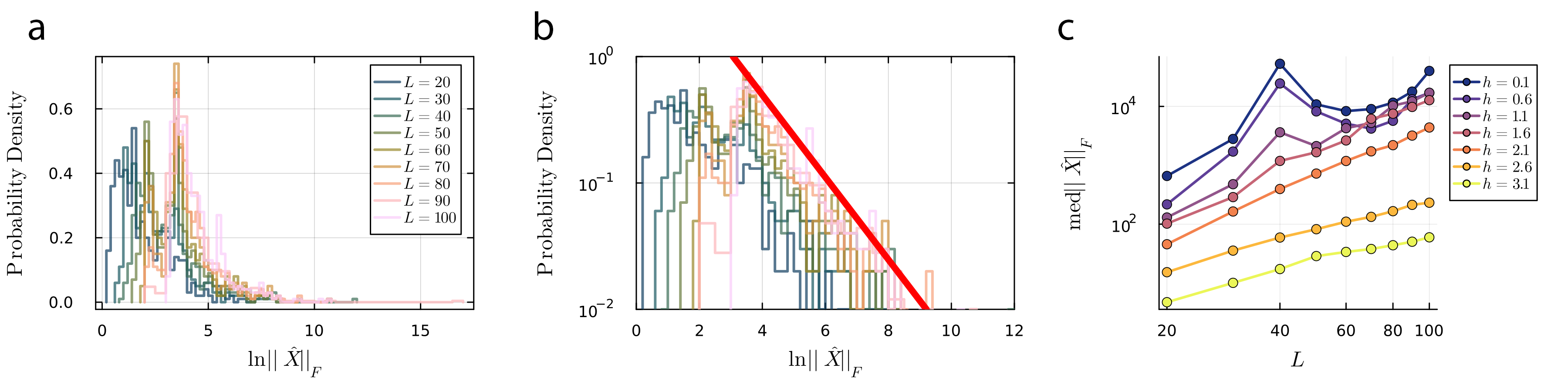}
    \caption{
    Analysis of the AGP Frobenius norm in the interacting quasiperiodic model. All three figures were generated based on 500 uniformly sampled phase shifts $\delta$ and open boundary conditions.
    We refer the reader to Sec.~\ref{sec:methods} for details on sampling for the quasiperiodic model.
    (a) Calculated probability distributions of $\ln\|\hat X\|_F$ with potential depth $h=3.1$.
    Each colored distribution represents varying system sizes $L$, from $L=20$ to $L=100$, in steps of 10. 
    (b) Same data as in (a) but with the vertical scale shown on a log scale.
    The red curve represents a polynomial $\|\hat X\|_F^{-1.75}$ fit to the tail of the probability distribution of $\|\hat X\|_F$---see text around Eq.~(\ref{eq:tail_behavior}) for details on the form plotted. 
    (c) Median of $\|\hat X\|_F$ as a function of system size $L$. Each colored curve represents a different potential depth $h$, from $h=0.1$ to $h=3.1$ in steps of 0.5.
    Here, we observe polynomial growth with $L$ of $\mathrm{med}\|\hat X\|_F$ and an anomalous peak at $L=40$ in the regime $h<2$. For a full discussion of these results, see Sec.~\ref{sec:agp_results_quasiperiodic}.}
    \label{fig:agp_quasiperiodic_figure}
\end{figure*}

We show calculations of the Frobenius norms of the AGP for the quasiperiodic model in Fig.~\ref{fig:agp_quasiperiodic_figure} in the single-particle localized regime, for the potential depth $h=3.1$. Calculations for other values of $h$ can be found in Appendix \ref{appendix_agp_full_results}.

The distributions over different phase shifts $\delta$ in Fig.~\ref{fig:agp_quasiperiodic_figure}(a) show a similar power-law tail behavior to that of the disordered model.
Interestingly, $P(\|\hat X\|_F)$ displays a slower-decaying tail, where the power-law tail has a lower coefficient when compared to the disordered model, with $P(\|\hat X\|_F)\propto \|\hat X\|_F^{-1.75}$.
A $\|\hat X\|_F^{-1.75}$ fit to the tail of the distribution of $\|\hat X\|_F$ is shown in Fig.~\ref{fig:agp_quasiperiodic_figure}(b).
A slower power-law tail universally appears in the quasiperiodic model for all the quantities analyzed in this work that depend on these four--energy denominators.
The scaling behavior of the tail here differs from that in the disordered case, likely due to the fractal structure of the density of states in the quasiperiodic model.
We show a few plots of the density of states in Appendix \ref{app:derivation_tails}.
Previous works have noted that this distinct spectral structure may underlie the different localization physics observed in the quasiperiodic model compared to that of the disordered case \cite{znidaric_interaction_2018}.

The scaling of the median of the Frobenius norm of the AGP is shown in Fig.~\ref{fig:agp_quasiperiodic_figure}(c).
Just as in the weakly-interacting Anderson model, when $h>2$, the system is in the localized phase, and $\mathrm{med}\|\hat X\|_F$ seems to scale as a polynomial with $L$.

When the states are delocalized at low disorders for $h<2$, one also observes polynomial growth of $\|\hat X\|_F$. 
Curiously, in this delocalized regime, $\mathrm{med}\|\hat X\|_F$ grows non-monotonically, and $\mathrm{med}\|\hat X\|_F$ spikes at certain system sizes. In the case with $k=\sqrt{2}$, this occurs at $L=40$.
These spikes shift to different system sizes when the wavenumber is changed to different irrational numbers, which we have separately checked for $k=\phi$, $k=\sqrt{3}$, and $k=\sqrt{6}$. Here, $\phi$ is the golden ratio.
Anomalous behavior of this kind is known to arise in the quasiperiodic model at certain integer distances. 
For example, Ref.~\cite{znidaric_interaction_2018} observed that single-particle resonances are enhanced at orbitals separated by distances equal to the Fibonacci numbers $F_n$, when one picks $k=\phi$.
Similarly, Ref.~\cite{thomson_local_2023} reported suppressed interaction between LIOMs separated at the Fibonacci numbers for $k=\phi$, and the corresponding integer sequences that relate to the silver and bronze ratios.
In our case, we find a more general condition that explains these anomalous system sizes, encompassing previous observations: they occur when the underlying potential is nearly commensurate with the lattice.
For open boundary conditions, this corresponds to when $k(L+1)\approx n$ for some $n\in \mathbb{Z}$, so $k\approx \frac{n}{L+1}$ for some integer $n$.
For example, in the calculation presented in Fig.~\ref{fig:agp_quasiperiodic_figure}(c), which was done for $k=\sqrt{2}$, there is a notable spike at $L=40$, and $\sqrt{2}\times 41\approx 58$. 
This observation helps explain why corresponding integer sequences arise for a given irrational number $k$, an example being the Fibonacci numbers when one chooses $k=\phi$, the golden ratio.
Here, we note that $41$ is a member of the integer sequence $\frac{1}{2}Q_n$, where $Q_n$ are the Pell-Lucas numbers. 
Just as successive ratios of the Fibonacci numbers form the best rational approximations for $\phi$ ($\lim_{n\rightarrow\infty}F_{n+1}/F_n =\phi$), similarly, $\lim_{n\rightarrow\infty}\frac{1}{2}Q_n/P_n =\sqrt{2}$, where $P_n$ here refers to the Pell numbers.
Thus, if $L+1=\frac{1}{2}Q_n$ for some $n$, then $(L+1)\times \sqrt{2}\approx m\in\mathbb{Z}$, leading to the potential being nearly commensurate with the lattice.
We confirm that for $k=\phi$, these spikes occur at system sizes corresponding to the Fibonacci numbers and other integers $m$ that happen to satisfy $m\times k\approx n$.
This underlying commensurability is likely the culprit that leads to increased resonances and other strange behavior in the spectrum, leading to a spike in the quantities calculated here. This would be expected as a periodic system exhibits vastly different physics from an incommensurate one.
\\

\subsection{Results on LIOM Corrections}\label{sec:liom_results}
\subsubsection{LIOM Corrections: Anderson Model}\label{sec:liom_results_disordered}

\begin{figure*}[t]
    \centering
    \includegraphics[width=\linewidth]{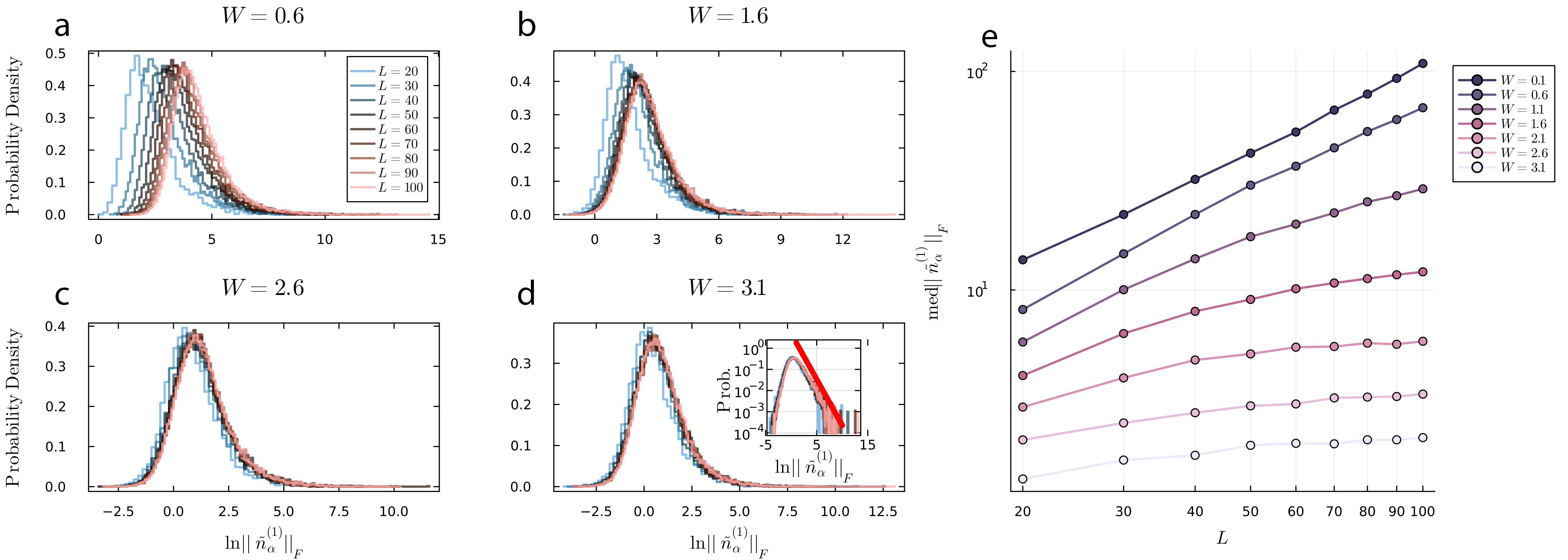}
    \caption{
    Analysis of the Frobenius norm of the first-order corrections to the noninteracting LIOMs, $\|\tilde{n}_\alpha^{(1)}\|_F$, in the weakly interacting Anderson model. Each plot displays all orbitals $\alpha = 1, \dots, L$ across the 500 disorder realizations used to generate the data shown in this figure.
    (a)-(d): Distributions of $\ln(\|\tilde{n}_\alpha^{(1)}\|_F)$ at  $W=0.6, 1.6, 2.6, 3.1$ and for $\alpha = 1,\dots, L$.
    Within each panel, each colored curve represents a different system size. 
    We show calculations between $L=20$ and $L=100$ in steps of $10$.
    The inset in (d) shows the same data as in the main (d) panel, but with the vertical scale shown on log-scale.
    The red line in the inset represents a polynomial $\|\tilde{n}_\alpha^{(1)}\|_F^{-2}$ fit to the tail of the probability distribution of $\|\tilde{n}_\alpha^{(1)}\|_F$--- see the text surrounding Eq.~(\ref{eq:tail_behavior}) for details on the form plotted.
    The convergence of the distributions for large enough $W$ and $L$ indicates the presence of an extensive number of quasiconservation laws after interactions are added, even in the thermodynamic limit.
    (e): Median $\mathrm{med}\|\tilde{n}_\alpha^{(1)}\|_F$ as a function of $L$ on a log-log scale. Each colored curve represents a different disorder strength $W$, from $W=0.1$ to $W=3.1$ in steps of $0.5$. One observes that the curves appear to saturate for large $L$, indicating that at least half of the LIOMs are conserved to second order when interactions are added. See Sec.~\ref{sec:liom_results_disordered} for a full discussion of the results.
    }
    \label{fig:liom_disordered_figure}
\end{figure*}

We present numerical calculations of the Frobenius norm of the LIOM corrections, $\|\tilde{n}^{(1)}_{\alpha}\|_F$, in Fig.~\ref{fig:liom_disordered_figure}, for the weakly--interacting Anderson model. For each $L$, we sample $500$ disorder realizations and record $\|\tilde{n}^{(1)}_{\alpha}\|_F$ for every orbital $\alpha=1, 2, ..., L$ in each disorder realization, giving a total of $500\times L$ datapoints in each distribution.\\

The data in Fig.~\ref{fig:liom_disordered_figure}(a)--(d) suggest that the LIOM corrections converge to a fixed distribution in the thermodynamic limit. At low disorder, finite-size effects play a role, and the eigenstates appear extended. In these cases, the distributions initially shift rightward, but appear to converge at large $L$. 
At high disorder, the wavefunctions are localized and the distributions are converged, even at low $L$. 
Additionally, Fig.~\ref{fig:liom_disordered_figure}(e) shows how the median of these distributions grows with $L$. Here, we find that at low disorder, the median of $\|\tilde{n}^{(1)}_{\alpha}\|_F$ grows as a polynomial with $L$. At higher disorders, we see similar growth of $\mathrm{med}\|\tilde{n}^{(1)}_{\alpha}\|_F$ for low $L$, but the curves appear to have practically saturated with increasing system size. For example, for $W=3.1$, the median of the LIOM correction norms only increased from $1.37$ to $2.11$ between $L=20$ and $L=100$. This, combined with the fact that the growth of the curves of $\mathrm{med}\|\tilde{n}^{(1)}_{\alpha}\|_F$ is clearly slowing down significantly at larger $L$, suggests that at least half of the noninteracting LIOMs admit bounded corrections in the thermodynamic limit for high enough disorder. 
This is also consistent with the fact that the LIOM correction distributions have sufficiently converged at $L=100$ for strong enough disorder.

\begin{figure}[t]
    \centering
    \includegraphics[width=\columnwidth]{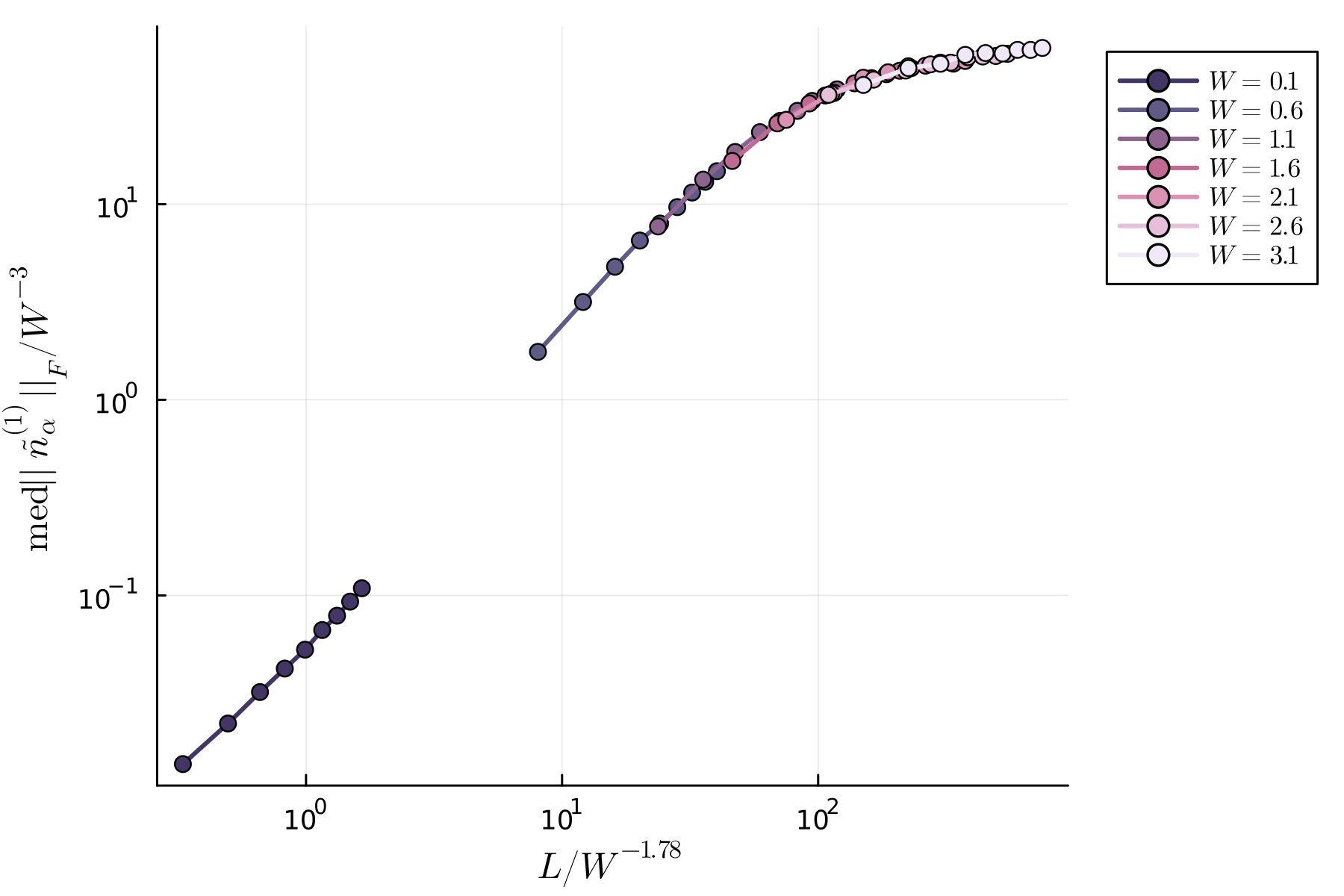}
    \caption{
    Same data as in Fig.~\ref{fig:liom_disordered_figure}(e), but with the system size on the x-axis scaled by the participation ratio, where $\mathrm{PR}\sim W^{-1.78}$ (see Appendix~\ref{app:localization_length} for details), and the y-axis scaled by a phenomenological scaling of $W^{-3}$. 
    The collapse of the curves suggests that the increase of $\mathrm{med}\|\tilde n_\alpha^{(1)}\|_F$ at small $L$ or low disorder strength $W$ is a finite-size effect.}
    \label{fig:collapsed_LIOM_correction}
\end{figure}

Furthermore, in Fig.~\ref{fig:collapsed_LIOM_correction}, we note that the $\mathrm{med}\|\tilde n_\alpha^{(1)}\|_{F}$ vs $L$ curves collapse when the x-axis is appropriately scaled by the participation ratio of the Anderson model, which scales approximately as $\mathrm{PR}(W) \propto W^{-1.78}$ in the range of $W$ considered here (see Appendix~\ref{app:localization_length} for a more detailed explanation) and when the y-axis is scaled phenomenologically with $\sim\mathrm{PR}^{1.7}$.
Here, the participation ratio is used as a proxy\footnote{Some caution is needed in using the median participation ratio as a proxy for the localization length in this analysis. Firstly, both the participation ratio and the localization length depend on the energy of the noninteracting orbital, while our analysis does not consider the energy resolution of the LIOM corrections. Moreover, the power-law dependence of the localization length on the disorder strength is expected not to hold at large $W$, where a more refined ansatz is needed \cite{colbois_breaking_2023}. See Appendix \ref{app:localization_length} for another discussion.} for the localization length $\xi$.
The collapse of these curves suggests that at low disorders, the growth of $\mathrm{med}\|\tilde n_\alpha^{(1)}\|_{F}$ is a finite-size effect.
If the system size $L$ exceeds the localization length, for any disorder strength $\mathrm{med}\|\tilde n_\alpha^{(1)}\|_{F}$ is expected to saturate to a finite value. Put together, these results suggest that even in the thermodynamic limit, a fraction of noninteracting LIOMs have finite and small corrections in the presence of interactions.
These LIOMs will survive for long times, supporting that thermalization can only happen on anomalously slow timescales in these systems: The rate of change of these approximate LIOMs can be rigorously bound by $\lambda^2 \|[\hat V, \hat n_\alpha ^{(1)}]\|$, and a less rigorous, FGR-based estimate suggests a thermalization timescale of at least $O(\lambda^{-4})$ (see Sec.~VII and Appendix F in Ref.~\cite{surace_weak_2023}).

Just as in the distributions of $\|\hat X\|_F$, the long $\|\hat X\|_F^{-2}$ power-law tails appear here for the same reason as they did in the AGP distributions [see the inset in Fig.~\ref{fig:liom_disordered_figure}(d)]. 
This shows that the perturbation theory breaks down for a fraction of the LIOMs, which receive large corrections due to resonances.

Resonances between noninteracting orbitals (not to be confused with many-body resonances) are known to play a big role in MBL, where different methods have been developed to treat them.
One example is provided in Ref.~\cite{de_roeck_rigorous_2023}.
In that work, the authors separate the Hamiltonian into resonating and non-resonating regions, where resonances involve small denominators past some cutoff $\delta$. 
They then treat these resonating regions nonperturbatively, which is possible given the small sizes of those resonating regions. 
In our work, we do not attempt to deliver any special treatment to these resonances since our work is, in spirit, perturbative.
Rather, our goal is to extract as much insight as possible from the perturbative expansion, while being aware of the presence of resonances and accounting for their effects in the naive perturbative treatment.

\subsubsection{Dependence of LIOM Corrections on Energy and Participation Ratio}\label{sec:liom_correction_dependence}
\begin{figure*}[t]
    \centering
    \includegraphics[width=\linewidth]{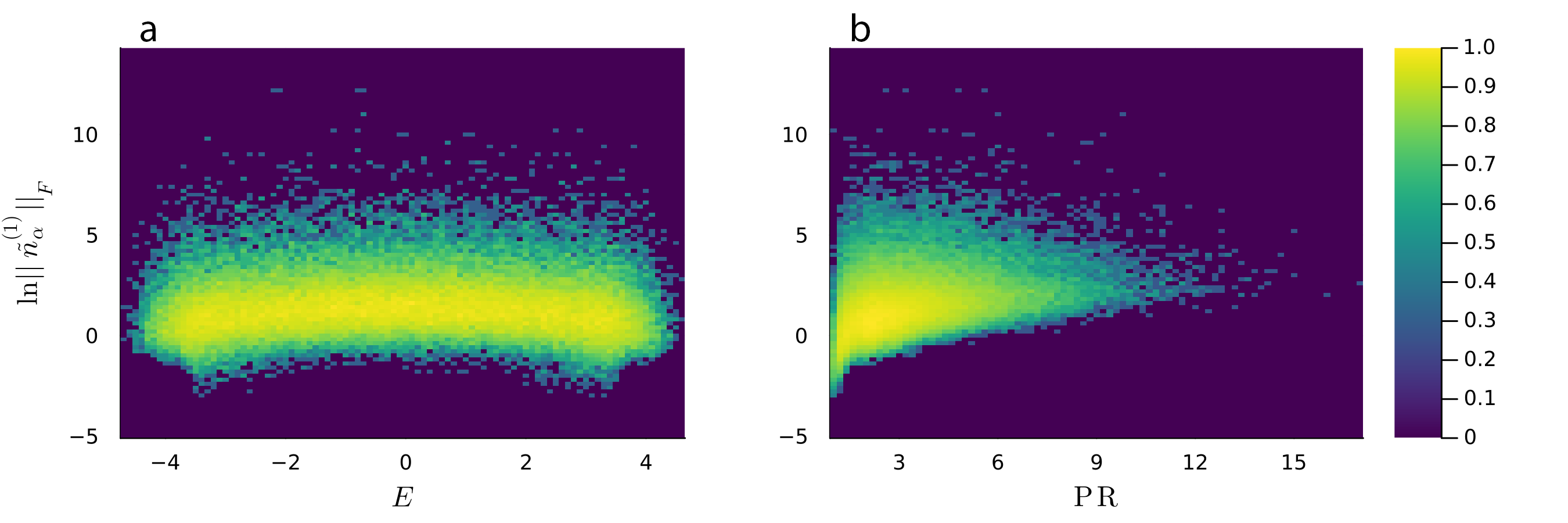}
    \caption{
    A density scatter plot of the natural log of the LIOM corrections $\ln\|\tilde n^{(1)}_\alpha\|_F$ at $W=3.1$ and $L=80$ against (a) the energy $E$ of its corresponding orbital $\alpha$ and (b) the participation ratio (PR) of the same orbital. Here, we use 500 different disorder realizations. For each disorder realization, we keep corrections to all orbitals $\alpha = 1, \dots, L$. Yellower colors indicate a higher density of points.
    }
    \label{fig:liom_correction_dependence_on_e_pr}
\end{figure*}
The defining characteristic of the distributions of $\|\tilde n^{(1)}_\alpha\|_F$ is the presence of slowly decaying power-law tails, which cause the mean to diverge. These tails also occur with large disorder strengths. To investigate the origin of these large corrections, we compute the dependence of the LIOM corrections on both the orbital energy and participation ratio when the disorder is larger at $W=3.1$, and show our results in Fig.~\ref{fig:liom_correction_dependence_on_e_pr}. One might intuitively expect that large corrections arise from rare thermal regions where the variation in the onsite potential is anomalously small. However, the data in Fig.~\ref{fig:liom_correction_dependence_on_e_pr} tells a more complex and partly counterintuitive story.

Figure~\ref{fig:liom_correction_dependence_on_e_pr}(a) shows that the \emph{typical} LIOM corrections are roughly correlated with energy, with larger corrections in the middle of the spectrum, although the correlation is not strong as long as one is away from the edges of the spectrum.
This is what one would expect, as states near the middle of the spectrum tend to be more delocalized. Delocalized orbitals typically produce larger numerators in the expression for the matrix elements of the AGP, Eq.~(\ref{eq:Xelements}). This trend is echoed in panel (b), where the typical and lower-bound LIOM corrections increase with the participation ratio (PR) of the corresponding orbital.

In contrast, the very large LIOM corrections that form the long tails tell a counterintuitive story.
In Fig.~\ref{fig:liom_correction_dependence_on_e_pr}(b), these large corrections appear anticorrelated with the participation ratio: they mostly occur with orbitals with a small PR, and seem less likely for those with large PR.
A possible explanation is that rare thermal regions with large participation ratios exhibit stronger level repulsion, which suppresses small denominators in Eq.~(\ref{eq:Xelements}). With this, it is reasonable to conjecture that the dominant mechanism behind the largest LIOM corrections is not rare thermal regions, but rather energy resonances that produce small four-energy denominators in Eq.~(\ref{eq:Xelements}). Since the numerators would be very small for orbitals that are spatially distant, we can further guess that these resonances likely occur between orbitals that are close in real space. This also provides another explanation for why the tails exist even for high disorder strengths. We will further reinforce this conjecture through calculations of the charge transport capacity in the second half of this work.

\subsubsection{LIOM Corrections: Quasiperiodic Model}\label{sec:liom_results_quasiperiodic}

\begin{figure*}[t]
    \centering
    \includegraphics[width=\linewidth]{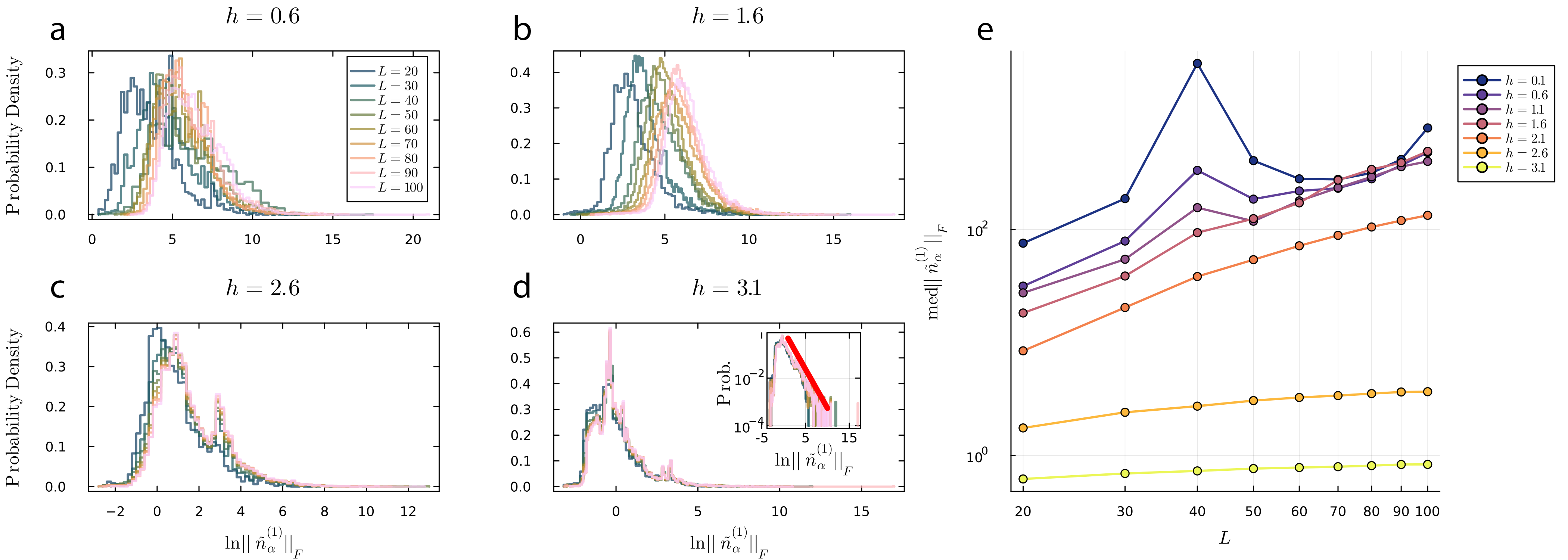}
    \caption{Analysis of the Frobenius norm of the first-order corrections to the noninteracting LIOMs, $\|\tilde{n}_\alpha^{(1)}\|_F$, for the weakly-interacting quasiperiodic model. 
    (a)-(d): Distributions of $\ln(\|\tilde{n}_\alpha^{(1)}\|_F)$ based on 500 phase shifts $\delta$, sampled at potential depths $h=0.6, 1.6, 2.6, 3.1$, with $\alpha=1\dots L$. Each curve represents the distribution calculated at different system sizes, from $L=20$ to $L=100$ in steps of 10. 
    The inset in (d) shows the same data as in the main (d) panel but with the vertical scale shown on a log scale.
    The red line in the inset represents a polynomial $\|\tilde{n}_\alpha^{(1)}\|_F^{-1.75}$ fit to the tail of the probability distribution of $\|\tilde{n}_\alpha^{(1)}\|_F$--- see the text surrounding Eq.~(\ref{eq:tail_behavior}) for details on the form plotted. Similarly to what was observed for the interacting Anderson model, the quasiperiodic model seems to admit an extensive number of quasiconservation laws for the localized phase $h>2$.
    (e): Median $\mathrm{med}\|\tilde{n}_\alpha^{(1)}\|_F$ as a function of $L$ on a log-log scale. Each colored curve represents a different potential depth 
    $h$, from $h=0.1$ to $h=3.1$ in steps of 0.5. One observes that the curves for $h<2$ grow with $L$, indicating that the corrections are, unsurprisingly, delocalized. 
    For $h>2$ the curves saturate, indicating that the noninteracting LIOMs persist to first order when interactions are added to the localized regime of the quasiperiodic model. For a full discussion of the results, see Sec.~\ref{sec:liom_results_quasiperiodic}}
    \label{fig:liom_quasiperiodic_figure}
\end{figure*}

The calculations of the Frobenius norm of the LIOM corrections for the interacting quasiperiodic model are shown in Fig.~\ref{fig:liom_quasiperiodic_figure}. 
The observed behavior of $\|\tilde n_\alpha^{(1)}\|$ here closely mirrors that of the interacting disordered model. 
In the localized regime ($h>2$), the data suggest the existence of quasiconserved quantities in the thermodynamic limit, where the distributions of the corrections converge for large $L$. 
The fit in the inset in  Fig.~\ref{fig:liom_quasiperiodic_figure}(d) indicates that these distributions, similar to those for the weakly-interacting Anderson model, feature slower-decaying power-law tails 
(We found a tail $\propto \|\tilde{n}_{\alpha}^{(1)}\|_F^{-1.75}$ fits the best for the LIOM corrections, just as with the AGP). 
This same scaling for the tail appeared in the AGP distributions, so, unsurprisingly, it shows up here as well.
Just as in the disordered case, the resonances here cause a breakdown of the perturbation theory for a portion of the LIOMs. 
For the delocalized regime, the data shows that $\|\tilde n_\alpha^{(1)}\|$ grows with $L$ for all the system sizes considered, suggesting that the correction to $\tilde n_\alpha^{(0)}$ is delocalized.
This is unsurprising behavior as the starting conserved quantities were already delocalized.

Figure~\ref{fig:liom_quasiperiodic_figure}(e) shows the growth of $\mathrm{med}\|\tilde{n}^{(1)}_{\alpha}\|_F$ with system size. For $h<2$, the quasiperiodic model features extended wavefunctions.
This is reflected in the power-law growth of $\mathrm{med}\|\tilde{n}^{(1)}_{\alpha}\|_F$. For $h>2$, the growth of $\mathrm{med}\|\tilde{n}^{(1)}_{\alpha}\|_F$ appears to be suppressed, reflecting the presence of a localized regime.
The separation of these curves is a clear indicator of the localization transition in the unperturbed model.

We note that the $\mathrm{med}\|\tilde{n}^{(1)}_{\alpha}\|_F$ curves exhibit a clear finite-size scaling collapse, analogous to the behavior observed in the Anderson model. For completeness, details of the scaling analysis for the quasiperiodic model across all quantities examined in this work are provided in Appendix~\ref{app:quasiperiodic_collapse}.

Finally, we remark that the calculated values for $\|\tilde n_\alpha^{(1)}\|_F$ have the same anomalous peaks as the AGP for the quasiperiodic model. This is unsurprising, since the same four-energy denominators appear and thus would be affected by the same commensuration issues discussed earlier in Sec.~\ref{sec:AGP_results}.

\subsubsection{LIOM Corrections: Conclusions}\label{sec:liom_conclusions}
Our numerical results show that, even when perturbed by weak interactions, the noninteracting LIOMs continue to exist to first order: specifically, we saw that the two localized models considered still preserve an $O(1)$ fraction of the noninteracting LIOMs up to first order. 
As previously mentioned, these LIOMs hence survive until time scales of $O(\lambda^{-4})$. 
From the perspective of integrability breaking, this makes interactions a ``weak'' perturbation to localized models-- consistent with the existence of MBL. However, the results raise many questions. In particular, the slow tails in the distributions of $\|\tilde n^{(1)}_\alpha\|_F$ show that perturbation theory breaks down for a finite fraction of the LIOMs--- the fate of those LIOMs remains unclear when interactions are added.
From just the norms of the corrections, it is unclear what the spatial structures of the resonances are that cause the breakdown of these LIOMs---in particular, do the resonances mix LIOMs locally, or nonlocally?
In principle, some such locality information is contained in the operator content of the corresponding corrections, but it is not easy to analyze.
In the next section, we introduce the \emph{charge transport capacity}, which can frame this question from a different and very physically relevant perspective.

\section{Charge Transport}\label{sec:transport}

In the rest of the paper, we turn our attention to the properties of charge transport in the localized models discussed in this work.
Transport suppression is a fundamental feature of MBL and the key reason why these systems fail to thermalize.
Unlike the LIOMs, which provide a useful framework for understanding the properties of MBL, but whose derivation from physical observables is not obvious, transport has a clear physical interpretation that makes it a more tractable quantity to study and interpret. 

We aim to argue that transport properties across a given link can be reliably studied even in the absence of full perturbative control over all LIOMs, provided that the surrounding regions remain perturbatively accessible.
As shown earlier, the first-order perturbation theory breaks down for a fraction of the LIOMs due to resonances.
One might worry that these uncontrolled corrections could destabilize the localized phase and lead to delocalization. 
However, one proposal---consistent with many-body localization---is that for strong disorder, such resonances have only local effects.
Rather than causing global delocalization, they signal the need for a non-perturbative reshuffling of LIOMs in the resonant region.

\begin{figure}[t]
    \centering
    \includegraphics[width=\columnwidth]{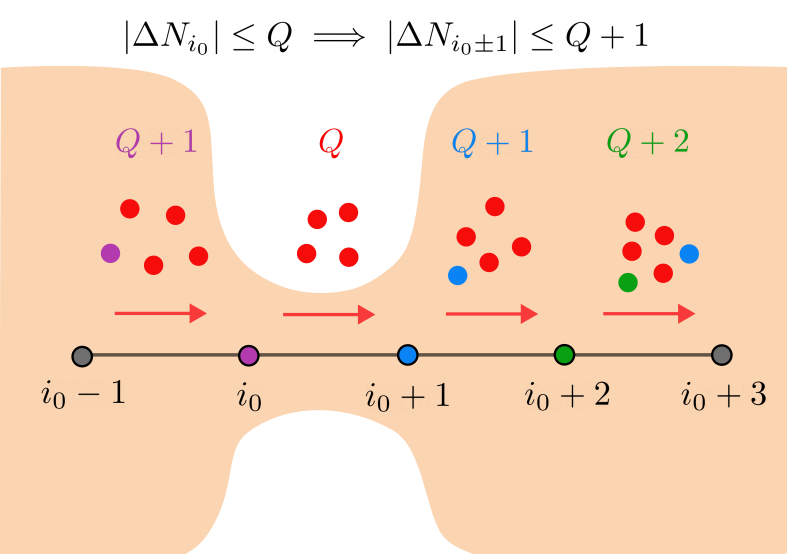}
    \caption{Illustration of the effect of bottlenecks in transport. $\Delta N_{i_0}$ represents the change in the total charge to the right of $i_0$, as in Eq.~(\ref{eq:change_in_charge}).
    If only Q charges can pass through the link that starts at $i_0$ (i.e., between $i_0$ and $i_0+1$), there cannot be more than $Q+1$ passing through the link at $i_0 \pm 1$, and so on. 
    Consequently, a single bottleneck---meaning a link permitting only a small number of particles to pass---can suppress transport throughout the entire lattice.
    }
    \label{fig:bottleneck_illustration}
\end{figure}

To probe localization more robustly, even in the presence of local resonances, we focus on a different quantity: the maximum total charge that can be transported across a given link, maximized over all evolution times and initial states, which we refer to as the \emph{charge transport capacity}.
If we can locally bound this quantity and demonstrate that, in the thermodynamic limit, there is a finite probability that it remains  $O(1)$ (i.e., does not scale with the system size), this indicates the presence of ``bottlenecks'' that effectively block transport through the entire system (see Fig.~\ref{fig:bottleneck_illustration}).
As we will argue in Sec.~\ref{subsec:pert_transp}, such a bound can be established within first-order perturbation theory, even when resonances are present, as long as they occur sufficiently far from the link in question. 
For strong enough disorder, we expect this bound to persist to all orders in perturbation theory and to represent true localization.

\begin{figure*}[t]
    \centering
    \includegraphics[width=300px]{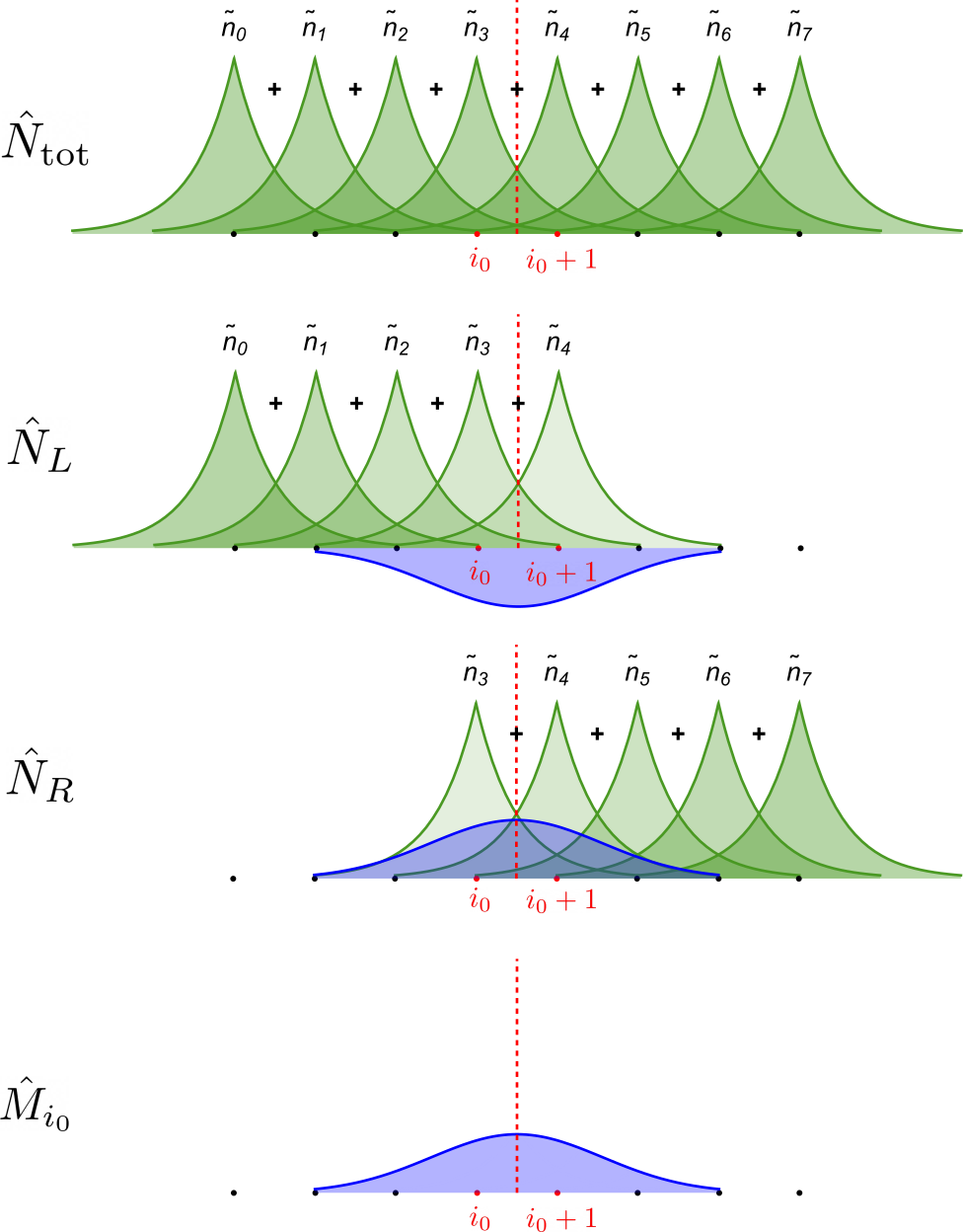}
    \caption{
    Illustration of the argument that 
$\hat M_{i_0} = \hat N_R - (\hat N_R)_\mathrm{diag} = -[\hat N_L - (\hat N_L)_\mathrm{diag}]$ only depends on degrees of freedom near $i_0$ deep in the MBL phase. 
    Here, the green objects represent the MBL LIOMs $\tilde n_{\alpha}$, which commute with the Hamiltonian, while the blue object represents noncommuting (off-diagonal) terms.
    The saturation of the green color represents the magnitude of $\mathrm{Tr}[\tilde n_\alpha \hat N_R]$. 
    If we can write $\hat N_{\text{tot}} = N_L + N_R$ as a sum of LIOMs $\tilde n_{\alpha}$, and $\hat N_L$ ($\hat N_R$) involves a sum of LIOMs and other off-diagonal operators mostly to the left (right) of the link at $i_0$, then it must follow that $\hat N_R$ with its diagonal component removed, $\hat M_{i_0}$, is only comprised of the terms (blue object) near the site $i_0$.
    }    \label{fig:transport_local_illustration}
\end{figure*}

\subsection{Charge Transport Capacity}
In this section, we introduce a measure of the charge transport capacity (for any process) across a link, inspired by the integrated energy current operator introduced in Ref.~\cite{de_roeck_rigorous_2023}.
We consider an OBC chain and fix a site $i_0$ where $1\leq i_0\leq L-1$. Here, we fix $t=1$.
We can define the charge current operator $\hat J_{i_0}$ across the link between $i_0$ and $i_0+1$, with rightward current being defined as positive:
\begin{equation}
\begin{aligned}
\label{eq:current}
    \hat{J}_{i_0} \equiv & -i[\hat{H}, \hat{N}_L (i_0)] = i[\hat H, \hat{N}_R(i_0)]  \\
    = &  -i\left(\hat{c}_{i_0}^\dagger \hat{c}_{i_0+1} - \hat{c}_{i_0+1}^\dagger \hat{c}_{i_0} \right).
\end{aligned}
\end{equation}
Here $\hat N_L(i_0) \equiv \sum_{i=1}^{i_0}\hat{c}_i^{\dagger}\hat{c}_{i}$ is the total number operator for all the sites to the left of the link. 
Similarly, $\hat N_R(i_0) = \sum_{i=i_0+1}^{L}\hat{c}_i^{\dagger}\hat{c}_{i}$ is the total number operator for all the sites to the right.
Note that in our model with only density-density interactions, the expression of the current operator $\hat J_{i_0}$ is the same for noninteracting or interacting models, independent of the interaction strength $\lambda$.

The maximal net charge transferred across $i_0$ over time $t$ (optimized over all possible initial states) is the operator norm of 
\begin{equation}\label{eq:change_in_charge}
    \Delta \hat N_{i_0} (t)\equiv  \hat N_R(i_0,t)-\hat N_R(i_0,0)=\int_0^t \hat J _{i_0}(s) ds,
\end{equation}
where $\hat N_R(i_0,t)\equiv e^{i\hat Ht}\hat N_R(i_0)e^{-i\hat H t}$ and $\hat{J}(s) \equiv e^{i\hat{H}s} \hat{J}_{i_0} e^{-i\hat{H} s}$ are time-evolved operators in the Heisenberg picture.

To make the charge transport practically computable and to quantify this intuition for the MBL phase, we note that the relation
\begin{equation}
\label{eq:defM}
    \hat J_{i_0} = i[\hat H, \hat M_{i_0}]
\end{equation}
is satisfied by any operator $\hat M_{i_0}$ of the form  $\hat M_{i_0} = \hat{N}_R(i_0) + \hat{D}$, where $\hat D$ is an operator that commutes with the Hamiltonian. It then follows that
\begin{equation}
    \left\|\Delta \hat N_{i_0}(t)\right\|_{\text{op}}=\left\|\hat M_{i_0}(t)-\hat M_{i_0}(0)\right\|_{\text{op}} \le 2\left\|\hat M_{i_0}\right\|_{\text{op}}.
\end{equation}
To tightly bound the charge transport, ideally, one would want to find $\hat D$ such that the operator norm of $\hat M_{i_0}$ is minimized.
In practice, minimizing the operator norm of $\hat M_{i_0}$ is challenging. Therefore, as we will discuss later, we instead choose to minimize its \emph{Frobenius norm}, which is much simpler. It does not provide an optimally tight upper bound to $\|\Delta\hat N_{i_0}(t)\|_{\text{op}}$, but gives a good probe of localization regardless.

Denoting many-body eigenstates/energies of $H$ as $\{\ket{\mu}, E_\mu \}$, and assuming non-degenerate spectrum (which is reasonable for a  finite-size disordered system), the matrix elements of the many-body operator $\hat{M}_{i_0}$ for $\mu \neq \nu$ are uniquely fixed
\begin{equation}
\big(\hat{M}_{i_0} \big)_{\mu\nu} = \frac{(\hat{J}_{i_0})_{\mu\nu}}{i(E_\mu - E_\nu)} = \big(\hat{N}_R \big)_{\mu\nu} ~.
\end{equation}
The minimal Frobenius norm is obtained by setting $\big(\hat{M}_{i_0} \big)_{\mu\mu} = 0$, $\forall \mu$, in which case we can also write as
\begin{equation}
\hat{M}_{i_0} = [\hat{N}_R(i_0)]_{\text{off-diag}} = \hat{N}_R - [\hat{N}_R]_{\text{diag}} ~.
\label{eq:NRoffdiag}
\end{equation}
Here, the subtracted ``diagonal part'' is with respect to the full Hamiltonian $\hat{H}$, which is how information about its ``dynamics'' enters the above expression.

From now on, we will consider $\hat{M}_{i_0}$ given by Eq.~(\ref{eq:NRoffdiag}) and we will call this operator the {\it charge transport capacity operator}, and its operator norm the charge transport capacity,
as norm of $\hat M$ provides a strict upper bound on the number of particles that can cross $i_0$, under \emph{any} process and at \emph{any} time. Consequently, if $\|\hat M\|_{\text{op}}$ remains finite in the thermodynamic limit, it serves as a clear signature of nonergodicity: regardless of how the system is initialized or how long it evolves, no more than a finite number of particles can ever be transported across $i_0$.
We also claim that localization (in the sense of the Hamiltonian being described by an l-bit model) implies that $\hat M$ is quasilocal around $i_0$.
This illustrates the power of $\hat M$ to describe localization: if its norm is finite in the thermodynamic limit, that would imply nonergodicity. Conversely, localization also implies that its norm would be finite.

Continuing with general discussion, we can now more formally argue that in the MBL phase, assuming the LIOM description holds, the above $\hat{M}_{i_0}$ is related to LIOMs and noncommuting partners (raising/lowering operators) localized near $i_0$ and is localized near $i_0$, and hence the charge transport capacity is bounded by a fixed number in the thermodynamic limit.
Indeed, let us take the LIOM description posited in Eq.~(\ref{eq:liom_equation}) and express the total charge operator $\hat N_{\text{tot}} = \sum_j \hat n_j$ in terms of $\tilde d_\alpha, \tilde d_{\alpha}^\dagger$.
Here, $\tilde d_\alpha, \tilde d_{\alpha}^\dagger$ are the annihilation and creation operators associated with the corresponding LIOM $\tilde n_\alpha$.
Since $\hat N_{\text{tot}}$ commutes with the Hamiltonian, and assuming the Hamiltonian spectrum is non-degenerate, it follows that $\hat N_{\text{tot}}$ can be expressed as a sum of terms containing only $\tilde n_\alpha$'s.\footnote{If we could in addition assume that there is an ``adiabatic-type'' one-to-one correspondence between the unperturbed LIOMs $\hat{d}_\alpha^\dagger \hat{d}_\alpha$ and the new LIOMs $\tilde{d}_\alpha^\dagger \tilde{d}_\alpha$, 
more specifically that there is a quasi-local unitary $U$ relating $\tilde{n}_\alpha \equiv \tilde{d}_\alpha^\dagger \tilde{d}_\alpha
= U \hat{d}_\alpha^\dagger \hat{d}_\alpha U^\dagger$, requiring also that $U$ commutes with $\hat{N}_{\text{tot}}$, we would conclude $\hat{N}_{\text{tot}} = \sum_\alpha \tilde{ n}_\alpha$.
However, this specific form is not used in the argument that follows.}

Consider now the parts $\hat{N}_L(i_0)$ and $\hat{N}_R(i_0)$, $\hat{N}_L(i_0) + \hat{N}_R(i_0) = \hat{N}_{\text{tot}}$.
The assumed quasi-locality of the LIOMs implies that any local operator---in particular, each $\hat{n}_j$---is expressible in terms of quasi-local strings containing products of $\tilde{d}_\alpha,\tilde{d}_\alpha^\dagger, \tilde{n}_\alpha$ localized near $j$.
For simplicity, let us assume that there is a maximal range $r_{\text{max}}$ for these strings.
Now recall that $\hat{N}_{\text{tot}}$ contains only strings of $\tilde{n}$'s.
Since any string in $\hat{N}_R(i_0)$ that starts farther than $r_{\text{max}}$ away from $i_0$ is not present in $\hat{N}_L(i_0)$ and hence must be the same as in $\hat{N}_{\text{tot}}$, such a string contains only $\tilde{n}$'s and hence commutes with $\hat{H}$.
Hence $\hat{M}_{i_0}$, which is the off-diagonal part of $\hat{N}_R(i_0)$, obtains contributions only from the strings in $\hat{N}_R(i_0)$ that are localized near $i_0$, so $\hat{M}_{i_0}$ is indeed localized near $i_0$. See Fig.~\ref{fig:transport_local_illustration} for a visual depiction of this argument.
We note that the restriction to strictly local strings (with size at most $r_{\text{max}}$) vs quasi-local (whose amplitudes decrease sufficiently fast with the string size) is not important for this argument.

Note that the above discussion relates $\hat{M}_{i_0}$ to strings of $\tilde d_\alpha, \tilde d_\alpha^\dagger, \tilde{n}_\alpha$ near $i_0$ and hence implicitly to the ``LIOM two-level systems'' near $i_0$, and we can view it as some cummulative property of the LIOMs near $i_0$.
However, we never need to identify LIOMs precisely to calculate $\hat{M}_{i_0}$, and its interpretation as the bound on the maximal charge transport capacity is very direct and physical, independent of the LIOM picture of the MBL phase.
This is the main appealing feature of this new characterization of the many-body system, potentially in the MBL phase. 

We note that the above argument relies on the validity of the LIOM model in the MBL phase.
However, the possible double-logarithmic growth of the number entropy in the MBL regime \cite{bardarson_unbounded_2012,kiefer-emmanouilidis_evidence_2020,kiefer-emmanouilidis_slow_2021,aceituno_chavez_ultraslow_2024} suggests that more subtle transport phenomena may persist even deep within the localized phase; whether this is true is not entirely resolved in the literature~\cite{luitz_absence_2020,ghosh_resonance-induced_2022,sierant_many-body_2025}.
Regardless, we emphasize that our calculations of the charge transport in this paper are fully independent of any assumptions about the form of the LIOM description in the localized regime.

We briefly note that the charge transport capacity introduced here does not have a straightforward relation to the more commonly-studied DC conductivity. For example, scaling analyses of conductivity in 1D localized systems were carried out in Ref.~\cite{anderson_new_1980}. We suspect that the two quantities capture related aspects of transport and may share a deeper connection, but we refrain from making any rigorous claims here.

Finally, we note that this operator is easy to calculate numerically if one knows the full spectrum of $\hat H$; this could be an interesting new measure to explore in the full interacting problem, although in this work we will primarily focus on perturbative characterizations.

\begin{figure}[t]
    \centering
    \includegraphics[width=\columnwidth]{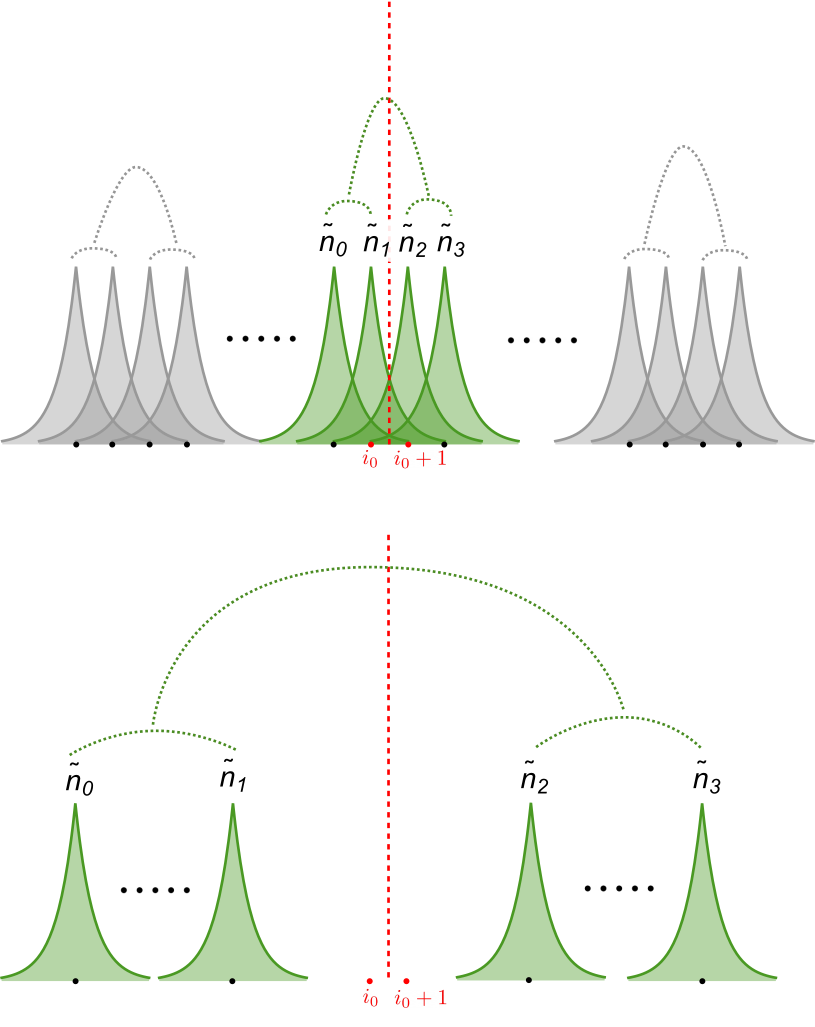}
    \caption{Heuristic illustration of how different spatial structures of resonances affect first-order transport in perturbation theory.
    The orbitals depicted represent \emph{noninteracting} LIOMs. 
    We expect that the charge transport across a given link is affected by resonances that go across the link. 
    Top: if only local resonances dominate, only orbitals that resonate near the link (green) will contribute, while those far away (gray) will not.
    Since the number of locally resonating LIOMs at a given link does not scale with $L$, the charge transport capacity should remain bounded in the thermodynamic limit.
    Bottom: If long-range resonances dominate, resonances between LIOMs depicted in green will influence the charge transport. 
    We expect that the number of cross-link resonances increases with system size $L$, leading to a charge transport capacity that grows with $L$.
    }
    \label{fig:resonating_transport_figure}
\end{figure}

\subsection{Analytical Forms of Noninteracting Transport and Its Perturbative Correction}
\label{subsec:pert_transp}

In the interacting case, the operator $\hat M_{i_0}$ with minimal Frobenius norm can, in principle, be obtained via exact diagonalization for small system sizes. However, our focus here is instead to compute this quantity perturbatively in the limit of weak interactions. This allows us to explore system sizes much larger than what is allowed by ED. To this end, we present the analytical forms of $\hat M_{i_0}^{(0)}$, which is $\hat M_{i_0}$ for the noninteracting case, and $\hat M_{i_0}^{(1)}$, which is the first-order correction to $\hat M_{i_0}^{(0)}$ when weak interactions are added.

To start, we want to find the operator $\hat M^{(0)}_{i_0}$ in the noninteracting case.
For the sake of concision, we omit the $i_0$ subscript in the quantities $\hat J_{i_0}$ and $\hat M_{i_0}$ below, and let the $i_0$ dependence be implicitly assumed.
We note that in all of the quantities calculated in this work, we fix $i_0=\left \lfloor\frac{L}{2}\right \rfloor$, unless specified.
In the rest of the paper, we define $\hat J_{\lfloor L/2 \rfloor}\equiv \hat J$, as we primarily focus on the transport capacity going through the middle link $i_0=\left \lfloor\frac{L}{2}\right \rfloor$.

The operator $\hat M^{(0)}$ is a solution of the equation
\begin{equation}
\label{eq:nonintM0}
    \hat J=i[\hat H_0,\hat M^{(0)}].
\end{equation}

We can write $\hat M^{(0)}$ as a fermion bilinear operator, with:
\begin{equation}
    \hat M^{(0)} = \sum_{\alpha, \beta} M_{\alpha\beta}^{(0)} \hat{d}^\dagger_{\alpha} \hat{d}_{\beta}.
\end{equation}
Then, from Eq.~(\ref{eq:nonintM0}), we can write the (off-diagonal) matrix elements $M^{(0)}_{\alpha\beta}$ in terms of the noninteracting eigenstates $\phi_\alpha$ and eigenvalues $\epsilon_\alpha$, as follows
\begin{equation}
\label{eq:M0albt}
     M^{(0)}_{\alpha\beta} = \frac{\phi_{i_0, \beta} \phi^*_{i_0+1, \alpha}-\phi^*_{i_0, \alpha} \phi_{i_0+1, \beta} }{\epsilon_\alpha - \epsilon_\beta}.
\end{equation}
To minimize the Frobenius norm, we set the diagonal matrix elements $M^{(0)}_{\alpha\alpha}=0$.
The full derivation of matrix elements can be found in Appendix \ref{app:a-derivation-transport}. We also note that the denominators here are not an issue, as one can simply rewrite these matrix elements in a form that does not involve any energy denominators. See Appendix \ref{app:noninteracting_transport_details} for more details and an analysis of the scaling of $\|\hat M^{(0)}\|_{F}$ and $\|\hat M^{(0)}\|_{\text{op}}$.

Now we consider the first-order perturbative correction to $\hat M^{(0)}$ for weak interactions.
As noted above, the current operator $\hat J$ does not change when we add density-density interactions.
We then compute the first-order correction $\hat M^{(1)}$ to the operator $\hat M^{(0)}$ such that Eq.~(\ref{eq:defM}) is valid up to higher order terms: 
\begin{equation*}
    \hat J = i\left[\hat H_0+\lambda \hat V, \hat M^{(0)}+\lambda \hat M^{(1)}\right]+O(\lambda^2).
\end{equation*}
We thus find that $\hat M^{(1)}$ has to satisfy the relation $[\hat H_0, \hat M^{(1)}]+[\hat V, \hat M^{(0)}]=0$. 
In this case, $\hat M^{(1)}$ is a four-fermion operator:
\begin{equation*}
\hat M^{(1)}=\sum_{\alpha\beta\gamma\delta}M^{(1)}_{\alpha\beta\gamma\delta} \hat{d}^{\dagger}_\alpha\hat{d}^{\dagger}_\beta\hat{d}_\gamma\hat{d}_\delta.
\end{equation*}
The ``matrix elements'' $M^{(1)}_{\alpha\beta\gamma\delta}$ can be derived exactly in terms of the interaction matrix elements $V_{\alpha\beta\gamma\delta}$ and the matrix elements of $\hat M^{(0)}$.
To minimize the Frobenius norm, we can set the ``diagonal''\footnote{Here we use ``diagonal'' to refer to terms that commute with $\hat H_0$.} terms  to zero, $M^{(1)}_{\alpha\beta\alpha\beta}=M^{(1)}_{\alpha\beta\beta\alpha}=0$, and uniquely determine all the other terms
\begin{align}\label{eq:M1}
M^{(1)}_{\alpha\beta\gamma\delta}&=\sum_{\mu}(V_{\alpha\beta\gamma\mu} M^{(0)}_{\mu\delta} + V_{\alpha\beta\mu\delta} M^{(0)}_{\mu\gamma} \nonumber \\ &- V_{\alpha\mu\gamma\delta} M^{(0)}_{\mu\beta} - V_{\mu\beta\gamma\delta} M^{(0)}_{\mu\alpha}) / (\epsilon_\alpha + \epsilon_\beta - \epsilon_\gamma - \epsilon_\delta).
\end{align}
The full derivations and formulas for these operators can be found in Appendix \ref{app:a-derivation-transport}.

We note that Eq.~(\ref{eq:M1}) shares structural features with other quantities computed in perturbation theory, particularly the presence of energy denominators of the form $\epsilon_\alpha + \epsilon_\beta - \epsilon_\gamma - \epsilon_\delta$, which signal the possibility of resonances.
However, one possibility is that $\hat M^{(1)}$ can remain finite even in the presence of resonances, provided these are local and occur far from the link.
This idea is illustrated in Fig.~\ref{fig:resonating_transport_figure}.
If the first-order resonances are non-local and involve orbitals spread throughout the system, they will contribute significantly to the transport.
In that case, the number of such resonances---and hence the charge transport capacity---is expected to grow with the system size. 
Conversely, if the resonances are predominantly local, then only orbitals near the link $i_0$ will significantly contribute to the transport capacity, and their number remains finite, leading to a charge transport capacity that does not scale with the system size.

Just as with the LIOM corrections, in principle, one could compute higher-order corrections to $\hat M$.
In practice, higher-order corrections are challenging to compute numerically due to unfavorable polynomial scaling with $L$. 
Thus, we do not attempt calculations of these quantities here and leave analysis of higher-order terms to future work.
We note, however, that $\hat M^{(1)}$ can be used to give rigorous bounds to the charge transported up to a finite time in the interacting case (see Appendix \ref{app:a-derivation-transport} for derivation):
\begin{align}
\|\Delta \hat N(t)\|\le 2\|\hat M^{(0)}\|+2\lambda \|\hat M^{(1)}\|+\lambda^2 t \left\|i[\hat V, \hat M^{(1)}]\right\|.
\end{align}

In the following sections, we analyze the behavior of $\hat{M}$ across both localized models studied in this work, considering both interacting and noninteracting cases.  
In the absence of interactions, we show that $\|\hat{M}^{(0)}\|_F$ saturates at large $L$, serving as an effective diagnostic for localization.  
Upon introducing interactions, we provide evidence that $\|\hat{M}^{(1)}\|_F$ remains finite in the localized regime even as $L \rightarrow \infty$.  
Together with the observation of low correlations between $\|\hat{M}^{(1)}\|_F$ at distant sites, this suggests that transport remains perturbatively bounded to first order in the interaction strength.

\subsection{Results on the Charge Transport Capacity in Noninteracting Systems}\label{sec:non_interacting_transport_results}

To gain some intuition on the behavior of these transport operators, we first analyze the behavior of the norm of the charge transport capacity operator, $\|\hat M^{(0)}\|_F$ for the two single-particle localization models considered in this paper, starting with the Anderson model.
In Appendix~\ref{app:noninteracting_transport_details}, we provide some additional details, including alternative expressions and bounds for the matrix elements of $\hat M^{(0)}$ and its Frobenius and operator norms, and semi-analytical understandings of scalings.

\subsubsection{Transport in the Single-Particle Anderson Model}\label{sec:non_interacting_transport_disordered}

\begin{figure*}[t]
    \centering
    \includegraphics[width=\linewidth]{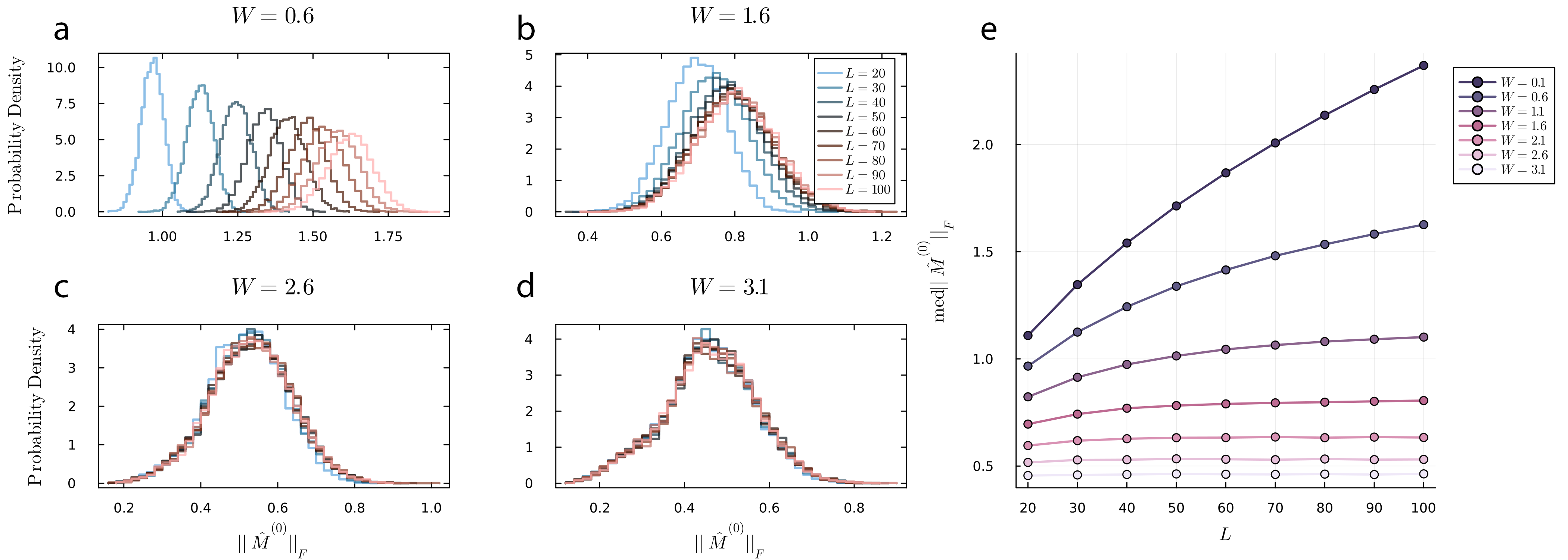}
    \caption{Analysis of $\|\hat M^{(0)}\|_F$ for the noninteracting Anderson Model based on 10000 disorder realizations, open boundary conditions, and with $i_0=\lfloor L/2\rfloor$. 
    (a)--(d): Calculated probability density functions of $\|\hat M^{(0)}\|_F$. 
    Each panel represents different disorder strengths $W=0.6,1.6,2.6,3.1$.
    Within each panel, each colored curve represents different system sizes from $L=20$ to $L=100$ in steps of $10$.
    The distributions are converged at large $L$, consistent with the presence of localization for arbitrarily weak disorder.
    (e): Medians of the distributions in (a)-(d), $\mathrm{med}\|\hat M^{(0)}\|_F$, as a function of $L$. 
    Each curve represents a different disorder strength. 
    Here, one sees that the curves saturate at large $L$, reflecting that, in the thermodynamic limit, charge cannot freely flow in the single-particle Anderson model due to localization. In the no disorder case $W=0$, the median scales as $\approx \frac{1}{4}\sqrt{L}$.
   }
\label{fig:noninteracting_transport_disordered_figure}
\end{figure*}

We analyze the behavior of the Frobenius norm of the charge transport capacity operator, $\|\hat M^{(0)}\|_F$, in the noninteracting disordered model\footnote{Although much larger system sizes are accessible in the noninteracting model, we restrict demonstrations to a maximum size $L=100$ for easier comparison with the weakly-interacting case, where we were only able to calculate quantities up to $L=100$.} and present our results in Fig.~\ref{fig:noninteracting_transport_disordered_figure}.
We refer the reader to Appendix \ref{app:operator_norm_data_noninteracting_transport} for additional calculations for the operator norm $\|\hat M^{(0)}\|_{\text{op}}$. 

When the disorder is turned on, all of the eigenstates of the Anderson Model are localized in the thermodynamic limit, and so the charge transport is suppressed. 
We see that this is indeed the case: the distributions of $\|\hat M^{(0)}\|_F$ in Figs.~\ref{fig:noninteracting_transport_disordered_figure}(a)-(d) appear to converge onto a fixed distribution when $L$ is large.
Finite-size effects affect the low-disorder cases, and when the localization lengths are large compared to the system size, this causes the states to behave as if they were extended. 
This leads to the distributions shifting rightward at low disorder strengths and small system sizes. 
These distributions appear to eventually converge at large $L$, especially evidenced by the $W=1.6$ and larger disorder cases.

The median of the distributions as a function of $L$ is plotted in Fig.~\ref{fig:noninteracting_transport_disordered_figure}(e). 
We note that in the absence of disorder, $\mathrm{med}\|\hat M^{(0)}\|_F = \frac{1}{4}\sqrt{L}$.
Furthermore, we see that $\mathrm{med}\|\hat M^{(0)}\|_{\text{op}} \sim L/4$ where $\|\cdot\|_{\text{op}}$ denotes the operator norm.
We note that these saturate the largest possible scaling with $L$ for these quantities in any system (i.e., with or without interactions), according to the bounds in Appendix \ref{app:bounds}.
This aligns with the interpretation of $\|\hat M^{(0)}\|_{\text{op}}$ being the charge transport capacity, where $\|\Delta N(t)\| \leq 2 \|\hat M^{(0)}\|_{\text{op}}$ means that at maximum, a total of $L/2$ particles can be transported, which is obviously the largest possible such $\|\Delta N(t)\|$.
This maximum is realized for an initial state with $L/2$ particles occupying the left half of the chain, where, in a clean system and after sufficient time, a recurrence-type behavior can occur in which all particles are found on the right half.
We provide a semi-analytical understanding of the scalings of these norms in Appendix~\ref{app:noninteracting_transport_details}.

\begin{figure}[t]
    \centering
    \includegraphics[width=\columnwidth]{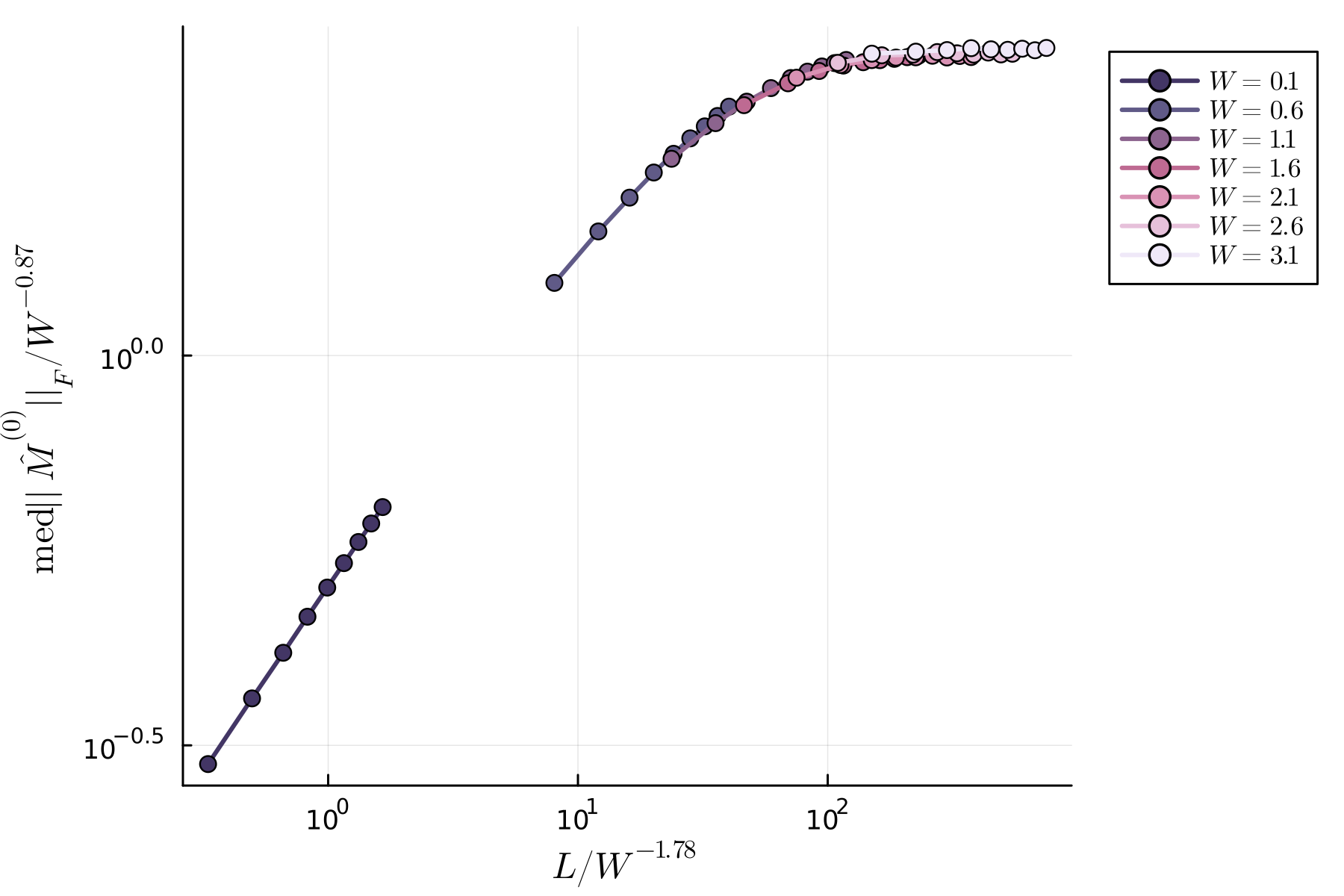}
    \caption{
    Same plot as in Fig.~\ref{fig:noninteracting_transport_disordered_figure}(e) on a log-log scale, but with the x-axis scaled by the participation ratio, as $\mathrm{PR}\sim W^{-1.78}$ (see Appendix~\ref{app:localization_length} for details), and the y-axis scaled roughly by $\sqrt{\mathrm{PR}}$ (with a phenomenological correction).
    The collapse of the curves suggests that the growth of transport for small system sizes is a finite-size effect, and the scaling of the y-axis confirms that the final saturated value for $\|\hat M^{(0)}\|_F$ scales roughly as $\sqrt{\mathrm{PR}}$, matching the $\sqrt{L}$ scaling of $\|\hat M^{(0)}\|_F$ in the no-disorder case.
    }
    \label{fig:noninteracting_transport_collapsed}
\end{figure}

Once we add the disorder, barring finite-size effects, the median of $\|\hat M^{(0)}\|_F$ saturates at large $L$ to finite values which decrease with increasing the disorder strength $W$, aligning with the expectation that there is zero transport in the Anderson model.
We expect that these final converged values for $\|\hat M^{(0)}\|_F$ are dependent on the localization length $\xi$  (approximated by the participation ratio $\mathrm{PR}$, cf.~Appendix \ref{app:localization_length}) roughly as $\propto \sqrt\xi$, in the same way that $\|\hat M^{(0)}\|_F$ scales proportionally to $\sqrt{L}$ in the no-disorder case.
Indeed, we observe in Fig.~\ref{fig:noninteracting_transport_collapsed} that the curves in Fig.~\ref{fig:noninteracting_transport_disordered_figure}(e) collapse when scaled on the y-axis approximately by $\sqrt{\xi}\approx (W^{-1.78})^{1/2}$, and on the x-axis by the localization length $\xi\approx W^{-1.78}$.

\subsubsection{Transport in the Single-Particle Quasiperiodic Model}\label{sec:non_interacting_transport_quasiperiodic}
\begin{figure*}[t]
    \centering
    \includegraphics[width=\linewidth]{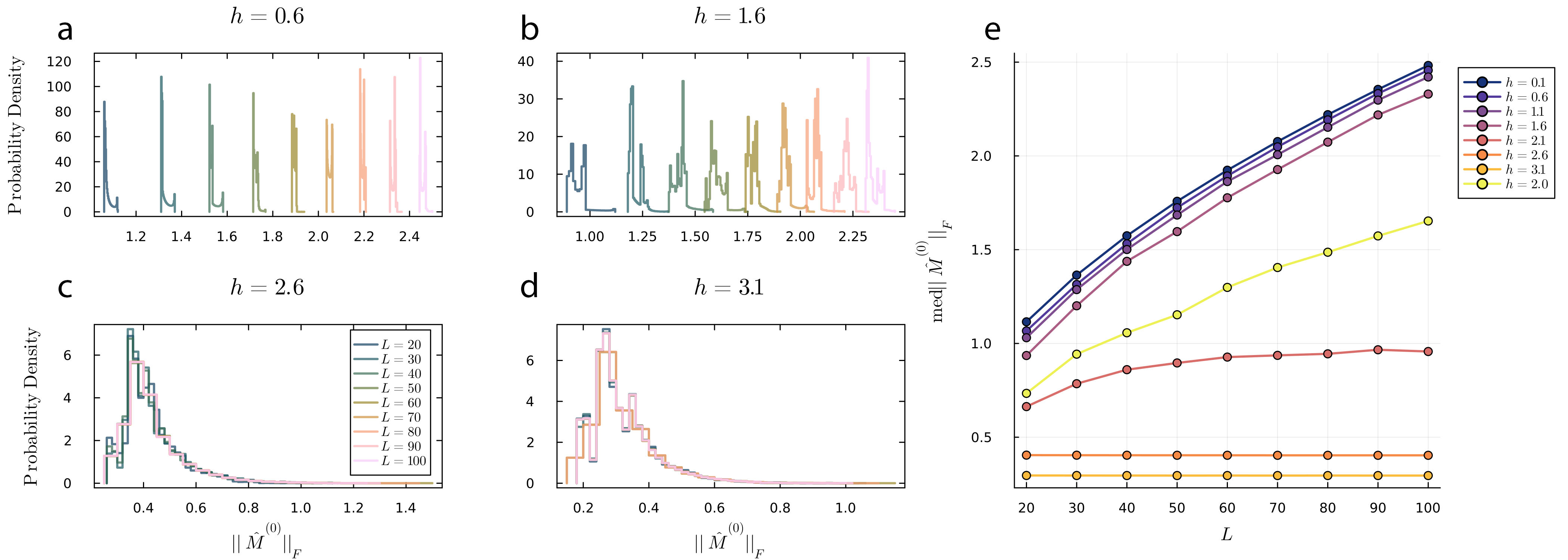}
    \caption{Analysis of the Frobenius norm of the charge transport capacity, $\|\hat M^{(0)}\|_F$, for the noninteracting quasiperiodic model with open boundary conditions. The analysis is done based on 10000 evenly spaced phase shifts $\delta$ between $\delta=0$ and $\delta=\frac{1}{k}$, with $i_0=\lfloor L/2\rfloor$, and wavenumber $k=\sqrt{2}$. 
    (a)--(d): Calculated probability density functions of $\|\hat M^{(0)}\|_F$.     Within each panel, each colored curve represents a different system size, going from $L=20$ to $L=100$, in steps of $10$.
    Panels (a) and (b) are in the delocalized phase for $h=0.6$ and $1.6$, where one sees narrow and rightwards shifting peaks. This is a signature of unbounded transport in the delocalized phase. The localized phase is represented in panels (c) and (d) for potential depths $h=2.6, 3.1$, which shows distributions converging sharply with increasing $L$. This reflects that transport is finite and limited in the localized phase.
    (e): We show the median of $\|\hat M^{(0)}\|_F$, $\mathrm{med}\|\hat M^{(0)}\|_F$, as a function of $L$.
    Each differently colored curve represents a different potential depth $h$, from $h=0.1$ to $h=3.1$ in steps of 0.5. We also show the curve at the transition $h=2$ (yellow). The curves in the extended phase $h<2$ show a square root scaling similar to that in the no-disorder case, while the curves in the localized phase for $h>2$ saturate to finite values at large $L$.
    }
    \label{fig:noninteracting_transport_quasiperiodic_figure}
\end{figure*}

In the quasiperiodic model, $\|\hat M^{(0)}\|_F$ exhibits distinct scaling behavior in the extended and localized phases, providing a clear signature of this localization transition.
We show the analysis of $\|\hat M^{(0)}\|_F$ in this case in Fig.~\ref{fig:noninteracting_transport_quasiperiodic_figure}.

For $h<2$, the eigenstates of this model are extended. For extended states, the current is free-flowing, and so the amount of charge transport capacity should generally increase with the system size. 
Indeed, the distributions of $\|\hat M^{(0)}\|_F$ in this regime, shown for $h=0.6$ and $h=1.6$, are narrow and shift rightward with increasing $L$, consistent with free flowing current.
For $h>2$, the system enters a localized phase, where we expect a limited amount of charge transport capacity.
Indeed, for $h=2.6,3.1$, the distributions of $\|\hat M^{(0)}\|_F$ appear sharply converged even at small system sizes, as shown in Figs.~\ref{fig:noninteracting_transport_quasiperiodic_figure}(c) and (d). This indicates that the distributions are fixed even in the thermodynamic limit, suggesting that the charge transport is small and bounded for $h>2$. 

The scaling of $\mathrm{med}\|\hat M^{(0)}\|_F$, presented in Fig.~\ref{fig:noninteracting_transport_quasiperiodic_figure}(e), offers a striking picture of the transition. 
For $h<2$, $\mathrm{med}\|\hat M^{(0)}\|_F$ scales as $\frac{1}{4}\sqrt{L}$---the same way as it did in the case without disorder, as discussed in the previous subsection.\footnote{Note that the relation $\mathrm{med}\|\hat M^{(0)}\|_F = \frac{1}{4}\sqrt{L}$ is true for zero potential when all orbitals have inversion symmetry in the middle of the chain, see Appendix~\ref{app:noninteracting_transport_details}. Furthermore, recall that we are discussing the case with even $L$ and $i_0 = L/2$.
The inversion symmetry is absent in the quasiperiodic model when $h \neq 0$, and we observe $O(1)$ deviations in the delocalized phase.}
For $h>2$, $\mathrm{med}\|\hat M^{(0)}\|_{F}$ saturates very quickly at sufficiently large system sizes to a small and finite value that depends on the potential depth $h$.
The saturation is very apparent at large potential values $h=2.6, 3.1$, where $\mathrm{med}\|\hat M^{(0)}\|_{F}$ appears to be almost independent of $L$.
This shows that only a small amount of charge can flow 
through the middle of the link, which is a strong indicator of localization.
At the transition $h=2$, $\|\hat M^{(0)}\|_{\text{op}}$ appears to scale linearly with $L$ but with a different slope from the delocalized phase (see Fig.~\ref{fig:noninteracting_transport_quasiperiodic_operator_norm} in Appendix~\ref{app:operator_norm_data_noninteracting_transport}), separating the curves in the extended regime and the curves for the localized regime.
For the quasiperiodic model, 
$\hat M$ proves to be
a diagnostic for localization: the stark contrast in the scaling of $\mathrm{med}\|\hat M^{(0)}\|_{F}$ between the localized and delocalized states offers a 
clear signature to characterize the underlying localization transition.

\subsection{Results on the Charge Transport Capacity in Interacting Systems}\label{sec:interacting_transport_results}

In this section, we analyze the behavior of the corrections to the Frobenius norm of the total charge transport capacity operator, $\|\hat M^{(1)}\|_F$, for weakly interacting localized models, starting with the interacting Anderson model and ending with the interacting quasiperiodic model.

\subsubsection{Anderson Model}\label{sec:interacting_transport_disordered_results}

\begin{figure*}[t]
    \centering
    \includegraphics[width=\linewidth]{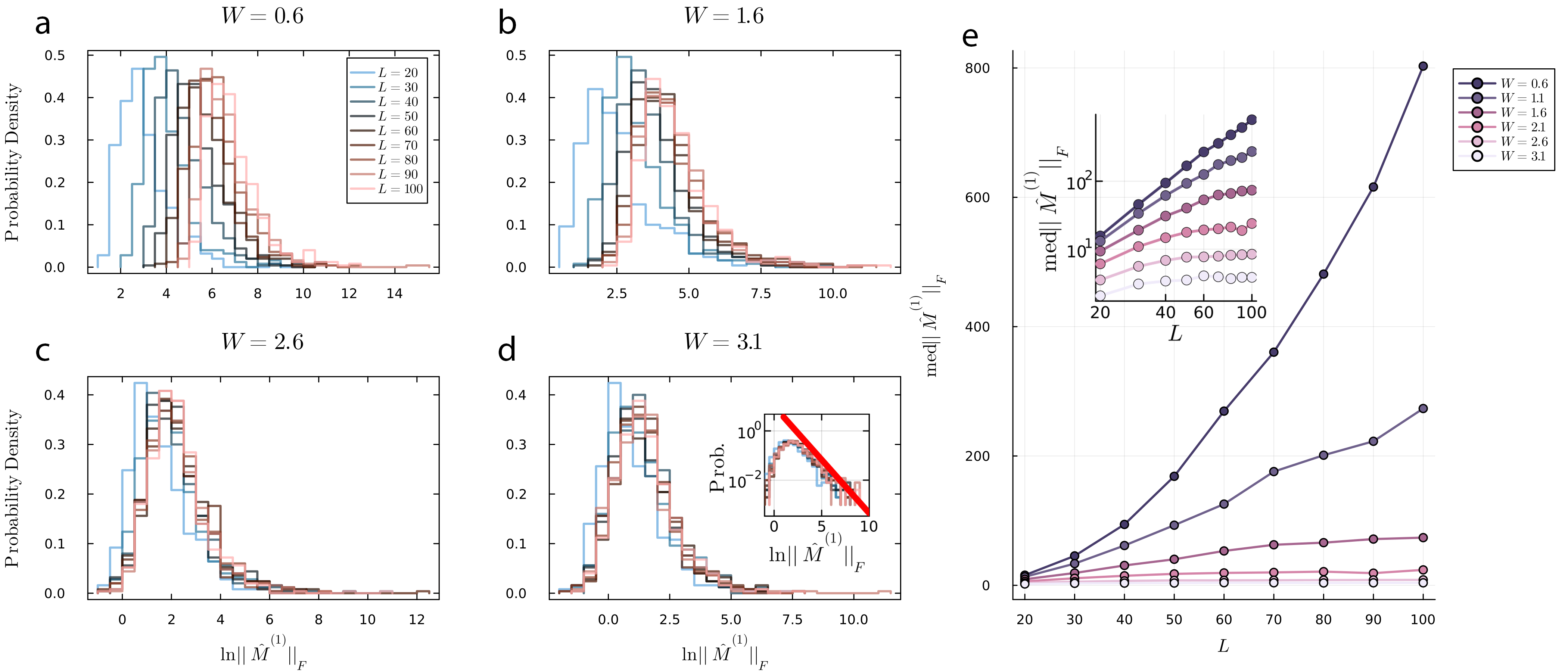}
    \caption{Analysis of the Frobenius norm of the first-order correction to the charge transport capacity, $\|\hat M^{(1)}\|_F$, for the weakly-interacting Anderson Model across the link with $i_0=\lfloor L/2\rfloor$ and open boundary conditions.
    (a)--(d): Calculated probability density functions of $\ln(\|\hat M^{(1)}\|_F)$ based on 1000 disorder realizations.
    Each panel shows a different disorder strength $W=0.6, 1.6, 2.6, 3.1$. 
    Within each panel, each colored curve represents a different system size, with system sizes from $L=20$ to $L=100$ in steps of 10.
    The inset in (d) shows a distribution of $\|\hat M^{(1)}\|_F$ at $h=3.1$ with a natural log scale on the y and x-axes, with a $\|\hat M^{(1)}\|_F^{-2}$ (red line) comparison to the tail--- see the text surrounding Eq.~(\ref{eq:tail_behavior}) for details on the form plotted.
    The convergence of the distributions at large disorder and $L$ indicates that the corrections to transport are bounded in the thermodynamic limit, suggesting that charge transport is suppressed in the presence of interactions at first-order in perturbation theory.
    (e): Plot of $\mathrm{med}\|\hat M^{(1)}\|_F$ as a function of $L$. 
    Each curve represents a different disorder strength $W$, from $W=0.6$ to $3.1$ in steps of 0.5.
    The inset in (e) shows the same data but plotted using a log-log scale. At large $L$, the transport is bounded, indicating the presence of bottlenecks even when interactions are added.
    }
\label{fig:disordered_interacting_transport_figure}
\end{figure*}

In Fig.~\ref{fig:disordered_interacting_transport_figure}, we show numerical results for $\|\hat M^{(1)}\|_F$ in the weakly-interacting Anderson model. Here, we omit results for $W=0.1$ for visual clarity in the plots.

\begin{figure}[t]
    \centering
    \includegraphics[width=\columnwidth]{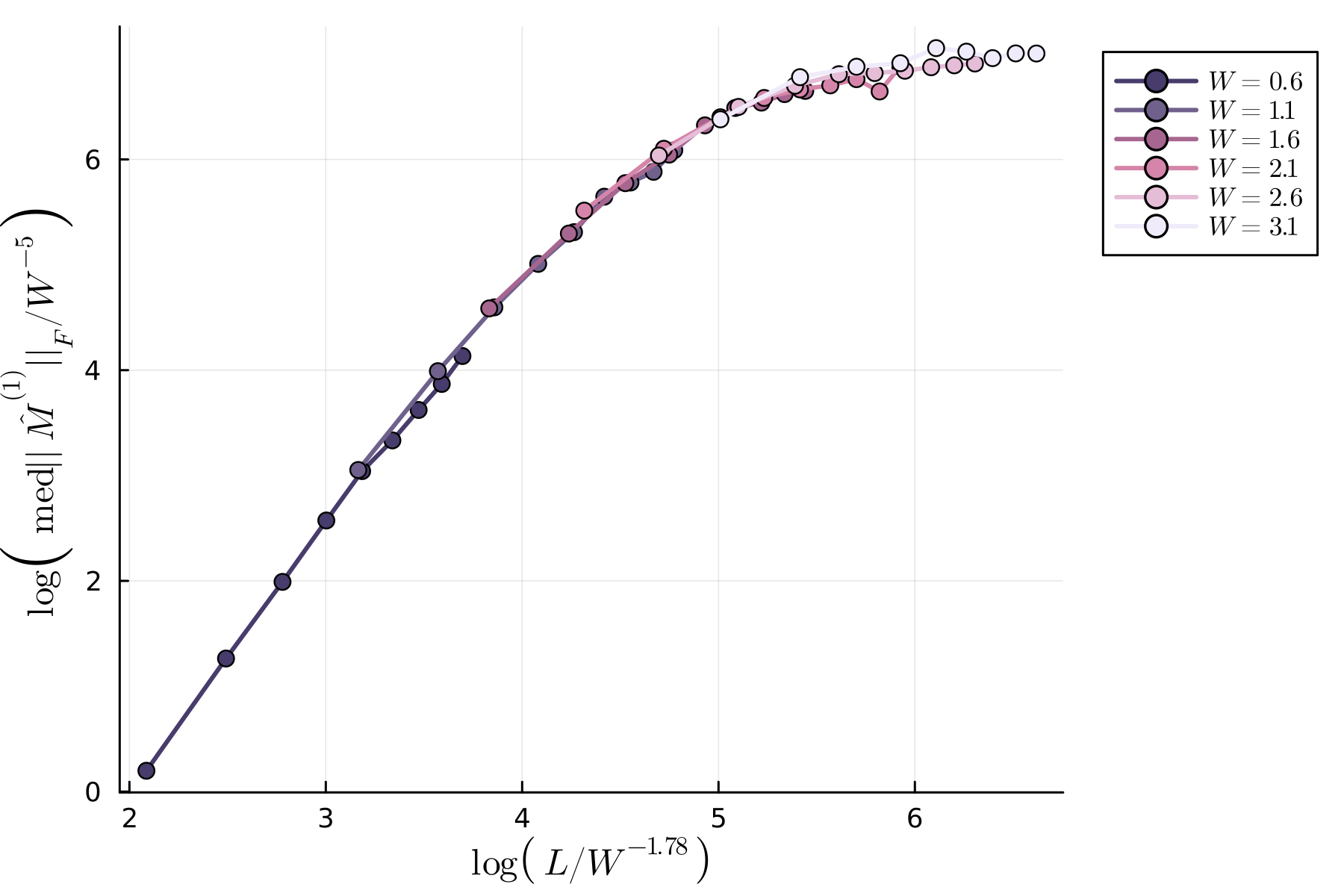}
    \caption{
    Same data of $\mathrm{med}\|\hat M^{(1)}\|_F$ as in Fig.~\ref{fig:disordered_interacting_transport_figure}(e), but with the x-axis scaled roughly by the participation ratio ($\mathrm{PR}\sim W^{-1.78}$, see Appendix \ref{app:localization_length} for details), and the y-axis scaled by a phenomenological scaling $W^{-5}$.
    The collapse of the curves here suggests that if one analyzes the corrections to large enough $L$, even the low disorder cases will have $\mathrm{med}{\|\hat M^{(1)}\|_F}$ saturate. 
    }
    \label{fig:scaled_disordered_interacting_transport}
\end{figure}

The most important takeaway is that the distributions of $\|\hat M^{(1)}\|_F$ shown in Figs.~\ref{fig:disordered_interacting_transport_figure}(a)--(d) appear to converge in the thermodynamic limit. 
Additionally, Fig.~\ref{fig:disordered_interacting_transport_figure}(e) shows that the median of these distributions converges at high disorder strengths or large system sizes. 
These curves appear to grow as a polynomial with $L$ when finite-size effects come into play, and plateau after the localization length is small enough compared to the system size. This is particularly clear in the inset in Fig.~\ref{fig:disordered_interacting_transport_figure}(e), which shows the same curves on a log-log scale and clearly illustrates that, for a given disorder strength $W$, these corrections appear to saturate if $L$ is large enough.
Just as the saturated values of $\|\hat M^{(0)}\|_F$ depend on $\xi$, we expect that the saturated values of $\|\hat M^{(0)}\|_F$ depend on $\xi$ as well. 
Similarly to what was done with the LIOM corrections in Fig.~\ref{fig:collapsed_LIOM_correction}, we show evidence in Fig.~\ref{fig:scaled_disordered_interacting_transport} that the $\mathrm{med}\|\hat M^{(1)}\|_{F}$ vs $L$ curves collapse when the x-axis is scaled with the participation ratio as $W^{-1.78}$ (see Appendix~\ref{app:localization_length}) and the y-axis is appropriately scaled phenomenologically. Here, we find this phenomenological scaling to be $\mathrm{PR}^{2.5}$. Like with the LIOM corrections, the curves of $\mathrm{med}\|\hat M^{(1)}\|_{F}$ collapse on top of each other nicely, similarly illustrating that its initial growth at low $L$ is a finite-size effect\footnote{We also note that this phenomenological scaling likely has an underlying physical origin, although we do not provide justification in this work.}.

These figures taken together show that even in the thermodynamic limit, for
a finite fraction
of the disorder realizations, transport is perturbatively finite through the middle of the lattice. 
Furthermore, as argued previously, the fact that the distributions converge in the thermodynamic limit means that the resonances contributing to the tails are primarily local. 
This is consistent with localization being preserved with added interactions.

Just as with the LIOM corrections, the distributions of $
\|\hat M^{(1)}\|_F$ suffer from slowly decaying inverse square power-law tails at large norms, implying that the mean of these distributions diverges.
These tails are caused by the same four-energy resonances that appear in the expressions for the AGP and LIOM corrections.
The existence of such tails is expected, and questions arise about their implications.
In a large enough lattice, one would certainly have a link that admits a large correction to $\hat M^{(0)}$.
One might worry that, reminiscent of the avalanche effect, a large correction at a single site could trigger extended large corrections throughout the entire lattice, restoring transport and suggesting that localization is not robust to interactions.
On the other hand, if localization is stable to perturbations, then these large corrections should remain uncorrelated at distant sites, implying that, in sufficiently large systems, large corrections to the charge transport capacity are only local, making the presence of bottlenecks inevitable even in the thermodynamic limit. 

\begin{figure}[t]
    \centering
    \includegraphics[width=\columnwidth]{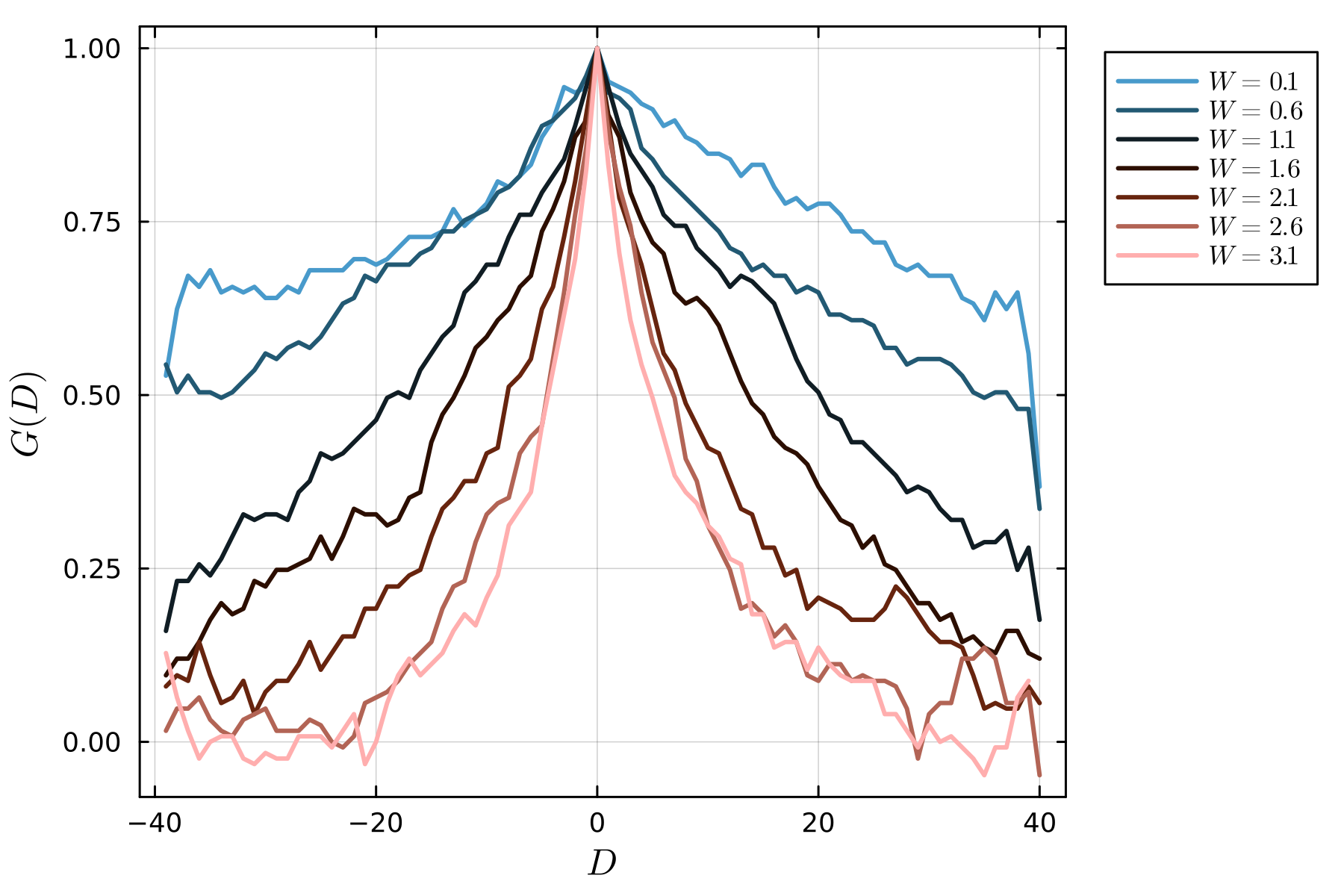}
    \caption{Calculations of the correlation coefficient $G(D)$, Eq.~(\ref{eq:correlation_function}), between $\Theta_{L/2}$ and $\Theta_{L/2+D}$, for 500 disorder realizations and $L=80$,  in the weakly-interacting Anderson model. Plotted here are cases out of those 500 disorder realizations where $\|\hat M_{L/2}^{(1)}\|_F>\mathrm{med}\|\hat M_{L/2}^{(1)}\|_F$.
    Each colored curve represents a different disorder strength.
    The center of the plot corresponds to the middle of the lattice, at $i_0=\lfloor L/2\rfloor$.
    We observe that if transport at the center exceeds the median, it appears uncorrelated with whether transport values further from the center are above or below the median.
    }
    \label{fig:covariance_figure}
\end{figure}

To check this, we numerically calculate whether or not transport corrections larger than the median at a site $i_0$ will continue to stay larger than the median as one travels away from $i_0$.
Thus, we define $\Theta_{i}\equiv \Theta\left(\|\hat M_{i}^{(1)}\|_F-\mathrm{med}\|\hat M_{i}^{(1)}\|_F\right)$.
Here, $\Theta$ refers to the Heaviside function---thus, $\Theta_{i}$ is 1 if $\|\hat M_{i}^{(1)}\|_F$ is larger than the median of the $\|\hat M_{i}^{(1)}\|_F$ distribution, and 0 otherwise.
We calculate the correlation coefficient $G(D)$ between $\Theta_{L/2}$ and $\Theta_{L/2+D}$ (for concision, we denote $\lfloor L/2\rfloor = L/2$, which is true in this case since we only perform calculations on lattices of even sizes), and limit ourselves to considering cases, out of $500$ disorder realizations, where $\|\hat M_{L/2}^{(1)}\|_F$ are \emph{large} charge transport corrections (defined as corrections greater than the median of $\|\hat M_{L/2}^{(1)}\|_F$ for those 500 disorder realizations) at the central link $i_0=\lfloor L/2\rfloor$. 
The full expression for this correlation function is
\begin{widetext}
\begin{align}\label{eq:correlation_function}
    G(D) &= \frac{\bigg\langle\left(\Theta_{L/2}-\left\langle\Theta_{L/2}\right\rangle\right)\left(\Theta_{L/2+D}-\left\langle \Theta_{L/2+D}\right\rangle\right)\bigg\rangle}{\sqrt{\left(\left\langle \Theta_{L/2}^2\right\rangle-\left\langle \Theta_{L/2}\right\rangle^2\right)\left(\left\langle \Theta_{L/2+D}^2\right\rangle-\left\langle \Theta_{L/2+D}\right\rangle^2\right)}}.
\end{align}
\end{widetext}
As shown in Fig.~\ref{fig:covariance_figure}, the correlation decreases to roughly half of its initial value at distances on the order of the localization length and further decreases to close to zero within another order of the localization length (see Appendix~\ref{app:localization_length} for a reference of the participation ratio, a proxy for the localization length, at different $W$ for large $L$).
These findings support the scenario where large corrections are uncorrelated at distant sites, suggesting that large charge transport corrections predominantly exert local influence. 
This suggests the existence of bottlenecks in the limit of large $L$.

Taken together, these numerical results suggest that transport remains bounded even in the presence of interactions. Despite the existence of these slow tails in the distributions, one can conclude that, to the first order, bottlenecks are certain to occur in a lattice of interacting fermions of size $L\rightarrow \infty$, bounding the charge transport capacity. This is consistent with localization being robust perturbatively, at least to the first order.

\subsubsection{Quasiperiodic Model}\label{sec:quasiperiodic_interacting_analysis}

\begin{figure*}[t]
    \centering
    \includegraphics[width=\linewidth]{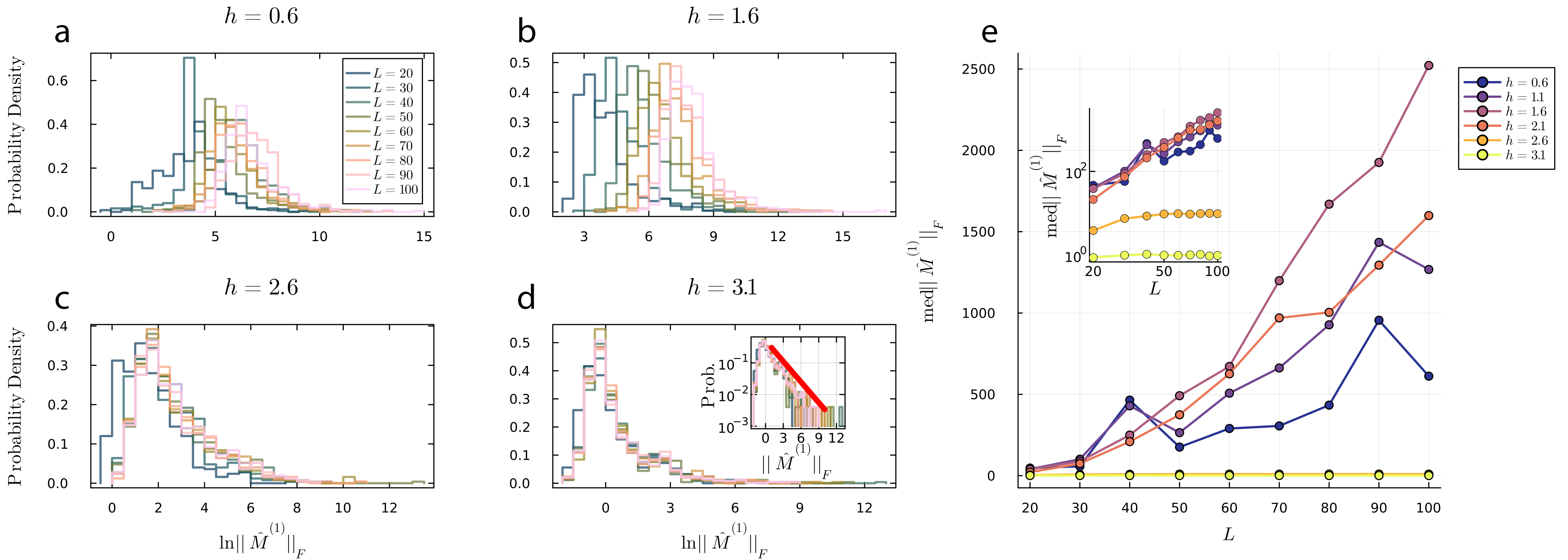}
    \caption{Numerical computations of the Frobenius norm of the first-order corrections to the charge transport capacity, $\|\hat M^{(1)}\|_F$, for the weakly-interacting quasiperiodic model with $i_0=\lfloor L/2\rfloor$ and open boundary conditions. Each calculation was done based on 500 phase shifts $\delta$ uniformly spaced between $0$ and $1/k$. (a)--(d): Calculated probability density functions of $\ln(\|\hat M^{(1)}\|_F)$. Each panel shows a different potential depth $h=0.6, 1.6, 2.6, 3.1$. Within each panel, each colored curve represents a different system size, between $L=20$ and $L=100$, in steps of 10. The inset in (d) shows a distribution of $\|\hat M^{(1)}\|_F$ at $h=3.1$ with a log-scale on the y and x-axes, with a $\|\hat M^{(1)}\|_F^{-1.5}$ (red line) comparison to the tail--- see the text surrounding Eq.~(\ref{eq:tail_behavior}) for details on the form plotted. (e): Plot of $\mathrm{med}\|\hat M^{(1)}\|_F$ as a function of $L$. Each curve represents a different potential depth $h$, from $h=0.6$ to $3.1$ in steps of 0.5. The inset in (e) shows the same plot but with a log-log scale.
    }
    \label{fig:quasiperiodic_interacting_transport_figure}
\end{figure*}

We show results for $\|\hat M^{(1)}\|_F$ in the interacting quasiperiodic model in Fig.~\ref{fig:quasiperiodic_interacting_transport_figure}.
Again, we omit results for $h=0.1$ for visual clarity in the plots.
For high potential strengths and large system sizes, the conclusions here are largely the same as those in the disordered case.
In panels (c) and (d), we see that the distributions of $\|M^{(1)}\|_F$ are converged at large $L$, indicating that the transport remains suppressed in the quasiperiodic model for high disorder.
 
Just as in the noninteracting case, the curves appear to exhibit different behavior for $h<2$ and $h>2$, possibly hinting at the presence of a localization transition.
Interestingly, for $h<2$, $\mathrm{med}\|\hat M^{(1)}\|_F$ appears to generally increase with $h$ at a fixed $L$.
This illustrates a breakdown of the perturbation theory as one nears the critical point.
In this delocalized regime, the curves of $\mathrm{med}\|\hat M^{(1)}\|_F$ seem to scale as a polynomial with $L$. 
We also observe that the distributions in Figs.~\ref{fig:quasiperiodic_interacting_transport_figure}(a) and (b) generally shift rightward.
These observations show that the corrections to the charge transport capacity are large, which is expected as a perturbed delocalized phase is likely to remain delocalized, leading to free-flowing current.

After $h=2$, $\mathrm{med}\|\hat M^{(1)}\|_F$ decreases with $h$, and the corrections sharply saturate when $L$ is large, corresponding to the distributions converging at large $L$ shown in Figs.~\ref{fig:quasiperiodic_interacting_transport_figure}(c) and (d). 
These show that bottlenecks are present in the quasiperiodic case when weak interactions are added.

We note that there is an anomalous peak in $\mathrm{med}\|\hat M^{(1)}\|_F$ at $L=90$, which does not appear in calculations of the norms of the AGP nor the LIOM corrections. 
At first glance, this seems at odds with our analysis of the anomalous spikes in Sec.~\ref{sec:AGP_results}, since $L=90$ is not nearly commensurate with the potential with $k=\sqrt{2}$. 
However, we conjecture that this spike has another simple underlying reason: it is likely related to commensuration of the wavefunction with the choice of the transport link $i_0=45$, in which $45+1\times \sqrt{2}\approx 65$.
Indeed, we separately verified that when the current is calculated through a different link that is not commensurate with the underlying Hamiltonian, the spike disappears.

\subsection{On the Avalanche Instability at Weak Disorder}\label{sec:avalanche_discussion}
We note that a limitation of this work is that it cannot detect any avalanche instability. 
At low order in perturbation theory, an ergodic seed can only delocalize neighboring regions, preventing the thermal region from spreading throughout the system.
From this perspective, avalanche instability is a fundamentally nonperturbative phenomenon.
Recent work in Ref.~\cite{colbois_interaction-driven_2024} provided numerical evidence that the Anderson model, perturbed by weak interactions, immediately delocalizes up to a critical disorder $W^*$ that corresponds to our $W$ between 4 and 6.
If avalanches indeed destabilize the localization at low $W$, our analysis did not detect such a transition.
In the case of the Anderson model, our expressions for the LIOM corrections and charge transport capacity do not leave any room to capture a transition behavior akin to avalanches.
The expressions for the AGP, LIOM corrections, and charge transport capacity operators, as in Eq.~(\ref{eq:Xabcd}), Eq.~(\ref{eq:liom_matrix_elements}), and Eq.~(\ref{eq:M1}), only depend on the exponentially localized single-particle orbitals and energy eigenvalues. 
For the single-particle Anderson model, the energy spectrum and wavefunctions have similar properties at any disorder.
Indeed, even at high disorders $W=10$ and $W=20$, we verified that the quantities we calculated show the same qualitative behavior.
In particular, the distributions still show the same slow power-law tails, and the distributions of the LIOM corrections and transport capacity converge at large enough $L$ for any disorder strength $W$.
From the above calculation, it is clear that all the quantities calculated here would be saturated at very large $W$. Therefore, we focus on the small disorder regime for easier finite-size scaling analysis.
In conclusion, at low disorder our perturbative treatment may fail to detect the avalanche instability, while it is expected to be valid at large disorder.

\section{Conclusion}\label{sec:conclusion}

In this work, we investigated the fate of one-dimensional localization under weak interactions using a first-order perturbative approach. 
We provided numerical evidence that introducing perturbative nearest-neighbor interactions to two paradigmatic single-particle localized models---the Anderson and quasiperiodic models---gives rise to an extensive number of quasiconservation laws, while disrupting a subset of them through locally-mixing resonances. 
These resonances can, in principle, be treated locally and allow for the extension of this perturbation theory to higher orders.
Specifically, the distributions of the Frobenius norms of the corrections to the LIOMs converge to a fixed distribution in the thermodynamic limit, which exhibits slow power-law decaying tails.
The prolonged thermalization times observed in certain perturbed integrable systems are typically attributed to the persistence of such approximate integrals of motion, which result from a special class of so-called weak integrability-breaking perturbations.
Our findings suggest that for these single-particle localized models, short-range interactions fall into this special class of perturbations. 

We also introduced a measure of the charge transport capacity across a link, $\hat M$.
In the noninteracting case, this measure is an interesting probe for localization.
For the localized regimes in the two single-particle models studied, we see that the Frobenius and operator norms of this charge transport operator have fixed distributions in the thermodynamic limit, clearly reflecting a lack of charge transport in these models.
Furthermore, the scaling of the median with $L$ sharply differs between the localized and delocalized phases in the quasiperiodic Aubry-André model, which cleanly indicates the localization transition in this model.
When perturbatively weak interactions are added, we show numerical evidence that the charge transport continues to be suppressed up to the first order. 
Specifically, the distributions of the Frobenius norm of the corrections to $\hat M^{(0)}$ converge in the thermodynamic limit.
While large corrections can occur at specific sites, they do not exhibit extended influence, pointing to the presence of bottlenecks in the weakly interacting models.
Combined with our analysis on the fate of the noninteracting LIOMs, we conclude that localization is robust to interactions up to the first order.

Many open questions remain.
Firstly, our analysis is limited to first-order $(l=1)$ in perturbation theory– it would be interesting to study what happens at higher orders $l$.
A challenge arises that with higher-order terms, the resonances become increasingly complex. 
The second-order term would involve six energy resonances, the third would have eight energy resonances, and so on. 
Is it possible that this series of corrections converges towards the MBL LIOMs for the single-particle LIOMs with small corrections and large enough disorder? 
This could be an interesting tool for constructing LIOMs or studying them analytically. 
For high enough disorder, we conjecture that this perturbation theory will extend to all orders and imply stability of MBL.
For low disorder, these increasingly complex resonances may result in large higher-order corrections for increasingly larger fractions of the original noninteracting LIOMs.
This would signify a breakdown in perturbation theory for low disorder if one wants to look past the first-order correction to the LIOMs, potentially supporting avalanche instability of the MBL in this regime while still exhibiting parametrically slow relaxation as discussed in Sec.~\ref{sec:liom_conclusions} and Sec.~\ref{sec:avalanche_discussion}.
On a similar note, we also leave open the question of the fate of LIOMs whose corrections are large and suffer from resonances. 
Here, we do not provide any special treatment for resonating regions of the Hamiltonian, such as what was done in Ref.~\cite{de_roeck_rigorous_2023}.
It is possible that something similar can be done here so that the perturbation theory can be performed on nonresonating regions only, ensuring its success.

Our analysis can be used to help answer some open questions in MBL.
For example, one question is whether only local interactions are classified as ``weak'' perturbations. 
Do these quasiconserved quantities cease to exist if the interactions are allowed to be long-ranged?
Many of the experimental platforms that realize MBL have degrees of freedom that are coupled by long-ranged couplings, such as power-law decaying interactions \cite{smith_many-body_2016}.
The properties of MBL are well studied in 1D in the context of strictly short-ranged interactions, yet its fate is unclear when the interactions are extended.
Some perturbative arguments in the past suggest that MBL is unstable in the presence of any interactions that exhibit slower than exponential decay \cite{de_roeck_stability_2017}.
On the other hand, experiments suggest that localization can persist in these systems \cite{smith_many-body_2016}, while other works have also suggested that a ``quasi-MBL'' phase exists for systems with long-range interactions, which features conserved quantities that decay algebraically with distance \cite{thomson_quasi-many-body_2020}.
It would be interesting to analyze the effect of these long-range interactions on the quantities studied in this work, namely the LIOMs and transport capacity. 
Additionally, it would be interesting to study a generalized charge transport capacity operator $\hat M$ that incorporates long-range interactions, which could be used to study transport in, for instance, power-law hopping models, which exhibit interesting localization transitions.

Another open question our analysis can help address is the existence of MBL in $d>1$ dimensions.
The seminal work of Ref.~\cite{de_roeck_stability_2017} argued that avalanches will inevitably destroy MBL in 2D and above. 
Despite this, experiments \cite{schreiber_observation_2015,choi_exploring_2016} and numerics \cite{wahl_signatures_2019,chertkov_numerical_2021,venn_many-body-localization_2024} have uncovered signatures of localization in 2D similar to those in the 1D case.
Furthermore, the avalanche arguments reconcile poorly with the perturbative arguments presented in Ref.~\cite{basko_metalinsulator_2006}, which predicted the existence of MBL in \emph{any} spatial dimension. 
It is fair to say that the existence of MBL in $d>1$ is still contested.
However, MBL in 2D is extremely difficult to study on all fronts. 
Signatures obtained in experiments can be difficult to distinguish from glassy phenomena, which can have similar experimental signatures \cite{yan_equilibration_2017}.
Furthermore, numerical studies have been largely impossible due to unfavorable scaling with system sizes. 
The analytical framework presented in this work can, in principle, be applied to any number of dimensions.
In particular, one may ask if this approach can be used to study quasiconserved quantities in interacting models in 2D and above with nearest-neighbor interactions.
The advantage of our method is the fact that larger system sizes are attainable (relative to what is achievable with exact diagonalization), and calculations scale as a polynomial with the number of sites $L$, as $O(L^4)$. 
In this work, we calculated quantities up to 100 sites (which could correspond to a $10\times 10$ lattice in 2D), but it is possible to go to larger sizes if one has the computational resources and time.
It would be extremely interesting to compare the behavior of quasiconserved quantities in the 2D case to those in the 1D case, as this may shed light on open questions about MBL in 2D.
The generalization of the charge transport capacity operator $\hat M$ to higher dimensions is nontrivial, but one could study a similar object where the current is defined through a general $(d-1)$-dimensional \emph{surface} cutting the $d$-dimensional lattice into two partitions.
In 1D, this 0-dimensional surface is simply a single link, which was studied in detail in this work.
In 2D, one would look at the current through a 1D \emph{line}.
Such a higher-dimensional charge transport capacity object could be an interesting probe for localization in higher dimensions.
If one finds that the first-order corrections to LIOMs in 2D are not finite or the charge--transport--capacity--type arguments for bottlenecks do not work, that could strongly hint that localization indeed breaks down in 2D.

While there are many open questions that our approach could help address, the common denominator is that our approach is a new and interesting avenue to study the robustness of localization.

\begin{acknowledgements}
We thank Liam O'Brien, Jeanne Colbois, Wojciech De Roeck, Nicolas Laflorencie, Gil Refael, and Marko Znidaric for useful discussion.
J.K.J.~is supported by the U.S. Department of Energy, Office of Science, Office of Advanced Scientific Computing Research, Department of Energy Computational Science Graduate Fellowship under Award Number(s) DE-SC0025528.
F.M.S.~acknowledges support provided by the U.S.~Department of Energy (DOE) QuantISED program through the theory consortium ``Intersections of QIS and Theoretical Particle Physics'' at Fermilab, and by Amazon Web Services, AWS Quantum Program.
A part of this work was done at the Erwin Schrödinger International Institute for Mathematics and Physics (ESI) of the University of Vienna.
O.I.M.~and J.K.J.~also acknowledge support by the National Science Foundation through grant DMR-2001186.

This report was prepared as an account of work sponsored by an agency of the United States Government. Neither the United States Government nor any agency thereof, nor any of their employees, makes any warranty, express or implied, or assumes any legal liability or responsibility for the accuracy, completeness, or usefulness of any information, apparatus, product, or process disclosed, or represents that its use would not infringe privately owned rights. Reference herein to any specific commercial product, process, or service by trade name, trademark, manufacturer, or otherwise does not necessarily constitute or imply its endorsement, recommendation, or favoring by the United States Government or any agency thereof. The views and opinions of authors expressed herein do not necessarily state or reflect those of the United States Government or any agency thereof.
\end{acknowledgements}

\appendix
\section{Derivation of the AGP and LIOM Corrections for Free Fermions}\label{app:appendix_agp_derivation}

In this section, we provide derivation details for the Adiabatic Gauge Potential (AGP) with a free fermion Hamiltonian, perturbed by nearest-neighbor density-density interactions. 

We wish to find the generator $\hat X$ such that 
\begin{equation*}
    i [\hat X,\hat H_0] = \hat V_{\text{off-diag}}.
\end{equation*}
For $\hat H_0$, we have a free-fermion Hamiltonian: 
\begin{equation}
    \hat H_0=\sum_{jk} A_{jk} \hat c^{\dagger}_{j}\hat c_{k}= \sum_{jk} \left(\sum_\alpha \phi_{j \alpha}\epsilon_\alpha \phi_{k \alpha}^* \right) \hat c_j^\dagger \hat c_k=\sum_\alpha \epsilon_\alpha \hat d_\alpha^\dagger \hat d_\alpha.
\end{equation}
Here, we have defined $\hat d_\alpha= \sum_k \phi_{k\alpha}^* \hat c_k$.

We consider a perturbation $\hat V$ consisting of nearest-neighbor interactions:
\begin{align}
    \hat V&=\sum_j \hat n_j \hat n_{j+1}\\
    &=\frac{1}{4}\sum_j (\hat c_{j+1}^\dagger \hat c_{j}^\dagger \hat c_j \hat c_{j+1}-\hat c_{j}^\dagger \hat c_{j+1}^\dagger \hat c_j \hat c_{j+1} \\&-\hat c_{j+1}^\dagger \hat c_{j}^\dagger \hat c_{j+1} \hat c_{j}+\hat c_{j}^\dagger \hat c_{j+1}^\dagger \hat c_{j+1} \hat c_{j}).
\end{align}
Here, we use this form because we aim to unambiguously associate a tensor $V_{ijkl}$ to the quartic operator $\hat V$, such that $\hat V = \sum_{ijkl} V_{ijkl} \hat c_i^\dagger \hat c_j^\dagger \hat c_k \hat c_l$.
To avoid the ambiguity associated with transformations of the form $\hat V \rightarrow \hat V+\gamma (\hat c_i^\dagger \hat c_j^\dagger+\hat c_j^\dagger \hat c_i^\dagger)\hat c_k \hat c_l = \hat V$, we require that $V_{ijkl} = -V_{jikl} = -V_{ijlk}$. 

Now, we rewrite $\hat V$ in terms of $\hat{d}_\alpha$, noting that $\hat c_j=\sum_\alpha \phi_{j\alpha} \hat d_\alpha$. With this transformation, $\hat V$ takes on the form:
\begin{equation}
    \hat V=\frac{1}{4}\sum_{\alpha \beta \gamma \delta}V_{\alpha \beta \gamma \delta} \hat d_\alpha^\dagger \hat d_\beta^\dagger \hat d_\gamma \hat d_{\delta},
\end{equation}
with matrix elements
\begin{multline}
    V_{\alpha \beta \gamma \delta}=\frac{1}{4}\sum_j(\phi_{j+1,\alpha}^* \phi_{j,\beta}^* \phi_{j,\gamma} \phi_{j+1,\delta}-\phi_{j,\alpha}^* \phi_{j+1,\beta}^* \phi_{j,\gamma} \phi_{j+1, \delta}\\
    -\phi_{j+1, \alpha}^* \phi_{j, \beta}^* \phi_{j+1, \gamma} \phi_{j, \delta}+\phi_{j, \alpha}^* \phi_{j+1,\beta}^* \phi_{j+1,\gamma} \phi_{j,\delta}).
\end{multline}
This tensor has the required property that $V_{\alpha\beta\gamma\delta} = -V_{\beta\alpha\gamma\delta} = -V_{\alpha\beta\delta\gamma}$.
Now, we consider the following ansatz for $\hat X$:
\begin{equation*}
    \hat X=\sum_{\alpha, \beta,\gamma,\delta }X_{\alpha\beta\gamma\delta}\hat{d}_\alpha^\dagger \hat{d}_\beta^\dagger \hat{d}_\gamma \hat{d}_\delta.
\end{equation*}
We want to express $X_{\alpha\beta\gamma\delta}$ in terms of $V_{\alpha \beta \gamma \delta}$. To do so, we invoke the following property:
\begin{align*}
\bigl[\hat{c}_i^\dagger\,\hat{c}_j^\dagger\hat{c}_k\,\hat{c}_l,
\hat{c}_n^\dagger\hat{c}_m\bigr]
=
\delta_{l n}\,\hat{c}_i^\dagger\hat{c}_j^\dagger\,\hat{c}_k\hat{c}_m
 - 
\delta_{i m}\,\hat{c}_n^\dagger\hat{c}_j^\dagger\hat{c}_k\hat{c}_l
\\ +
\delta_{k n}\hat{c}_i^\dagger\hat{c}_j^\dagger\hat{c}_m\,\hat{c}_l
 -
\delta_{j m}\,\hat{c}_i^\dagger\hat{c}_n^\dagger\,\hat{c}_k\hat{c}_l.
\end{align*}
From this, one finds that:
\begin{align}
    V_{\alpha \beta \gamma \delta} &=i \sum_{\sigma} (A_{\sigma\delta} X_{\alpha \beta \gamma \sigma} - A_{\alpha \sigma} X_{\sigma \beta \gamma \delta} \\
    &\qquad + A_{\sigma \gamma} X_{\alpha \beta \sigma \delta} - A_{\beta \sigma} X_{\alpha \sigma \gamma \delta})\\
    &=i(\epsilon_\delta-\epsilon_\alpha+\epsilon_\gamma-\epsilon_\beta)X_{\alpha\beta\gamma\delta}.
\end{align}
We thus define
\begin{equation}
    X_{\alpha\beta\gamma\delta}=
    \begin{cases}
        i\frac{V_{\alpha\beta\gamma\delta}}{\epsilon_\alpha+\epsilon_\beta-\epsilon_\gamma-\epsilon_\delta} & \epsilon_\alpha+\epsilon_\beta-\epsilon_\gamma-\epsilon_\delta\neq 0 \\
        0 & \epsilon_\alpha+\epsilon_\beta-\epsilon_\gamma-\epsilon_\delta=0.
    \end{cases}
\end{equation}
We can now compute the correction to the $\alpha$-th localized orbital:
\begin{equation}
    \tilde{n}_{\alpha}^{(1)} \equiv i[\hat X, \hat{d}_\alpha^\dagger \hat{d}_\alpha]=\sum_{\mu\nu\rho\sigma} f^\alpha_{\mu\nu\rho\sigma}\hat{d}_\mu^\dagger \hat{d}_\nu^\dagger \hat{d}_\rho \hat{d}_\sigma
\end{equation}
One can compute the coefficients $f^\alpha_{\mu\nu\rho\sigma}$ by evaluating the commutator:

\begin{equation}
f^\alpha_{\mu\nu\rho\sigma}=i(\delta_{\alpha \sigma} +\delta_{\alpha \rho} -\delta_{\alpha \mu} -\delta_{\alpha \nu}) X_{\mu\nu\rho \sigma}.
\end{equation}
We can also go back to real space to get:
\begin{equation}
\tilde{n}_{\alpha}^{(1)}=\sum_{jklm} g^\alpha_{jklm} \hat{c}_j^\dagger \hat{c}_k^\dagger \hat{c}_l \hat{c}_m,
\end{equation}
With coefficients:
\begin{equation}
    g^\alpha_{jklm}=\sum_{\mu\nu\rho\sigma} f^\alpha_{\mu\nu\rho\sigma} \phi_{j\mu}\phi_{k\nu}\phi_{l\rho}^* \phi_{m\sigma}^*.
\end{equation}

\section{Hilbert-Schmidt product and Frobenius Norm of Fermionic Operators}\label{app:a_frobenius_norm_derivation}

In this section, we show how to compute the Hilbert-Schmidt inner product of two fermionic operators $\hat A, \hat B$, defined as $\braket{\hat A,\hat B}_{HS}\equiv \frac{1}{\mathcal D}\Tr[\hat A^\dagger \hat B]$, where $\mathcal D$ is the Hilbert space dimension.
We will use that $\frac{1}{\mathcal D}\Tr[\dots]$ is the thermal expectation value at infinite temperature, and Wick's theorem can be applied to express expectation values of products of four or more fermionic operators in terms of expectation values of fermion bilinears \cite{surace_fermionic_2022}, of the form $\braket{\hat c_i^\dagger \hat c_j}_{\beta=0}=\frac{1}{2}\delta_{ij}$ and $\braket{\hat c_i \hat c_j^\dagger}_{\beta=0}=\frac{1}{2}\delta_{ij}$.

\subsection{Quadratic Operators} \label{quadratic}
Let us consider the case of quadratic operators of the form
\begin{equation}
    \hat O_M = \sum_{ij}M_{ij}\hat c_i^\dagger \hat c_j.
\end{equation}
The Hilbert-Schmidt product of two operators of this form reads
\begin{align}
    \braket{\hat O_M, \hat O_N}_{HS}&=\braket{\hat O_M^\dagger \hat O_N}_{\beta=0}\nonumber\\
    &=\sum_{ijkl}M_{ij}^*N_{kl} \braket{\hat c^\dagger_j \hat c_i\hat c_k^\dagger \hat c_l }_{\beta=0}.
\end{align}
We now evaluate
\begin{equation}
\braket{\hat c^\dagger_j \hat c_i\hat c_k^\dagger \hat c_l }_{\beta=0}=\frac{1}{4}(\delta_{ij}\delta_{kl}+\delta_{jl}\delta_{ik}),    
\end{equation}
and we obtain
\begin{align}
 \braket{\hat O_M, \hat O_N}_{HS}&=  \frac{1}{4}\sum_{ij}(M_{ii}^*N_{jj}+M^*_{ij}N_{ij})\nonumber\\
 &=\frac{1}{4}(\Tr[M^\dagger]\Tr[N]+\Tr[M^\dagger N] ).
\end{align}

\subsection{Quartic Operators} \label{quartic}
We now consider the case of quartic operators of the form
\begin{equation}
    \hat Q_M
=\sum_{ijkl} M_{ijkl}\hat c_i^\dagger \hat c_j^\dagger \hat c_k \hat c_l
\end{equation}
with $M_{ijkl}=-M_{jikl}=-M_{ijlk}=M_{jilk}$ (note that the tensor $M$ can always be written in this form).

We can compute now the Hilbert-Schmidt product as
\begin{align}
    &\braket{\hat Q_M, \hat Q_N}_{HS}=\braket{\hat Q_M^\dagger \hat Q_N}_{\beta=0}\nonumber\\
    &\qquad =\sum_{ijkl}\sum_{pqrs}M_{ijkl}^*N_{pqrs}\braket{\hat c_l^\dagger \hat c_k^\dagger \hat c_j \hat c_i\hat c_p^\dagger \hat c_q^\dagger \hat c_r \hat c_s}_{\beta=0}.
\end{align}
The infinite temperature correlator can be computed using Wick's theorem, obtaining a sum over all of the 24 possible contractions (with signs).
The calculation can be simplified by noting that many of the contractions give the same results, thanks to the antisymmetric properties of the tensors. 
This produces three independent contractions, and we finally obtain
\begin{align}
    \braket{Q_M^\dagger, Q_N}_{HS}=& \frac{1}{4}\left(\sum_{ij}M^*_{ijij}\right)\left(\sum_{pq}N_{pqpq}\right) \nonumber\\
    &+\sum_{lkpq}M^*_{lklq}N_{pqpk}+\frac{1}{4}\sum_{lkpq}M^*_{lkpq}N_{pqlk}.
\end{align}

\section{Additional Results for the AGP}\label{appendix_agp_full_results}

\begin{figure*}[t]
    \centering
    \includegraphics[width=\linewidth]
    {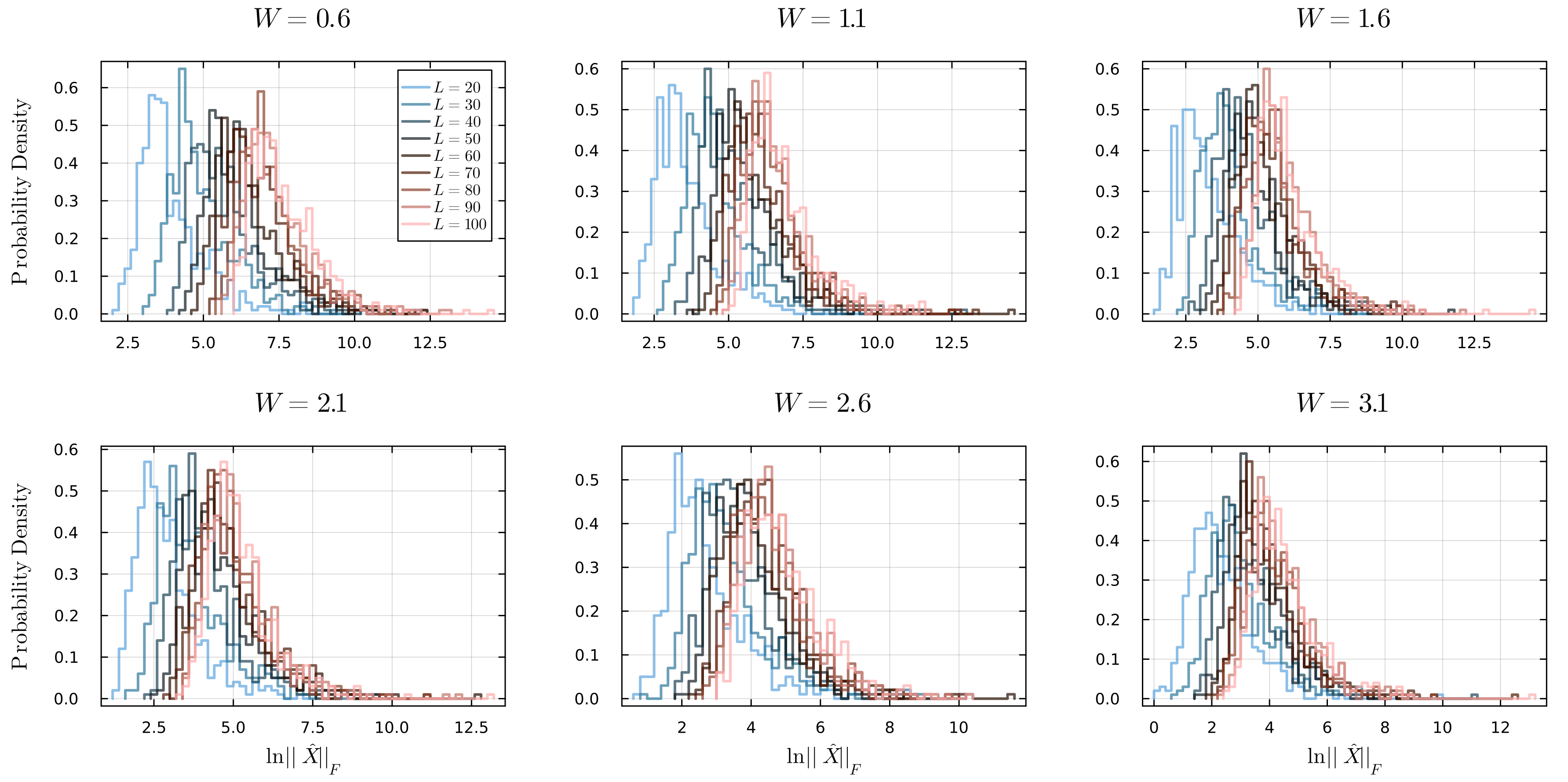}
    \caption{Calculated probability density functions of the logarithm of the Frobenius norm of the AGP, $\ln\|\hat X\|_F$, similar to Fig.~\ref{fig:agp_disordered_figure} (repeating panel with $W=2.1$ for easy comparisons here), for the weakly-interacting Anderson model. Each distribution was generated based on 500 disorder realizations with open boundary conditions. Each panel represents a different disorder strength $W$, where disorders from $W=0.6$ to $W=3.1$ are shown, in steps of $0.5$. Within each panel, each curve represents a distribution for a different system size. Here, we show $L=20$ to $L=100$ in steps of $10$.
    }
    \label{fig:appendix_agp_disordered}
\end{figure*}

\begin{figure*}[t]
    \centering
    \includegraphics[width=\linewidth]
{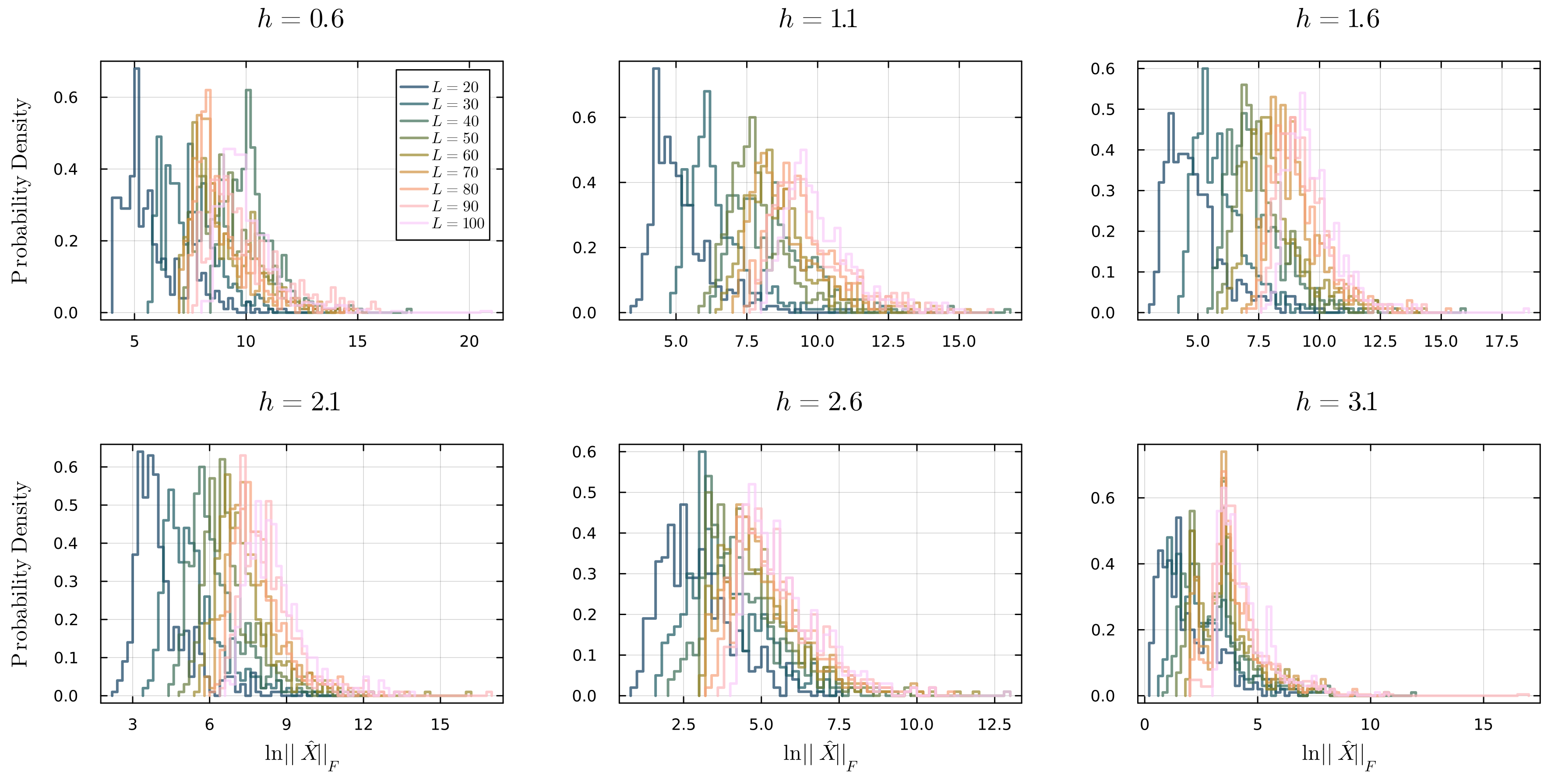}
    \caption{Calculated probability density functions of the Frobenius norm of the logarithm of the AGP, $\ln\|\hat X\|_F$, similar to Fig.~\ref{fig:agp_quasiperiodic_figure} (repeating panel with $h=3.1$ for easy comparisons here), for the weakly-interacting $k=\sqrt{2}$ quasiperiodic model, over 500 evenly spaced phase shifts $\delta$ within a period. Each panel represents different potential depths $h$, from $h=0.6$ to $h=3.1$ in steps of 0.5, read from top left to bottom right. Within each panel, each curve represents a distribution for a different system size, which we calculate from $L=20$ to $L=100$ in steps of 10.}
    \label{fig:appendix_agp_quasiperiodic}
\end{figure*}

In the main text, we present the AGP results for a single value of $W$ in the Anderson model or $h$ in the quasiperiodic model, as the overall behavior of the AGP remains qualitatively similar across different disorder strengths or potential depths. 
In particular, the same tail behavior is consistently observed, and the median of the AGP exhibits polynomial growth with system size $L$.
For completeness, Fig.~\ref{fig:appendix_agp_disordered} displays additional probability density functions of $\ln\|\hat X\|_{F}$ for the weakly interacting Anderson model at all values of $W$ calculated, while Fig.~\ref{fig:appendix_agp_quasiperiodic} shows corresponding results for the weakly interacting quasiperiodic model at different $h$ values.

\section{Probability Density of Denominators}\label{app:derivation_tails}
\subsection{Uniform Distribution of Eigenvalues}
Let the disorder strength be $W$ and the hopping parameter be $t$. We wish to find the probability density of $\frac{1}{\epsilon_\alpha+\epsilon_\beta-\epsilon_\gamma-\epsilon_\delta}$. We assume that the eigenvalues $\epsilon_\alpha$ are sampled independently from a uniform distribution $\mathrm{Unif}(-W', W')$. Here, $W'\in \mathbb{R}^+$ is upper bounded by $W+2t$. 

\begin{figure}[t]
    \centering
    \includegraphics[width=\columnwidth]{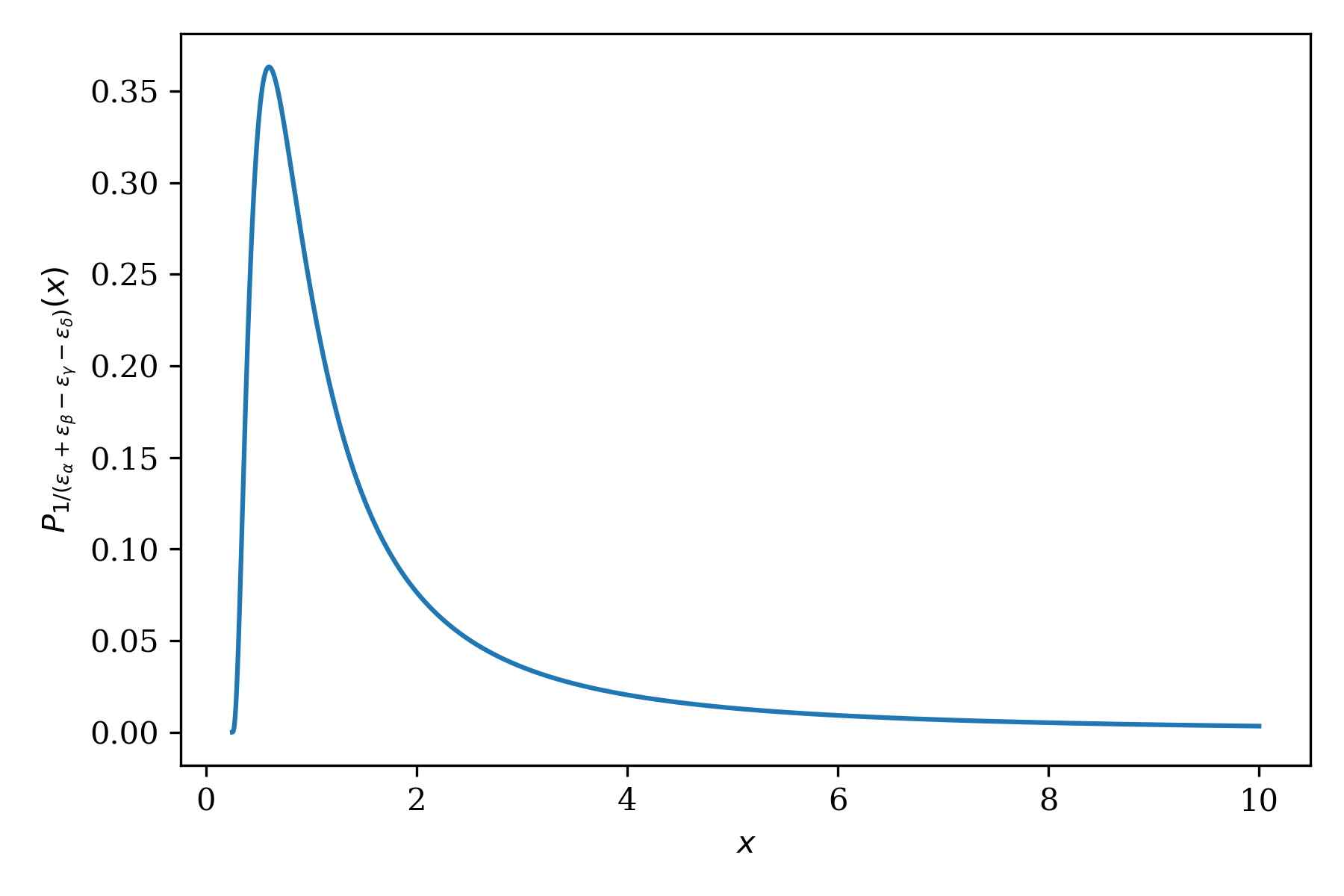}
    \caption{Probability density function of $x\equiv 1/(\epsilon_\alpha+\epsilon_\beta-\epsilon_\gamma-\epsilon_\delta)$ calculated from Eq.~(\ref{probability_density_calculated}). For this plot, $\epsilon_\alpha$ are assumed to have uniform distributions between $-1$ and $1$.}
    \label{fig:probably_reciprocal_distribution_figure_denominators}
\end{figure}

The probability density of $\epsilon_\alpha+\epsilon_\beta-\epsilon_\gamma-\epsilon_\delta$, $P_{\epsilon_\alpha+\epsilon_\beta-\epsilon_\gamma-\epsilon_\delta}(z)$ can be obtained by using convolution. One first obtains the distribution of $\epsilon_\alpha+\epsilon_\beta$ by observing that
\begin{equation}
    P_{\epsilon_\alpha+\epsilon_\beta}(z)=\int_{x\in X} P_{\epsilon_\alpha} (x)P_{\epsilon_\beta}(z-x)dx,
\end{equation}
where $X$ is the state space of $\mathrm{Unif}(-W', W')$. One can then build up convolutions to find $P_{\epsilon_\alpha+\epsilon_\beta-\epsilon_\gamma}$, then $P_{\epsilon_\alpha+\epsilon_\beta-\epsilon_\gamma-\epsilon_\delta}$ to obtain:
\begin{widetext}
\begin{equation}\label{probability_density_calculated}
    P_{\epsilon_\alpha+\epsilon_\beta-\epsilon_\gamma-\epsilon_\delta}(z) = 
\begin{cases}
\frac{(4W'+z)^3}{96W'^4} & -4W'\leq z\leq-2W'\\[6pt]
\frac{1}{3W'}-\frac{z^2}{8W'^3}-\frac{z^3}{32W'^4} & -2W'< z\leq0\\[6pt]
\frac{1}{3W'}-\frac{z^2}{8W'^3}+\frac{z^3}{32W'^4} & 0< z\leq 2W'\\[6pt]
-\frac{(-4W'+z)^3}{96W'^4} & 2W'<z\leq 4W'
\end{cases}
\end{equation}
\end{widetext}
Because $1/x$ is a monotonic function, the reciprocal distribution of $P$ can be given by:
\begin{equation}
     P_{\frac{1}{\epsilon_\alpha+\epsilon_\beta-\epsilon_\gamma-\epsilon_\delta}}(x) = \frac{1}{x^2}P_{\epsilon_\alpha+\epsilon_\beta-\epsilon_\gamma-\epsilon_\delta}\left(\frac{1}{x}\right).
\end{equation}
The reciprocal distribution is shown in Fig.~\ref{fig:probably_reciprocal_distribution_figure_denominators}. We focus on the tail behavior in the section $\frac{1}{2W'}\leq z< \infty$. In this sector, we have
\begin{equation}
     P_{\frac{1}{\epsilon_\alpha+\epsilon_\beta-\epsilon_\gamma-\epsilon_\delta}}(x) = \frac{1}{3W'}\frac{1}{x^2}-\frac{1}{8W'}\frac{1}{x^4}+\frac{1}{32W'^4}\frac{1}{x^5}.
\end{equation}

At large $x$, the $\frac{1}{x^2}$ behavior dominates. This provides a simple explanation for the tails observed in the quantities calculated in this work. 

\subsection{Density of States of the Quasiperiodic Model}
\begin{figure*}[t]
    \centering
    \includegraphics[width=\linewidth]
    {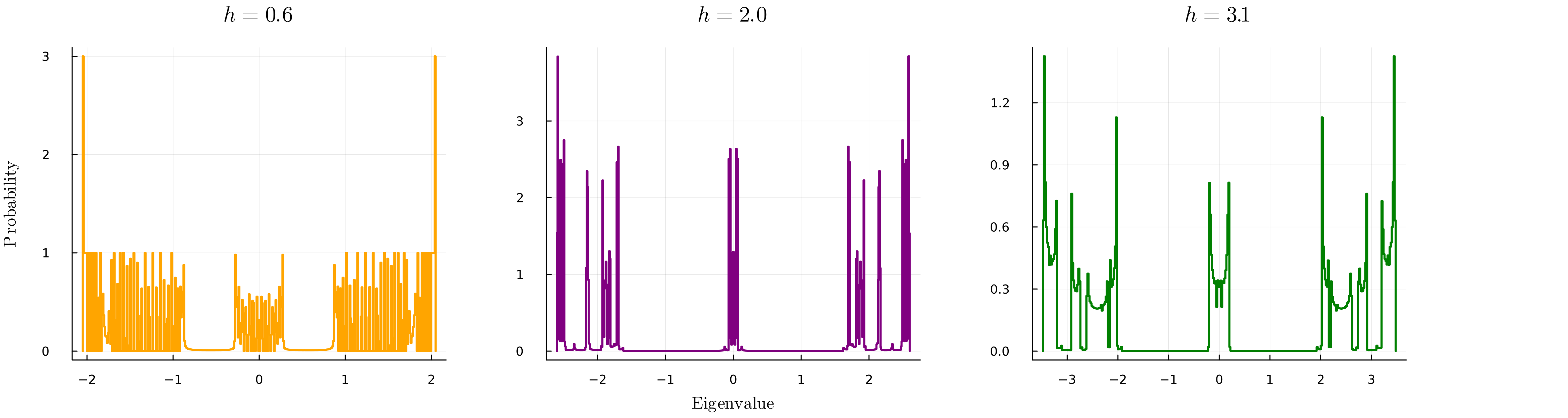}
    \caption{Calculated density of states of the quasiperiodic model, where onsite potential values are given by Eq.~(\ref{eq:quasiperiodic_potential}). Here, we fix $k=\sqrt{2}$ and $L=100$. The DOS is calculated over the whole spectrum of each of $\hat H$ with $500$ different phases $\delta$ between $\delta\in\left[0,\frac{1}{k} \right]$.}
\label{fig:density_of_states_quasiperiodic}
\end{figure*}

In our work, we observe slower tails ($x^{-1.5}$ and $x^{-1.75}$) in various quantities associated with the quasiperiodic model, when compared to the Anderson model.
Similar to how the $\frac{1}{x^2}$ tails in the Anderson model can be traced back to the distribution of energy denominators, a comparable mechanism is at play here.
However, in the quasiperiodic case, the distribution of combinations like $\epsilon_\alpha + \epsilon_\beta - \epsilon_\gamma - \epsilon_\delta$ is highly nontrivial, making it difficult to carry out an analysis analogous to that in Appendix~\ref{app:derivation_tails}. 
Therefore, for reference, we present the density of states (DOS) for the single-particle quasiperiodic model in Fig.~\ref{fig:density_of_states_quasiperiodic}, computed at $k = \sqrt{2}$ and $L = 100$, for values of $h$ below, at, and above the transition point $h = 2$.
These DOS plots are generated over the spectra of 500 uniformly spaced phase offsets $\delta$ in the interval $[0, \frac{1}{k}]$.
The energy levels for a given $\delta$ follow a devil’s staircase-like structure (Cantor function) \cite{azbel_quantum_1979}, resulting in a highly complex and nontrivial density of states. 
This complex structure in the DOS likely underlies the emergence of the observed form of the slower-decaying tails.

\section{Participation Ratio Data}\label{app:localization_length}

\begin{figure}[h]
    \centering
    \includegraphics[width=\columnwidth]
    {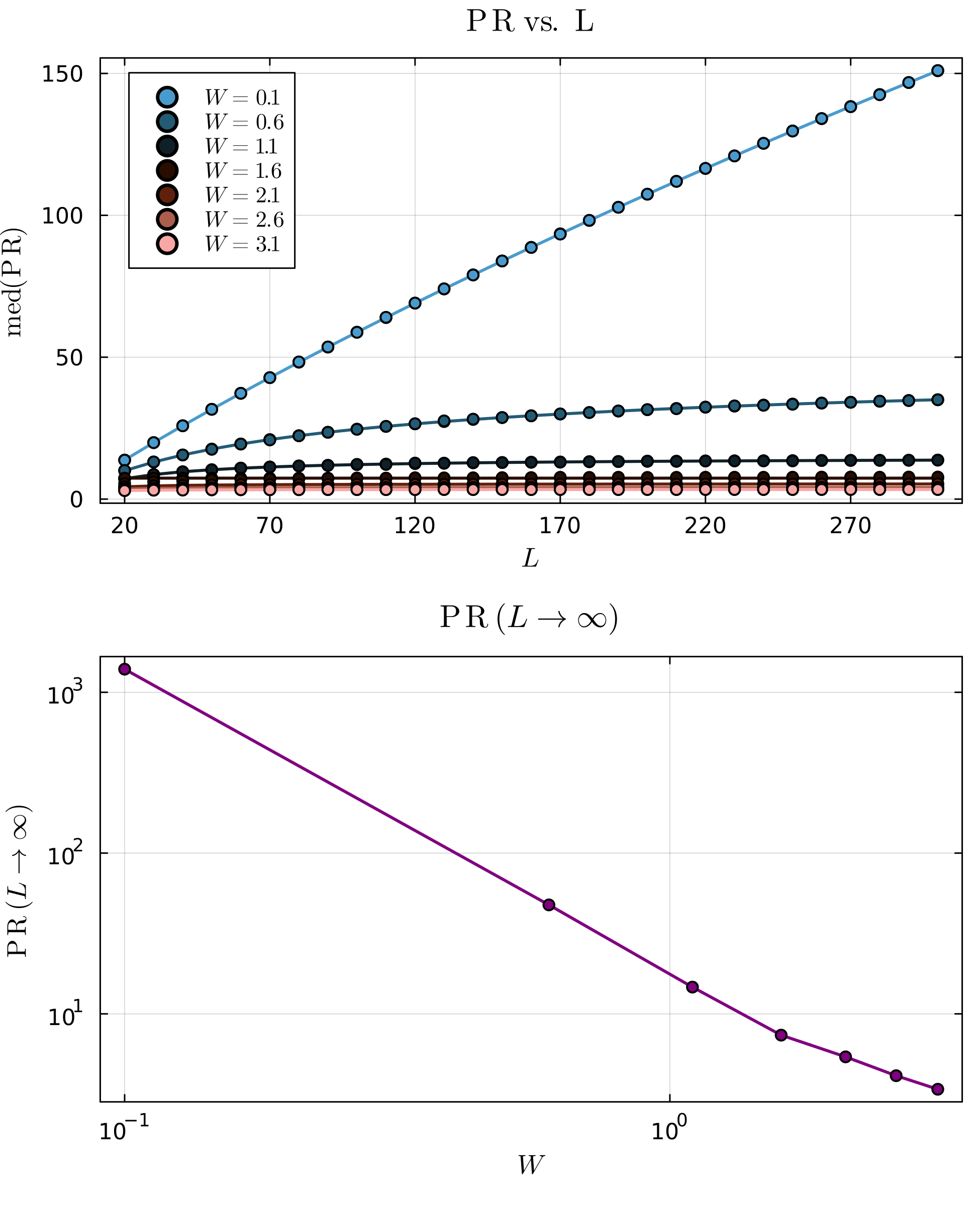}
    \caption{
    (a) Median participation ratio (PR) plotted against $L$.
    The median is taken over 500 disorder realizations, where $\mathrm{PR}$ is calculated for the eigenstate in the middle of the spectrum for each of the disorder realizations.
    Each colored curve represents a different disorder strength $W$.
    The dashed line represents the best fit of the form $\mathrm{PR}(L)=A-\frac{A}{1-BL^\alpha}$, where $A$ represents the converged $\mathrm{PR}$ value, $\lim_{L\rightarrow\infty}\mathrm{med(PR)}$, and $B, \alpha$ are fitting parameters.
    (b) Fitted values of $A(W)=\lim_{L\rightarrow\infty}\mathrm{med(PR)}$ versus the disorder strength $W$.
    A naive polynomial fit to this curve gives a power-law description $\text{PR} \sim W^{-1.78}$.
    }
    \label{fig:ipr_vs_L}
\end{figure}

Understanding the localization length $\xi$ as a function of $W$ is crucial to understanding finite-size effects and provides a useful reference point for diagnosing localization.
The participation ratio (PR) is a proxy for the localization length and is thus useful to understand.
For a normalized orbital $\phi_\alpha(j)$, the participation ratio is
\begin{equation}
\text{PR} = \frac{1}{\sum_j |\phi_\alpha(j)|^4} ~.
\end{equation}
For an orbital with a large localization length in 1D, $\text{PR} \sim \xi$~\footnote{
The participation ratio, while a proxy for the localization length, is not equivalent to it.
The localization length $\xi$ is used to characterize the exponential decay of the orbitals $\phi_i$, where $\phi_i(j)\sim\exp[-|i-j|/\xi]$
On the other hand, the participation ratio measures the extent of the wavefunction $\phi$.
Extracting $\xi$ directly from the PR is not straightforward---see, for instance, Ref.~\cite{laflorencie_entanglement_2022}.
For our purposes, it is sufficient to look at the participation ratio rather than trying to directly calculate $\xi$.}.

For quick reference, in this section, we show data on how the median of the participation ratio grows with system size in Fig.~\ref{fig:ipr_vs_L}(a), at different disorder strengths $W$.
Here, the median is taken over $500$ disorder realizations, with the participation ratio being calculated for the eigenstate in the middle of the spectrum for each disorder realization. 
We are particularly interested in the behavior of the $\mathrm{PR}$ in the thermodynamic limit $L \rightarrow \infty$. 
At small $L$, the curves exhibit an initial power-law growth before saturating to a disorder-dependent plateau that decreases with increasing $W$.
Moreover, the PR must vanish as $L \rightarrow 0$. 
These features motivate the use of the following fitting form for each $\mathrm{med[PR]}$ 
\begin{equation*}
    \mathrm{med[PR](L)} = A - \frac{A}{1 - B L^\alpha}.
\end{equation*}
Here, $A$ represents the value of the participation we desire, $A = \lim_{L\rightarrow\infty}{\mathrm{med[PR](L)}}$.
$\alpha$ is the power-law exponent and $B$ is a fitting parameter to enforce that $\mathrm{PR}$ is 0 when $L=0$.
Note that this form does not have the anticipated exponential decay at very large $L \gg \xi$; however, much of our data is not deep in this regime, and the above form with few parameters is adequate for our purposes.
We then show these converged $\mathrm{PR}$ values in Fig.~\ref{fig:ipr_vs_L}(b).
From this, one can approximately obtain the scaling of the localization length with $W$.
For our purposes, a simple power-law approximation suffices.
However, we note that this is an oversimplification, and the scaling actually differs in the low versus high disorder regimes \cite{miniatura_ultracold_2011}.
Regardless, we find $\lim_{L \rightarrow \infty} \mathrm{med[PR](L)} \sim W^{-1.78}$.
This scaling allows us to rescale the median values of other observables by the localization length.
When one does so, the curves collapse when the y-axis is appropriately scaled by some power of the disorder strength as well, highlighting the role of finite-size effects.

\section{Derivation of Charge Transport Operator}\label{app:a-derivation-transport}

In this section, we present more details on the derivation of $\hat M^{(0)}$, the charge transport capacity in the noninteracting case, and $\hat M^{(1)}$, the correction to $\hat M^{(0)}$ when nearest-neighbor interactions are added.
First, we wish to find a $\hat M^{(0)}$ such that:
\begin{equation}\label{current_operator_equation}
    \hat J_{i_0} = i\left[\hat H_0, \hat M^{(0)}_{i_0}\right].
\end{equation}
Here, $\hat H_0 = \sum_{ij} A_{ij}\hat{c}^{\dagger}_i\hat c_j=\sum_{\alpha}\epsilon_\alpha \hat{d}^\dagger_{\alpha}\hat d_{\alpha}$ is our free-fermion Hamiltonian. Also, $\hat J_{i_0}$ is the charge current operator:
\begin{equation*}
    \hat J_{i_0} = -i\left(\hat c_{i_0}^\dagger \hat{c}_{i_0+1}-\hat{c}_{i_0+1}^\dagger \hat{c}_{i_0}\right).
\end{equation*}
We can rewrite $\hat J_{i_0}$ in the orbital basis:
\begin{equation}\label{orbital_basis_current}
    \hat J_{i_0} = i \sum_{\alpha, \beta} (  \phi_{i_0, \beta} \phi^*_{i_0+1, \alpha}-\phi^*_{i_0, \alpha} \phi_{i_0+1, \beta})\hat d^{\dagger}_\alpha\hat d_\beta.
\end{equation}

We consider the following 2-fermionic operator ansatz for $\hat M^{(0)}$:
\begin{equation}
    \hat M^{(0)} = \sum_{\alpha, \beta}M^{(0)}_{\alpha\beta} \hat{d}^\dagger_{\alpha}\hat d_{\beta}.
\end{equation}
With this, we evaluate the commutator on the right-hand side of Eq.~(\ref{current_operator_equation}):
\begin{equation*}
    \hat J_{i_0} = i\left[\hat H_0, \hat M^{(0)}_{i_0}\right] = i\sum_{\alpha,\beta}M^{(0)}_{\alpha\beta}(\epsilon_\alpha-\epsilon_\beta) \hat{d}^\dagger_{\alpha}\hat d_{\beta}.
\end{equation*}
We can directly compare this with Eq.~(\ref{orbital_basis_current}) to obtain the matrix elements $M^{(0)}_{\alpha\beta}$:
\begin{equation}
 M^{(0)}_{\alpha\beta} = \frac{\phi_{i_0, \beta} \phi^*_{i_0+1, \alpha}-\phi^*_{i_0, \alpha} \phi_{i_0+1, \beta}}{\epsilon_\alpha-\epsilon_\beta}.
\end{equation}
Now, when we add interactions, we enforce that $\hat J$ has the same expression and consider finding a first-order correction $\hat M^{(1)}$ such that:
\begin{equation*}
    \hat J_{i_0} = i\left[\hat H_0 + \lambda \hat V, \hat M^{(0)}+\lambda\hat M^{(1)}\right].
\end{equation*}
This is equivalent to solving:
\begin{equation}
    \left[\hat H_0, \hat M^{(1)}\right] = -\left[\hat V, \hat M^{(0)}\right].
\end{equation}
Evaluating the right-hand side:
\begin{align}\label{commuator_transport}
    \left[\hat V, \hat M^{(0)}\right]&=\sum_{\alpha\beta\gamma\delta\mu}(V_{\alpha\beta\gamma\mu}M^{(0)}_{\mu\delta} +V_{\alpha\beta\mu\delta}M^{(0)}_{\mu\gamma} \nonumber \\ &- V_{\mu\beta\gamma\delta}M^{(0)}_{\alpha\mu}-V_{\alpha\mu\gamma\delta}M^{(0)}_{\beta\mu}) \hat{d}^{\dagger}_\alpha\hat{d}^{\dagger}_\beta\hat{d}_\gamma\hat{d}_\delta.
\end{align}
For the left-hand side, we create a four-fermionic operator ansatz for $\hat M^{(1)}$:
\begin{equation*}
\hat M^{(1)}=\sum_{\alpha\beta\gamma\delta}M^{(1)}_{\alpha\beta\gamma\delta} \hat{d}^{\dagger}_\alpha\hat{d}^{\dagger}_\beta\hat{d}_\gamma\hat{d}_\delta.
\end{equation*}
With this ansatz, we find that the left-hand side evaluates to:
\begin{equation*}
\left[\hat H_0, \hat M^{(1)}\right]=\sum_{\alpha\beta\gamma\delta}M^{(1)}_{\alpha\beta\gamma\delta} (\epsilon_\alpha+\epsilon_\beta-\epsilon_\gamma-\epsilon_\delta)\hat{d}^{\dagger}_\alpha\hat{d}^{\dagger}_\beta\hat{d}_\gamma\hat{d}_\delta.
\end{equation*}
Comparing this with the left-hand side, Eq.~(\ref{commuator_transport}), we find the matrix elements $M^{(1)}_{\alpha\beta\gamma\delta}$ by direct comparison:
\begin{align}
M^{(1)}_{\alpha\beta\gamma\delta}&=\sum_{\mu}(V_{\alpha\beta\gamma\mu} M^{(0)}_{\mu\delta} + V_{\alpha\beta\mu\delta} M^{(0)}_{\mu\gamma} \nonumber \\ &- V_{\alpha\mu\gamma\delta} M^{(0)}_{\mu\beta} - V_{\mu\beta\gamma\delta} M^{(0)}_{\mu\alpha}) / (\epsilon_\alpha + \epsilon_\beta - \epsilon_\gamma - \epsilon_\delta).
\end{align}
With these operators, one can provide a rigorous upper bound to the total amount of charge transported.
The difference $\Delta\hat N(t)=\hat N_R(t)-\hat N_R(0)$ can be expressed as (using Heisenberg time-evolved operators with respect to the full Hamiltonian $H_0 + \lambda V$)
\begin{align}
    \Delta \hat N(t)&=\int_0^t \hat J(s)ds \nonumber\\
    &= \int_0^t \left(i[\hat H_0+\lambda \hat V, \hat M^{(0)}(s)+\lambda M^{(1)}(s)]\right. \nonumber\\
    &\qquad \left.-i\lambda^2[\hat V(s), M^{(1)}(s)]\right)ds \nonumber\\
    & =\hat M^{(0)}(t)+\lambda\hat M^{(1)}(t)-\hat M^{(0)}(0)-\lambda\hat M^{(1)}(0) \nonumber \\
    & \qquad -\lambda^2\int_0^t i[\hat V(s), \hat M^{(1)}(s)]ds.
\end{align}
We thus obtain the following bound
\begin{equation}
    \|\Delta \hat N(t)\|  \le 2 \|\hat M^{(0)}\|+2\lambda \|\hat M^{(1)}\|+\lambda^2 t \|i[\hat V, \hat M^{(1)}]\|.
\end{equation}

\section{General Bounds on the Charge Transport Capacity}
\label{app:bounds}
 In this Appendix, we derive a general bound on the Frobenius norm of $\hat M_{i_0}$. We note that  
\begin{equation}
    \|\hat M_{i_0}\|_F \le \|\hat N_R -s \hat N_\text{tot}\|_F
\end{equation}
for an arbitrary constant $s$, and $\hat N_\text{tot}=\sum_j \hat n_j$.
This follows from Eq.~(\ref{eq:NRoffdiag}) using that $[\hat N_\text{tot}]_\text{off-diag}=0$, because of the total charge conservation $[\hat N_\text{tot}, \hat H]=0$.
The Frobenius norm $\|\hat N_R -s\hat N_\text{tot}\|_F$ can be computed as described in Appendix~\ref{app:a_frobenius_norm_derivation} for arbitrary bilinear operators:
\begin{align}
    \|\hat N_R -s\hat N_\text{tot}\|_F^2=&\frac{1}{4}[(-s)i_0+(1-s)(L-i_0)]^2 \\
    &+\frac{1}{4}[s^2 i_0+(1-s)^2(L-i_0)]\\
    =&\frac{1}{4}[L(L+1)s^2-2(L-i_0)(L+1)s\\
    &+(L-i_0)(L-i_0-1)].
\end{align}
To find the optimal bound, we minimize over $s$.
The minimum is obtained for $s=(L-i_0)/L$, yielding:
\begin{equation}
    \|\hat M_{i_0}\|_F^2 \le \|\hat N_R - \frac{L-i_0}{L}\hat N_\text{tot}\|_F^2 = \frac{1}{4}\frac{i_0(L-i_0)}{L}.
    \label{eq:genboundFnormM}
\end{equation}
This bound holds in general, as it only uses that $\hat H$ is charge-conserving.
In particular, in Appendix~\ref{app:noninteracting_transport_details} we will argue that these bounds are approximately saturated for the clean non-interacting chain.

It is interesting to note that the value $s = i_0/L$ minimizes also the operator norm of $\hat N_R - s\hat N_\text{tot}$:
\begin{align}
    \|\hat N_R-\frac{L-i_0}{L}\hat N\|_\text{op}&=\max \left(i_0\left(1-\frac{i_0}{L}\right), \frac{i_0}{L}(L-i_0)\right)\nonumber\\
    &=\frac{i_0(L-i_0)}{L}.
\end{align}
This represents an upper bound for the operator norm of $\Delta \hat N_{i_0}(t)$ for arbitrary times: 
\begin{equation}
    \| \Delta \hat N_{i_0}(t)\|_\text{op} \le 2\frac{i_0(L-i_0)}{L}.
\end{equation}

However, this bound on the operator norm is not very useful in practice, as it is looser than (or equivalent to, for $i_0=L/2)$ the more intuitive bound\footnote{
This bound can be obtained by noting that
\begin{equation}
    \|\Delta \hat N_{i_0}(t)\|_\text{op} \le 2\|\hat N_R -\frac{L-i_0}{2}\hat I\|_\text{op}=L-i_0,
\end{equation}
and
\begin{equation}
    \|\Delta \hat N_{i_0}(t)\|_\text{op} \le 2\|\hat N_R -\hat N+\frac{i_0}{2}\hat I\|_\text{op}=i_0,
\end{equation}
where $\hat I$ is the identity operator.}

\begin{equation}
    \|\Delta \hat N_{i_0}(t)\|_\text{op} \le \min(i_0, L-i_0).
\end{equation}

In the absence of further assumptions about the Hamiltonian, the bound $\|\Delta \hat N_{i_0}(t)\| \le \min(i_0, L-i_0)$ is optimal. A simple example of a model Hamiltonian that saturates the bound is the following site-dependent hopping model
\begin{equation}
    \hat H_\text{hop}=-\sum_{j=1}^{L-1} t_j (\hat c_{j+1}^\dagger\hat c_j+\text{H.c.}),
\end{equation}
with $t_j=\sqrt{j(L-j)}$. This model realizes perfect state transfer \cite{Christandl2004}, in which a particle initialized at site $j_0$ is perfectly transferred to site $L+1-j_0$ at time $\tau=\pi/2$. 
By preparing an initial state where all sites in the smaller of the two regions (left or right of the link) are occupied, one finds that all particles have crossed the link at time $\tau$, thus saturating the bound.

\section{Details About the Charge Transport Capacity Matrix Elements and Norms in the Noninteracting Case} \label{app:noninteracting_transport_details}
Here we discuss in more detail the matrix elements, Eq.~(\ref{eq:M0albt}), of the charge transport capacity operator in the free-fermion case.
We first note that in general, there are no issues with small denominators in Eq.~(\ref{eq:M0albt}).
To show this, we can start with the eigenvalue equation for each orbital $\alpha$:
\begin{equation}
\epsilon_\alpha \phi_{j\alpha} = h_j \phi_{j\alpha} - \phi_{j+1,\alpha} - \phi_{j-1,\alpha} ~, \quad j = 1, \dots, L ~.
\end{equation}
Here we specialize to our model in Eq.~(\ref{eq:hamiltonian}) setting the hopping amplitude $t = 1$, and use sentinels $\phi_{0,\alpha} = \phi_{L+1,\alpha} = 0$ for compact writing in the OBC chain with $L$ sites.
Since the Hamiltonian is real-valued, we can take the orbitals $\phi_{j\alpha}$ to be real-valued and assume this as well as their orthonormality throughout.
Taking such equations for two orbitals $\alpha \neq \beta$, multiplying the one for the $\alpha$ orbital by $\phi_{j\beta}$ and the one for the $\beta$ orbital by $\phi_{j\alpha}$, and subtracting one equation from the other, we get after simple reorganization
\begin{align}
(\epsilon_\alpha - \epsilon_\beta) \phi_{j\alpha} \phi_{j\beta} =& 
\phi_{j\alpha} \phi_{j+1,\beta} - \phi_{j\beta} \phi_{j+1,\alpha} \nonumber \\
& - \left[\phi_{j-1,\alpha} \phi_{j\beta} - \phi_{j-1,\beta} \phi_{j\alpha} \right] ~.
\end{align}
This gives
\begin{equation}
\left( M_{j-1}^{(0)} \right)_{\alpha\beta} -\left( M_j^{(0)} \right)_{\alpha\beta} = \phi_{j\alpha} \phi_{j\beta} ~.
\end{equation}
Iterating starting from the leftmost site $j = 1$, we obtain
\begin{equation}
\label{eq:M0albt_nodenom}
\left( M_{i_0}^{(0)} \right)_{\alpha\beta} = -\sum_{j=1}^{i_0} \phi_{j\alpha} \phi_{j\beta} ~.
\end{equation}
(This result can be also derived more directly from $\hat{M}_{i_0} = [\hat{N}_R]_{\text{off-diag}} = -[\hat{N}_L]_{\text{off-diag}}$).
Thus, we have found a rewriting of Eq.~(\ref{eq:M0albt}) which has no energy denominators, demonstrating that no divergences can arise from them.
This agrees with the fact that the charge transport capacity is upper-bounded by $\min\{i_0, L-i_0\}$ independent of any other details of the system, since one cannot change the charge on any site by more than one.

We can extract further insights from the above expression for the matrix elements.
For $\alpha \neq \beta$, we can use the orthogonality of the orbitals to equivalently write
\begin{equation}\label{eq:m_0_right}
\left( M_{i_0}^{(0)} \right)_{\alpha\beta} =\sum_{j=i_0+1}^{L} \phi_{j\alpha} \phi_{j\beta} ~.
\end{equation}
Furthermore, applying the Cauchy–Schwarz inequality to the above expressions, we can obtain the following bounds on the matrix elements:
\begin{equation}
\begin{aligned}
&\Big|\left( M_{i_0}^{(0)} \right)_{\alpha\beta} \Big| \leq \min\Bigg\{ 
\Bigg(\sum_{j=1}^{i_0} \phi_{j\alpha}^2\Bigg)^{1/2} \Bigg(\sum_{j=1}^{i_0} \phi_{j\beta}^2 \Bigg)^{1/2} \!\!, \\
&\qquad \Bigg(\sum_{j=i_0+1}^L \phi_{j\alpha}^2\Bigg)^{1/2} \Bigg(\sum_{j=i_0+1}^L \phi_{j\beta}^2 \Bigg)^{1/2}
\Bigg\} ~.
\end{aligned}
\label{eq:boundM0albt}
\end{equation}

\subsection{Estimates of the Matrix Elements and the Frobenius and Operator norms in the Localized Phase}
Let us consider the above matrix elements when all orbitals are localized (either in the Anderson insulator or in the localized phase of the Aubry-André model).
In this case, we see that $\big( M_{i_0}^{(0)} \big)_{\alpha\beta}$ is significant only if both $\alpha$ and $\beta$ orbitals reside near $i_0$, within a distance of order the localization length $\xi$.
Indeed, if both of them are far to the left of $i_0$, then the corresponding weights in the region to the right of $i_0$ are small, and the matrix element is small because of the second line of Eq.~(\ref{eq:boundM0albt}).
On the other hand, if both of them are far to the right of $i_0$, we can use the smallness of their weights to the left of $i_0$ and the first line of Eq.~(\ref{eq:boundM0albt}).
In fact, from these expressions, it is clear that we just need one of the orbitals to be far away from $i_0$, localized either to the left or to the right, for the matrix element to be small.
We thus conclude that only about $O(\xi \times \xi)$ matrix elements are significant.

In the regime when $\xi$ is much larger than the lattice spacing, we can further estimate the scaling of these matrix elements with $\xi$ as follows:
For orbitals $\alpha \neq \beta$ localized near $i_0$, within $O(\xi)$, we estimate magnitudes of their amplitudes as roughly $|\phi_{j\alpha}|, |\phi_{j\beta}| \sim O(1/\sqrt{\xi})$ and signs as roughly random from each other (since these orbitals need to be orthogonal to each other and all the other orbitals nearby).
Hence, we can make a random-walk-like estimate of the r.h.s.\ in Eq.~(\ref{eq:M0albt_nodenom}):
\begin{equation}
\left|\left( M_{i_0}^{(0)} \right)_{\alpha\beta} \right| \sim O\left(\frac{1}{\sqrt{\xi}} \right)^2 \times O\left(\sqrt{\xi} \right) \;\sim\; O\left(\frac{1}{\sqrt{\xi}} \right) ~.
\end{equation}
Hence, for the squared Frobenius norm of the charge transport capacity operator $\hat{M}_{i_0}^{(0)}$, which up to a factor of $1/4$ coincides with the standard squared Frobenius norm of the matrix $M_{i_0}^{(0)}$ (cf.\ Appendix~\ref{app:a_frobenius_norm_derivation}), we obtain
\begin{equation}
\big\|\hat{M}_{i_0}^{(0)}\big\|_F^2 \sim O\left(\xi \times \xi \right) \times O\left(\frac{1}{\sqrt{\xi}} \right)^2 \sim O(\xi) ~.
\end{equation}
We will confirm this scaling using a more precise argument later, which will also indirectly support the above estimate of the matrix element scaling.

For the operator norm of $\hat{M}_{i_0}^{(0)}$ in the many-body Hilbert space, we first need to estimate eigenvalues of the corresponding $L \times L$ matrix $M_{i_0}^{(0)}$.
Assuming that the only significant part of the matrix is the discussed $O(\xi \times \xi)$ block formed by orbitals within $O(\xi)$ from $i_0$, with entries of magnitude $O(1/\sqrt{\xi})$, treating these as uncorrelated and applying Random Matrix Theory, we estimate that there are $O(\xi)$ significant eigenvalues with a typical magnitude of order $O(\sqrt{\xi}) \times O(1/\sqrt{\xi}) \sim O(1)$.
Roughly half of them are positive and half are negative.
Adding up only positive (or only negative) ones then gives an estimate of the operator norm
\begin{equation}\label{eq:scaling_operator_norm}
\big\|\hat{M}_{i_0}^{(0)}\big\|_{\text{op}} \sim O(\xi) ~.
\end{equation}

From the scalings of the Frobenius and operator norms of $\hat{M}_{i_0}^{(0)}$ with $\xi$, we can also estimate the scalings of these norms with the system size $L$ in the delocalized regime, by simply replacing $\xi \to L$:
\begin{equation}
\big\|\hat{M}_{i_0}^{(0)}\big\|_F \sim O(\sqrt{L})~, \qquad
\big\|\hat{M}_{i_0}^{(0)}\big\|_{\text{op}} \sim O(L) ~.
\end{equation}
These scalings are supported by our numerical calculations as well as arguments from different perspectives in the main text and in the following subsection.

\subsection{Concise Expression for the Frobenius Norm and More Accurate Estimates}

Returning to the general case, we can calculate the Frobenius norm of $\hat{M}_{i_0}^{(0)}$ as follows
(reminding that we use the many-body Hilbert space Frobenius norm, see Appendix~\ref{app:a_frobenius_norm_derivation}):
\begin{equation}
\begin{aligned}
\|\hat{M}_{i_0}\|_F^2 &= \frac{1}{4} \sum_{\alpha \neq \beta} \left( M_{i_0}^{(0)} \right)_{\alpha\beta}^2~, \\
\sum_{\alpha \neq \beta} \left( M_{i_0}^{(0)} \right)_{\alpha\beta}^2 &= \sum_{\alpha,\beta} (1 - \delta_{\alpha,\beta}) \sum_{j,j'=1}^{i_0} \phi_{j\alpha} \phi_{j\beta} \phi_{j'\alpha} \phi_{j'\beta} \\
& = \sum_{j,j'=1}^{i_0} \delta_{jj'} - \sum_\alpha \left(\sum_{j=1}^{i_0} \phi_{j\alpha}^2 \right)^2 \\
&= i_0 - \sum_\alpha \left(\sum_{j=1}^{i_0} \phi_{j\alpha}^2 \right)^2 ~,
\end{aligned}
\end{equation}
where we have used $\sum_\alpha \phi_{j\alpha} \phi_{j'\alpha} = \delta_{jj'}$ from the orthonormality of the $L\times L$ matrix $\phi_{j\alpha}$.
From this and identical analysis applied starting from Eq.~(\ref{eq:m_0_right}), we have trivial upper bounds $\|\hat{M}_{i_0}\|_F^2 \leq \frac{1}{4} \min\{i_0, L-i_0\}$.
However, we already have better and more general bounds in Eq.~(\ref{eq:genboundFnormM}) in Appendix~\ref{app:bounds}.
The utility of the above expression is that we can also understand the behavior of the Frobenius norm in the delocalized and localized regimes.

In the delocalized regime, assuming $i_0$ is a finite fraction of $L$, for large enough $L$ and for most of the orbitals we can estimate the fraction of the weight on the sites up to $i_0$ as $\sum_{j=1}^{i_0} \phi_{j\alpha}^2 \approx \frac{i_0}{L}$.
Hence we obtain $\|\hat{M}^{(0)}_{i_0}\|_F^2 \approx \frac{1}{4}\frac{i_0 (L - i_0)}{L}$.
This scaling saturates the general bound in Eq.~(\ref{eq:genboundFnormM}), which can be understood from the extended nature of the orbitals, where, except for a vanishing fraction of them, the weight is distributed essentially uniformly across the system. 
As an example, for $i_0 = L/2$ we have $\|\hat{M}^{(0)}_{i_0}\|_F^2 \approx \frac{1}{4}\frac{L}{4}$; 
For $L$ even and assuming inversion symmetry relative to bond center $L/2,L/2+1$, this estimate is, in fact, exact since we can choose orbitals to have a definite inversion symmetry eigenvalue and hence have precisely half of the weight on the half-system.

On the other hand, when orbitals are localized, $w_\alpha[1,\dots,i_0] = \sum_{j=1}^{i_0} \phi_{j\alpha}^2 \approx 1$ for orbitals localized far to the left of $i_0$; $w_\alpha[1,\dots,i_0] \approx 0$ for orbitals localized far to the right of $i_0$; and $w_\alpha[1,\dots,i_0] \approx O(1)$ for orbitals near $i_0$, within $\xi$ from $i_0$.
Hence, $\|\hat{M}^{(0)}_{i_0}\|_F^2 \approx O(\xi)$.

We can combine this result with the picture that only $O(\xi \times \xi)$ matrix elements are important to estimate that those matrix elements must typically take values $O(1/\sqrt{\xi})$.
This is then a more precise argument for the scaling of these matrix elements than the schematic random-walk-like estimate given earlier.

\section{Operator Norm Data for $\hat M^{(0)}$}\label{app:operator_norm_data_noninteracting_transport} 

In the noninteracting case, we can easily calculate the operator norm of the charge transport capacity operator $\hat{M}^{(0)}$.
First, we diagonalize the corresponding matrix $M^{(0)}_{\alpha\beta}$, obtaining eigenvalues $\{m_\kappa, \kappa=1,\dots,L\}$.
Since $\hat{M}$ is a fermion bilinear operator and we are interested in its operator norm in the full many-body Hilbert space, we can calculate it as the largest in absolute value eigenvalue of $\hat{M}^{(0)}$:
\begin{equation}
\| M^{(0)} \| = \max \left\{ \sum_{\kappa, m_\kappa > 0} m_\kappa, \left|\sum_{\kappa, m_\kappa < 0} m_\kappa \right| \right\}.
\end{equation}

\begin{figure*}[t]
    \centering
    \includegraphics[width=\linewidth]{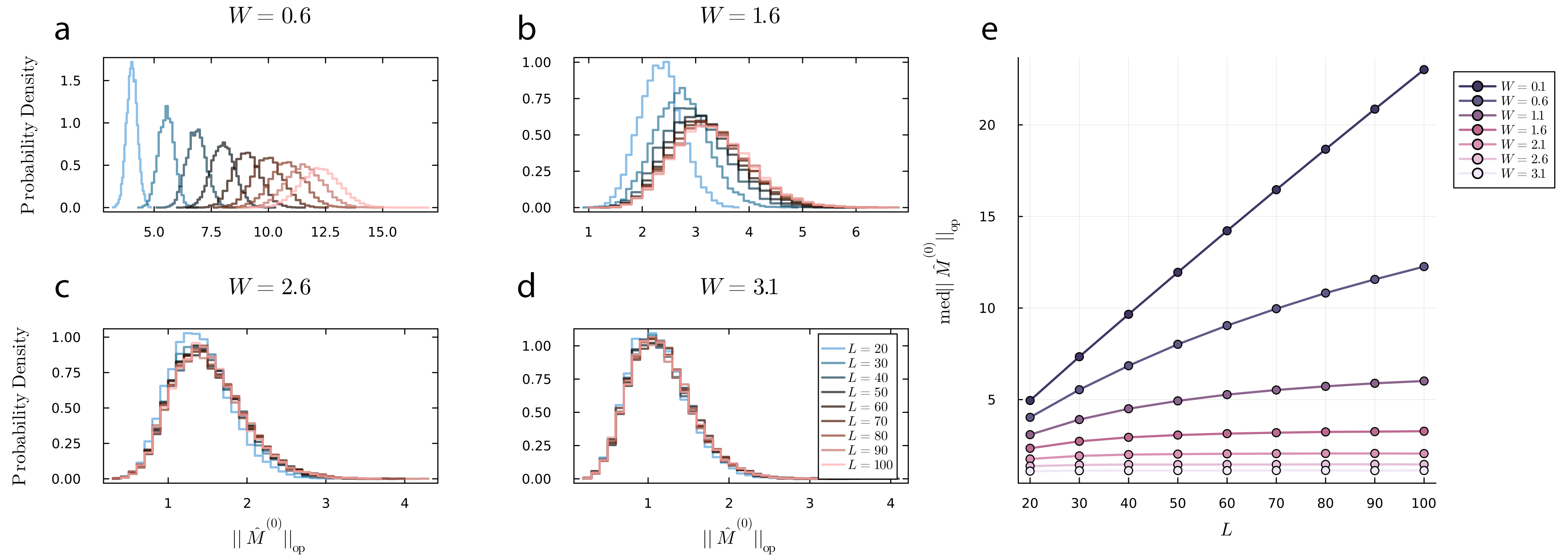}
    \caption{Analysis of the operator norm of the charge transport capacity, $\|\hat M^{(0)}\|_{\mathrm{op}}$, for the noninteracting Anderson Model with open boundary conditions with $i_0=\lfloor L/2\rfloor$. The calculations here are done based on 10000 disorder realizations.
    (a)--(d): Calculated probability density functions of $\|\hat M^{(0)}\|_{\mathrm{op}}$.
    Each panel represents different disorder strengths at $W = 0.6, 1.6, 2.6, 3.1$. 
    Within each panel, each colored curve represents different system sizes between $L=20$ and $L=100$, in steps of 10. 
    (e): $\mathrm{med}\|\hat M^{(0)}\|_{\mathrm{op}}$ as a function of $L$.
    Each curve represents a different disorder strength.
    In the low disorder limit, the operator norm of $\hat M^{(0)}$ approaches the curve $\mathrm{med}\|\hat M^{(0)}\|_{\mathrm{op}} \approx L/4$ while $L \ll \xi$, showing different scaling from that of the Frobenius norm, which was shown in the main text in Fig.~\ref{fig:noninteracting_transport_disordered_figure}.}
    \label{fig:noninteracting_transport_disordered_operator_norm}
\end{figure*}

\begin{figure*}[t]
    \centering
    \includegraphics[width=\linewidth]{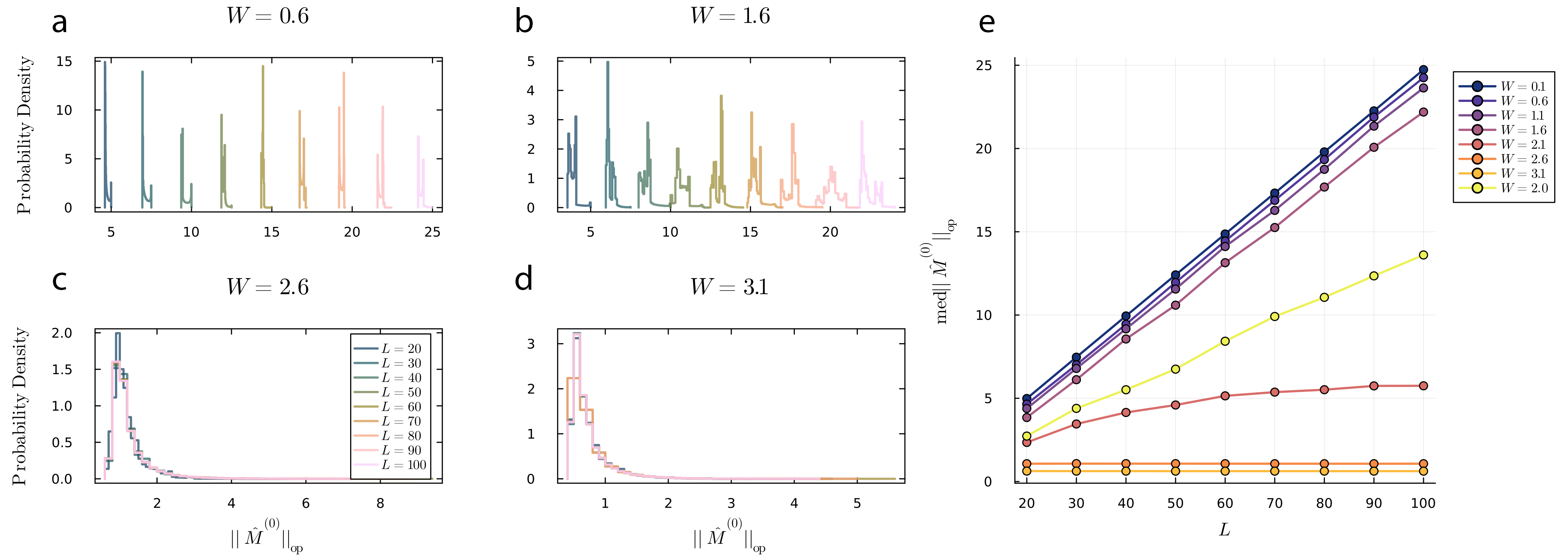}
    \caption{Analysis of the operator norm of the charge transport capacity, $\|\hat M^{(0)}\|_{\mathrm{op}}$, for the noninteracting quasiperiodic model with open boundary conditions based on 500 evenly spaced phase shifts $\delta$ between $\delta=0$ and $\delta=\frac{1}{k}$. Here, $i_0=\lfloor L/2\rfloor$ and $k=\sqrt{2}$. (a)--(d): Calculated probability density functions of $\|\hat M^{(0)}\|_{\mathrm{op}}$. Each panel shows distributions at different potential depths, at $h=0.6,1.6,2.6,3.1$. Within each panel, each colored curve represents a different system size. We calculate distributions between $L=20$ and $L=100$ in steps of 10. (e): $\mathrm{med}\|\hat M^{(0)}\|_{\mathrm{op}}$ as a function of $L$. Each curve represents a different potential depth $h$, ranging from $h=0.1$ to $h=3$ in steps of 0.5. We also show the curve at the transition $h=2$ (yellow).}
    \label{fig:noninteracting_transport_quasiperiodic_operator_norm}
\end{figure*}

Here, we present calculations of $\|\hat M^{(0)}\|_\text{op}$ for the Anderson Model in Fig.~\ref{fig:noninteracting_transport_disordered_operator_norm}, and for the quasiperiodic model in Fig.~\ref{fig:noninteracting_transport_quasiperiodic_operator_norm}. 
As explained in the main text, in the clean model or the delocalized phase of the Aubry-André model, the median of the operator norm of $\hat M^{(0)}$ scales linearly with system size, as $\approx L/4$.
This contrasts with the behavior of the Frobenius norm, which scales as $\approx \frac{1}{2} \sqrt{L/4}$.
On the other hand, in the localized phases, the operator norm saturates to a finite disorder-dependent value, which scales with the localization length consistent with Eq.~(\ref{eq:scaling_operator_norm}) 
We refer the reader to the main text of Sec.~\ref{sec:non_interacting_transport_results} and Appendix~\ref{app:noninteracting_transport_details} for a more detailed discussion.

\section{Transport in the Interacting Quasiperiodic Model for $k=\sqrt{3}$}\label{app:additional_quasiperiodic_wavenumbers}

\begin{figure*}[t]
    \centering
    \includegraphics[width=\linewidth]{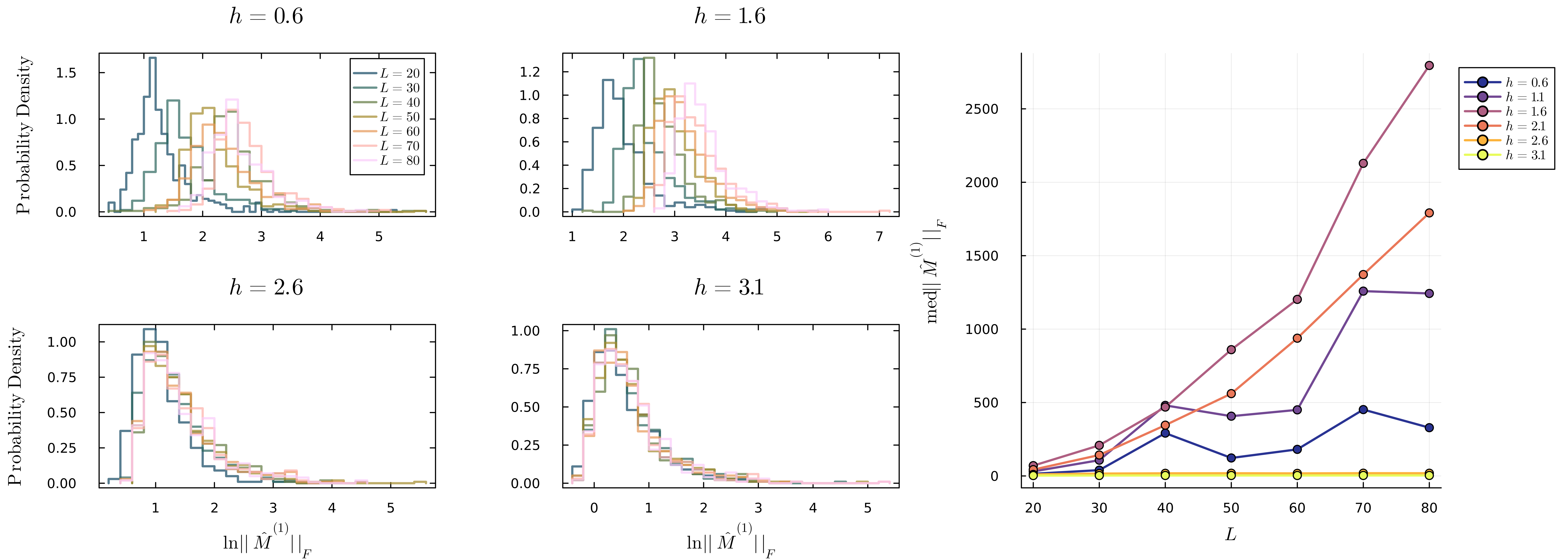}
    \caption{
    (a)-(d): Probability density functions of $\|\hat M^{(1)}\|$ in the interacting quasiperiodic model with $k = \sqrt{3}$, based on 500 evenly spaced phase shifts $\delta$ between $\delta=0$ and $\delta=1/k$. 
    Within each panel, each distribution represents a different system size $L$.
    (e): Scaling of $\mathrm{med}\|\hat M^{(1)}\|$ with $L$.
    Each curve represents a different potential depth $h$.
    One sees that the spikes occur at different locations compared to the case $k = \sqrt{2}$ considered in the main text, notably at $L = 40$ and 70. Those system sizes are approximately commensurate with the underlying potential, as $\sqrt{3}\times 41\approx 71$ and $\sqrt{3}\times 71 \approx 123$.}
\label{fig:appendix_interacting_transport_quasiperiodic_sqrt3}
\end{figure*}

In this work, we use the wavenumber $k = \sqrt{2}$ as the default for calculations involving the quasiperiodic potential. 
A natural question is whether this particular choice is in any way ``special'' and influences the final results.
To address this, in Fig.~\ref{fig:appendix_interacting_transport_quasiperiodic_sqrt3}, we present results for the Frobenius norm of the correction to the charge transport capacity operator, $\|\hat M^{(1)}_\alpha\|_F$, in the interacting quasiperiodic model using a different wavenumber, $k = \sqrt{3}$. 
These calculations are performed for various values of $h$, both below and above the transition point, and system sizes up to $L = 80$.

We find that the qualitative behavior seen in the $k = \sqrt{2}$ case, described in Sec.~\ref{sec:quasiperiodic_interacting_analysis}, persists in the $k = \sqrt{3}$ case as well.
This consistency supports the idea that the choice of irrational wavenumber $k$ does not qualitatively affect the key quantities studied in this work.

Changing the wavenumber, however, does shift which lattice sizes happen to be very close to commensurate with the underlying potential.
We note the pronounced peaks at $L = 40$ and $L = 70$, where the $L = 70$ peak is absent from the $k = \sqrt{2}$ case.
These peaks can be attributed to the same mechanism discussed in Sec.~\ref{sec:liom_results_quasiperiodic}.
Specifically, $(L+1)\times k = 41 \times \sqrt{3} \approx 71$ and $71 \times \sqrt{3} \approx 123$, both of which are close to integers.
This supports our interpretation that such peaks arise from near-commensurability of the quasiperiodic potential at these specific system sizes.

We also note that the use of the golden ratio is the most common in the literature.
In this work, we separately calculated quantities relating to the quasiperiodic model with the wavenumber given by the golden ratio, and we do not observe any differences in our conclusions.

\section{Finite-Size Scaling in the Quasiperiodic Model}\label{app:quasiperiodic_collapse}

\begin{figure}[t]
    \centering
    \includegraphics[width=\columnwidth]{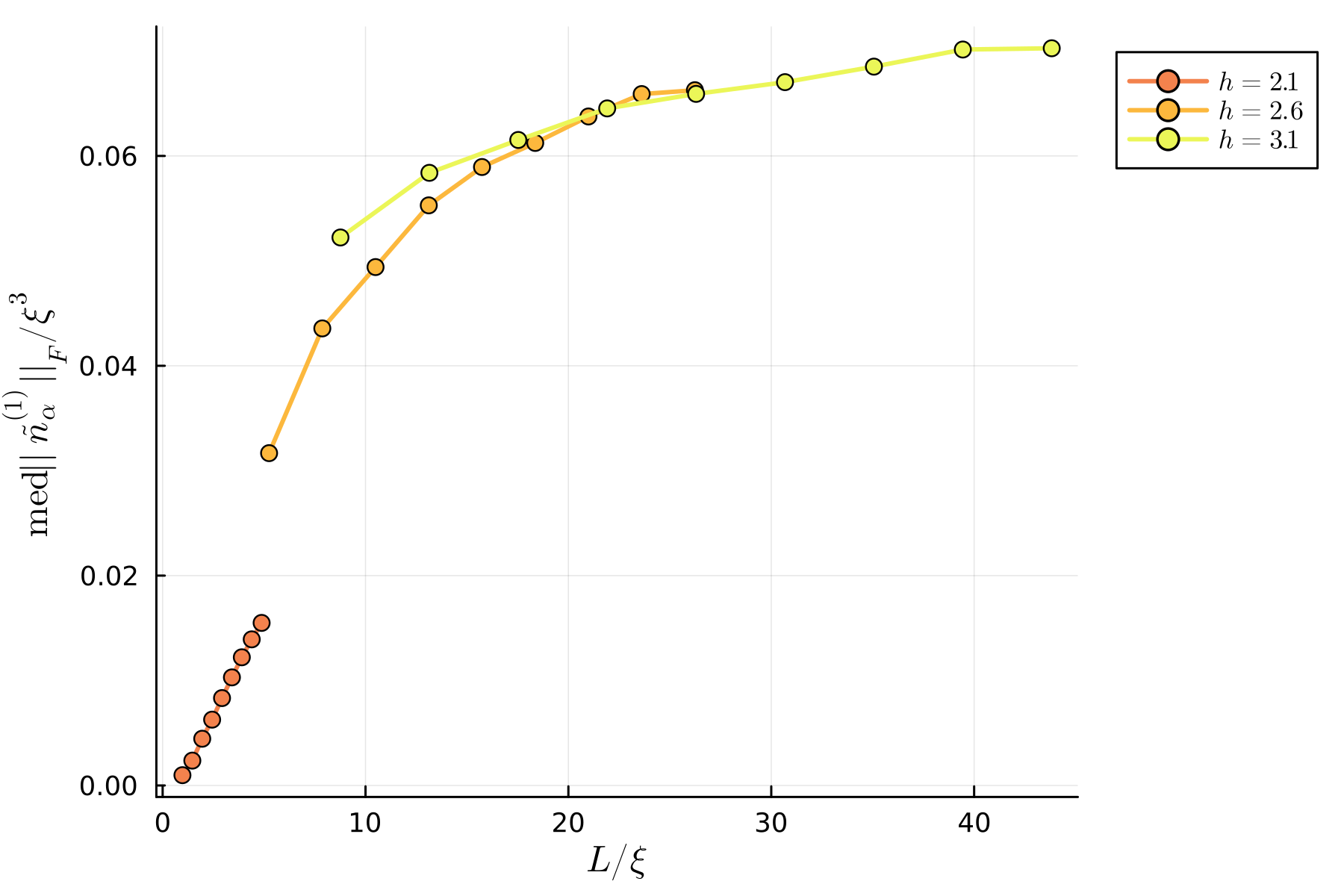}
    \caption{Same data as in Fig.~\ref{fig:liom_quasiperiodic_figure}(e) in the localized regime $h > 2$, but with the system size on the x-axis scaled by the localization length $\xi=1/\log(h/2)$, and the y-axis scaled by a phenomenological scaling of $\xi^{-3}$.}
    \label{fig:collapsed_LIOM_correction_quasiperiodic}
\end{figure}

\begin{figure}[t]
    \centering
    \includegraphics[width=\columnwidth]{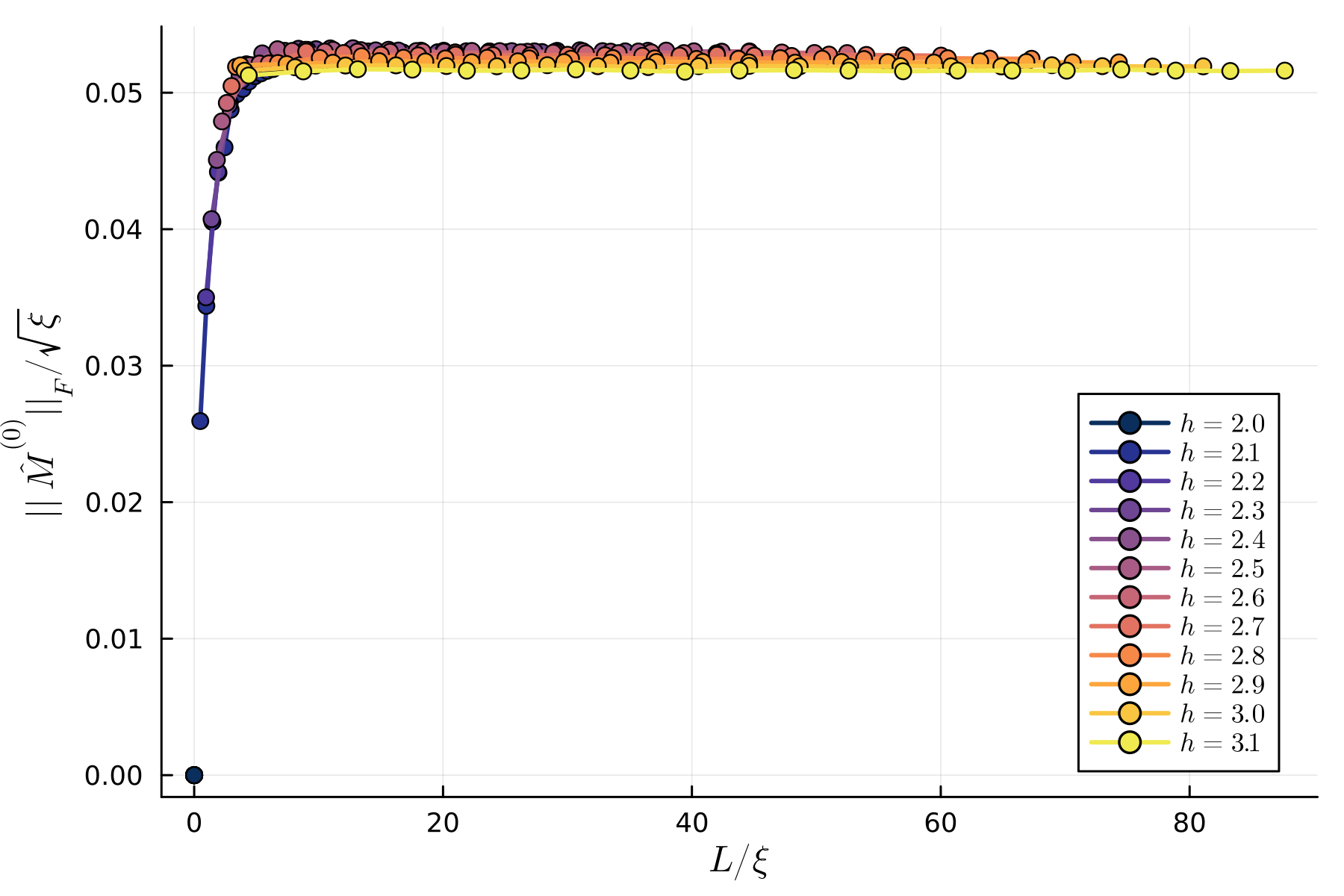}
    \caption{
    Median of $\|\hat M^{(0)}\|_F$ in the quasiperiodic model, with the x-axis scaled by the localization length $\xi=1/\log(h/2)$, and the y-axis scaled roughly by $\sqrt{\xi}$ (which is a phenomenologically expected scaling).
    The data is the same as in Fig.~12(e), showing only the localized regime $h > 2$, with more added $h$ values for more thorough scaling collapse analysis, which is easy to do in the non-interacting case.
    The collapse of the curves suggests the finite-size scaling $\mathrm{med}(\|\hat M^{(0)}\|_F)\sim \sqrt{\xi}f(L/\xi)$.}
\label{fig:noninteracting_transport_collapsed_quasiperiodic}
\end{figure}

\begin{figure}[t]
    \centering
\includegraphics[width=\columnwidth]{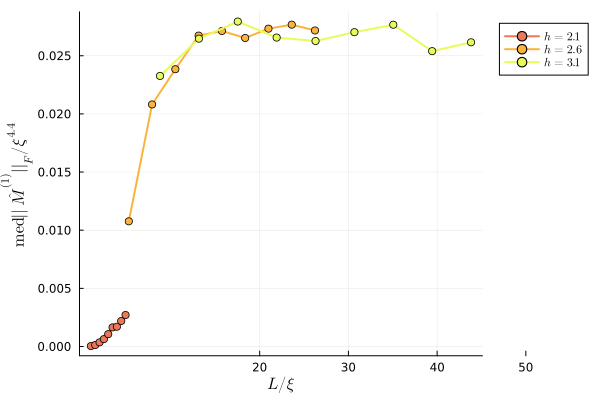}
    \caption{
    Same data of $\mathrm{med}\|\hat M^{(1)}\|_F$ as in Fig.~\ref{fig:quasiperiodic_interacting_transport_figure}(e), but with the x-axis scaled by the localization length $\xi=1/\log(h/2)$, showing data only in the localized regime $h > 2$ of the non-interacting starting model. The y-axis is scaled by a phenomenological scaling $\xi^{4.4}$.}
    \label{fig:collapsed_quasiperiodic_first_order_transport}
\end{figure}

In this section, we briefly show a finite-size scaling analysis for quantities relating to the quasiperiodic Aubry-André model, specifically in the localized regime, similarly to the analysis in the Anderson model in the main text.
The analysis is easier in the Aubry-André model, as it admits an exact formula for the localization length $\xi$ \cite{aubry_analyticity_1980}:
\begin{equation}
    \xi=1/\log(2/h).
\end{equation}
First, we look at the LIOM corrections in the quasiperiodic model, and wish to collapse the $\mathrm{med}\|\tilde n^{(1)}\|_F$ curves, just like in the case with the LIOM corrections of the Anderson model. 
We scale the x-axis by $\xi$, and phenomenologically scale the y-axis by a power of $\xi$.
The results of this collapse are shown in Fig.~\ref{fig:collapsed_LIOM_correction_quasiperiodic}.

We next consider the noninteracting charge transport capacity $\|\hat M^{(0)}\|_F$.
Inspired by the case for the Anderson model, a natural finite-size scaling ansatz to assume is:
\begin{equation}
    \mathrm{med}\|\hat M^{(0)}\|_F\sim\sqrt{\xi} f(L/\xi).
\end{equation}
We show the collapse of the curves using this ansatz in Fig.~\ref{fig:noninteracting_transport_collapsed_quasiperiodic}, supporting the ansatz and illustrating the finite-size effects.

Finally, we study the interaction correction to the charge transport capacity $\hat M^{(1)}$ in the localized phase While we do not have as much data as for $\hat M^{(0)}$, we can still attempt to 
collapse the $\mathrm{med}\| \hat M^{(1)}\|_F$ curves, just like in the case with the charge transport capacity corrections of the Anderson model-- see Fig.~\ref{fig:collapsed_quasiperiodic_first_order_transport} for the results.

\bibliography{references,morereferences}

\end{document}